\documentclass[sigconf, nonacm]{acmart}

\usepackage[noend]{algorithmic}
\usepackage{array,multirow}
\usepackage{xspace}
\usepackage{pifont}
\usepackage{adjustbox}
\usepackage{xcolor}
\usepackage{colortbl}
\usepackage{graphicx}
\usepackage{arydshln}
\usepackage[small]{subfigure}
\usepackage{bm}
\usepackage{amsmath}
\usepackage{makecell}
\usepackage{url}
\usepackage{stackengine}
\usepackage[normalem]{ulem}

\newcommand{\vsshrink}{\vspace{-0.3em}}

\newcommand{\vlshrink}{\vspace{-1.2em}}

\newcommand{\sstitle}[1]{\vspace{0.6ex}\noindent\underline{\textbf{#1}}}

\newcommand{\Card}{{\sf C}\xspace}
\newcommand{\TCard}{\widetilde{\sf C}\xspace}
\newcommand{\CE}{{\sf CardEst}\xspace}
\newcommand{\LAT}{{\sf Latency}\xspace}
\newcommand{\PC}{{\sf PlanCost}\xspace}
\newcommand{\PV}{{\sf PlanVal}\xspace}

\newcommand{\PRR}{{\sf PlanEmb}\xspace}
\newcommand{\CP}{{\sf CmpPlan}\xspace}
\newcommand{\WP}{{\rm Wins}\xspace}
\newcommand{\RWP}{{\rm RandomWins}\xspace}

\newcommand{\ltr}{{\em learning-to-rank}\xspace}

\newcommand{\eat}[1]{}

\newcommand{\rzhu}[1]{\begin{color}{purple}#1\end{color}}

\newcommand{\highlight}[1]{\textcolor{blue}{#1}}

\newcommand{\revise}[1]{\begin{color}{blue}{#1}\end{color}}
\newcommand{\delete}[1]{{\begin{color}{red}{#1}\end{color}}}

\newcommand{\rao}{\textsf{Lero}\xspace}

\mathchardef\mhyphen="2D

\newtheorem{proposition}{Proposition}

\DeclareMathOperator*{\argmax}{arg\,max}
\DeclareMathOperator*{\argmin}{arg\,min}

\newcommand{\pr}[2][]{\operatorname*{Pr}_{#1}\hspace{-0.06cm}\left[#2\right]\xspace}

\newcommand{\ep}[2][]{\mathbb{E}_{#1}\hspace{-0.06cm}\left[#2\right]\xspace}

\newcommand{\indicator}[1]{\mathbb{I}_{#1}\xspace}

\newcommand{\squishlist}{
	\begin{list}{$\bullet$}{
		\setlength{\itemsep}{0pt}
		\setlength{\parsep}{3pt}
		\setlength{\topsep}{3pt}
		\setlength{\partopsep}{0pt}
		\setlength{\leftmargin}{1.0em}
		\setlength{\labelwidth}{1em}
		\setlength{\labelsep}{0.5em}
   }
}

\newcommand{\squishenum}{
	
	\begin{list}{\usecounter{scount}}{
		\setlength{\itemsep}{0pt}
		\setlength{\parsep}{3pt}
		\setlength{\topsep}{3pt}
		\setlength{\partopsep}{0pt}
		\setlength{\leftmargin}{1.2em}
		\setlength{\labelwidth}{1em}
		\setlength{\labelsep}{0.5em}
	}
}

\newcommand{\squishend}{
	\end{list}
}

\definecolor{mygrey}{RGB}{230,230,240}
\definecolor{myblue}{RGB}{175, 238, 235}

\allowdisplaybreaks

\newcommand\vldbdoi{XX.XX/XXX.XX}
\newcommand\vldbpages{XXX-XXX}
\newcommand\vldbvolume{16}
\newcommand\vldbissue{6}
\newcommand\vldbyear{2023}
\newcommand\vldbauthors{Rong Zhu, Wei Chen, Bolin Ding, Xingguang Chen, Andreas Pfadler, Ziniu Wu, Jingren Zhou}
\newcommand\vldbtitle{\shorttitle} 
\newcommand\vldbavailabilityurl{https://github.com/Blondig/Lero-on-PostgreSQL}
\newcommand\vldbpagestyle{empty} 

\begin{document}

\title{\rao: A Learning-to-Rank Query Optimizer}
%    \title{\rao: Learning to Compare Plans in Query Optimizer} \vspace{-0.5em} 
% \author{Rong Zhu$^{1, \#}$,  Wei Chen$^{1, \#}$, Bolin Ding$^1$,  Xingguang Chen$^{1, 2}$, \break Andreas Pfadler$^1$, Ziniu \!Wu$^{3}$, Jingren Zhou$^{1, *}$}
% %	\vspace{0.5em}
% 	\affiliation{%
% 	\institution{\LARGE{\textit{$^1$Alibaba Group}, \textit{$^2$The Chinese University of Hong Kong}, \textit{$^3$Massachusetts Institute of Technology}} \\
% 	\textsf{$^1$\{red.zr, wickeychen.cw, bolin.ding, andreaswernerrober, jingren.zhou\}@alibaba-inc.com} \\
% 	\textsf{$^2$xgchen@link.cuhk.edu.hk} \hspace{1em} \textsf{$^3$ziniuw@mit.edu} 
% 	}
% 	}

\author{Rong Zhu$^{\#}$, Wei Chen$^{\#}$}
\affiliation{%
  \institution{Alibaba Group}
  \city{Hangzhou}
  \state{China}
}
\email{{red.zr, wickeychen.cw}@alibaba-inc.com}

% \author{Wei Chen}
% \affiliation{%
%   \institution{Alibaba Group}
%   \city{Hangzhou}
%   \state{China}
% }
% \email{wickeychen.cw@alibaba-inc.com}

\author{Bolin Ding}
\affiliation{%
  \institution{Alibaba Group}
  \city{Hangzhou}
  \state{China}
}
\email{bolin.ding@alibaba-inc.com}

\author{Xingguang Chen}
\affiliation{%
  \institution{Alibaba Group, The Chinese University of Hong Kong}
  \city{Hangzhou}
  \state{China}
}
\email{xgchen@link.cuhk.edu.hk}

\author{Andreas Pfadler}
\affiliation{%
  \institution{Alibaba Group}
  \city{Hangzhou}
  \state{China}
}
\email{andreaswernerrober@alibaba-inc.com}

\author{Ziniu Wu}
\affiliation{%
  \institution{Massachusetts Institute of Technology}
  \city{MA}
  \state{USA}
}
\email{ziniuw@mit.edu}

\author{Jingren Zhou$^{*}$}
\affiliation{%
  \institution{Alibaba Group}
  \city{Hangzhou}
  \state{China}
}
\email{jingren.zhou@alibaba-inc.com}

\begin{abstract}
% Query optimizer (QO) is the core part, as well as the most challenging problem, in DBMS. Recently, a variety of works try to apply machine learning (ML) techniques to learn end-to-end QO systems. Although they use advanced ML models, rather than estimated cost, to approach the exact latency of plans, their performance is still not fully favorable. The inherent hardness of the latency prediction problem gives rise to lots of limitations to existing learned QO systems, including but not limited to unstable performance, high training cost and low update speed.
% 
%\ziniu{The title is a bit misleading. Learning to rank plans in QO is a studied problem~\cite{ma2020active, li2012robust,ding2018plan,marcus2019plan}, but there is no learned end-to-end QO that is designed based on rank plans. We should emphasize that in the title, something like: ``Auncel: a plan-rank-based robust query optimizer''}
A recent line of works apply machine learning techniques to assist or rebuild cost-based query optimizers in DBMS. While exhibiting superiority in some benchmarks, their deficiencies, e.g., unstable performance, high training cost, and slow model updating, stem from the inherent hardness of predicting the cost or latency of execution plans using machine learning models.
In this paper, we introduce a {\em \underline{le}arning-to-\underline{r}ank} query \underline{o}ptimizer, called \rao, which builds on top of a native query optimizer and continuously {learns} to improve the optimization performance. The key observation is that the relative order or {\em rank} of plans, rather than the exact cost or latency, is sufficient for query optimization. \rao employs a {\em pairwise} approach to train a classifier to compare any two plans and tell which one is better. Such a binary classification task is much easier than the regression task to predict the cost or latency, in terms of model efficiency and accuracy. Rather than building a learned optimizer from scratch, \rao is designed to leverage decades of wisdom of databases and improve the native query optimizer. With its non-intrusive design, \rao can be implemented on top of any existing DBMS with {minimal} integration efforts. 
% In such way, \rao could keep track of data changes while does not disturb other services. 
% 
We implement \rao and demonstrate its outstanding performance using PostgreSQL. In our experiments, \rao achieves near optimal performance on several benchmarks. 
It reduces the plan execution time of the native optimizer in PostgreSQL by up to $70\%$ and other learned query optimizers by up to $37\%$. Meanwhile, \rao continuously learns and automatically adapts to query workloads and changes in data.
% 
% In this paper, we pave a new way for learned QO. We witness that the relative order (or \emph{rank}) of plans suffices to QO. Learning the rough rank scores is much easier than the unique latency value. To this end, we design \rao, a new learned QO system following the rank-based paradigm. \rao could be easily deployed on any existing QO through system provided interfaces but significantly enhances its performance. Specifically, for each query, \rao guides the existing QO to generate a number of diversified, but possible better, plans by an effective cardinality guided policy. A rank score model is then applied to select the best plan for execution. To train this model, we propose an automatic pairwise learning method to learn adaptive rank scores that maximally preserve the relative order of plans. In background, an experience collector runs candidate plans on idle workers to collect training data and update the model periodically. In such way, \rao could keep track of data changes while does not disturb other services. Extensive experiments exhibit that \rao attains near optimal performance on several benchmarks. It outperforms traditional open-source system by up to $3.3 \times$, sophisticated commercial system by up to $2.9\times$ and state-of-the-art learned QO system by up to $1.6\times$. Meanwhile, it adapts much faster to dynamic data and/or query workload.
\end{abstract}

\maketitle

\pagestyle{\vldbpagestyle}
\begingroup\small\noindent\raggedright\textbf{PVLDB Reference Format:}\\
\vldbauthors. \vldbtitle. PVLDB, \vldbvolume(\vldbissue): \vldbpages, \vldbyear.\\
\href{https://doi.org/\vldbdoi}{doi:\vldbdoi}
\endgroup

\begingroup
\renewcommand\thefootnote{}\footnote{
\noindent
$\#$ Equal contribution. $*$ Corresponding author. \\
\rule{\linewidth}{0.2pt}
This work is licensed under the Creative Commons BY-NC-ND 4.0 International License. Visit \url{https://creativecommons.org/licenses/by-nc-nd/4.0/} to view a copy of this license. For any use beyond those covered by this license, obtain permission by emailing \href{mailto:info@vldb.org}{info@vldb.org}. Copyright is held by the owner/author(s). Publication rights licensed to the VLDB Endowment. \\
\raggedright Proceedings of the VLDB Endowment, Vol. \vldbvolume, No. \vldbissue\ %
ISSN 2150-8097. \\
\href{https://doi.org/\vldbdoi}{doi:\vldbdoi} \\
}\addtocounter{footnote}{-1}\endgroup

\ifdefempty{\vldbavailabilityurl}{}{
\vspace{.3cm}
\begingroup\small\noindent\raggedright\textbf{PVLDB Artifact Availability:}\\
The source code, data, and/or other artifacts have been made available at \url{https://github.com/Blondig/Lero-on-PostgreSQL}.
\endgroup
}

% !TeX spellcheck = en_US

%\newpage

% \begin{table}[t]
%     \centering
%     \caption{Quality of query optimizers on benchmarks.}
%     \vspace{-1.0em}
%     \scalebox{0.87}
%     {
%     \begin{tabular}{|c|c|ccc|}
%     \hline
%     \rowcolor{mygrey}
%      & & \multicolumn{3}{c|}{\text{\textsf{Execution Time (h)}}} \\ \cline{3-5} 
%      \rowcolor{mygrey}
%      \multirow{-2}{*}{\text{\textsf{System}}} & \multirow{-2}{*}{\text{\textsf{Optimizer Category}}} & \text{\textsf{STATS}} & \text{\textsf{IMDB}} & \text{\textsf{TPC-H}} \\ \hline
     
%      {PostgreSQL} & {Traditional cost-based} & {20.19} & {1.15} & {0.94} \\     \hline
%      {Commercial DBMS A} & {Traditional cost-based} & {17.16} & {1.01} & {1.61} \\     \hline
%      {PostgresSQL with Bao} & {Cost-based learning} & {15.32} & {0.47} & {1.17} \\ \hline
%      {PostgresSQL with \rao} & {Learning-to-rank} & {\bf 11.32} & {\bf 0.35} & {\bf 0.74} \\ \hline 
%     \multicolumn{2}{|c|}{\bf Optimal} & \large \bf 10.73 & \large \bf 0.19 & \large \bf 0.72 \\ \hline
%     \end{tabular}
%     }
%     \label{tab:exp-first}
%     \vspace{-2.1em}
% \end{table}

%\vspace{-1em}
\section{Introduction}
\label{sec: intro}

Query optimizer plays one of the most significant roles in databases. It aims to select an efficient execution plan for each query. Improving its performance has been a longstanding problem. 
% in the field of database systems.
% 
Traditional {\em cost-based} query optimizers~\cite{selinger1989access} find the plan with the minimum estimated {\em cost}, which is a proxy of execution latency or other user-specified metrics about resource consumption. Such cost models contain various formulas to approximate the actual execution latency, whose magic constant numbers are exhaustively and extensively tuned based on engineering practice.
Recent works~\cite{marcus2019neo,yang2022balsa,marcus2021bao} refine traditional cost models and plan enumeration algorithms with machine learning techniques. Although some progress has been made, they still suffer from deficiencies caused by the intrinsically difficult latency prediction problem.
% (see Section~\ref{sec: intro-challenge}).

In this paper, we propose \rao, a \ltr query optimizer which features a new lightweight {\em pairwise} machine learning model for query optimization. \rao adopts a non-intrusive design, with minimal modification to the existing system components in DBMS. Instead of building from scratch, \rao is designed to leverage decades of wisdom of query optimizers without extra, potentially significant, cold-start learning costs. \eat{By employing \ltr paradigm, }\rao can quickly improve quality of query optimization by judiciously exploring different optimization opportunities and {\em learning to rank} them more accurately.

% 
% (see Section~\ref{sec: intro-solution}).
% 
%Table~\ref{tab:exp-first} gives an overview of experimental results on the comparison between \rao and existing systems.
%\highlight{Needs to summarize the results. How to calculate the optimal?}

% \highlight{
% \% Section 1.1: Existing Query Optimizer: Status and Challenges \\
% 1. Traditional query optimizer: enumerating all plans by dynamic programming, compare their goodness using a cost model, the cost model using magic number(hyper-parameters) + experience-driven rules on inaccurate estimated cardinality to approach the execution latency --> leading high errors \\
% 2. Later, learned query optimizers, including Neo, Balsa and Bao, using ML models to replace cost model to approach the latency. But still many problems as they try to solve a very hard problem (latency prediction). \\
% }

%\subsection{Costing and Selecting Plans: from Human Experience to Machine Learning Models}

\vspace{-1em}
\subsection{From Heuristic-Based Costing Models to Machine Learning Models}
\label{sec: intro-challenge}
% \sstitle{Traditional Query Optimizer.}
% The pipeline of traditional cost-based query optimizers~\cite{selinger1989access} is illustrated in Figure~\ref{fig: CostArch}(a), with three major components: 
Traditional cost-based query optimizers~\cite{selinger1989access} have three major components: 
{\em cardinality estimator}, {\em cost model}, and {\em plan enumerator}. For an input query $Q$, cardinality estimator can be invoked to estimate the {\em cardinality}, i.e., the number of tuples in the output, for each sub-query of $Q$. The {\em cost} of a plan $P$ (with physical operators, e.g., {merge join} and {hash join}) for the query $Q$ is a proxy of latency or other user-specified metrics regarding the efficiency of executing $P$. The cost model $\PC(P)$ estimates $P$'s cost and is usually a function of estimated cardinalities of $Q$'s sub-queries. The plan enumerator considers valid plans of $Q$ in its search space and returns the one with the minimum estimated cost for execution.

Various heuristics were essential in developing these components. For example, independence between attributes across tables is assumed and utilized for estimating cardinalities of joins of multiple tables~\cite{tzoumas2011lightweight, leis2015good}. Magic constant numbers are prevalent in cost models. They are often calibrated and tuned over years to ensure that the estimated cost matches the plan's performance well empirically, under certain system and hardware configurations though. It is realized that such heuristics are not always reliable for varying data distributions or system configurations. As a result, cost models may produce significant errors and the plan generated from the traditional query optimizer may have poor quality~\cite{tzoumas2011lightweight, leis2015good, han2021CEbenchmark, DDH08identifying}.  

% It applies some exhaustively tuned rules using estimated cardinality and some magic numbers to derive a proxy of the plan's quality, e.g., its execution latency, denoted as $\LAT(P)$.  
% The simple statistical models and unrealistic independence assumptions for cardinality estimation~\cite{tzoumas2011lightweight, leis2015good} may produce huge errors. When coupled with the inaccurate, experience-driven cost model, the generated plan of traditional query optimizer often has poor quality~\cite{leis2015good, han2021CEbenchmark}.

\eat{
\begin{figure}[!t]
	\includegraphics[width=0.9\linewidth]{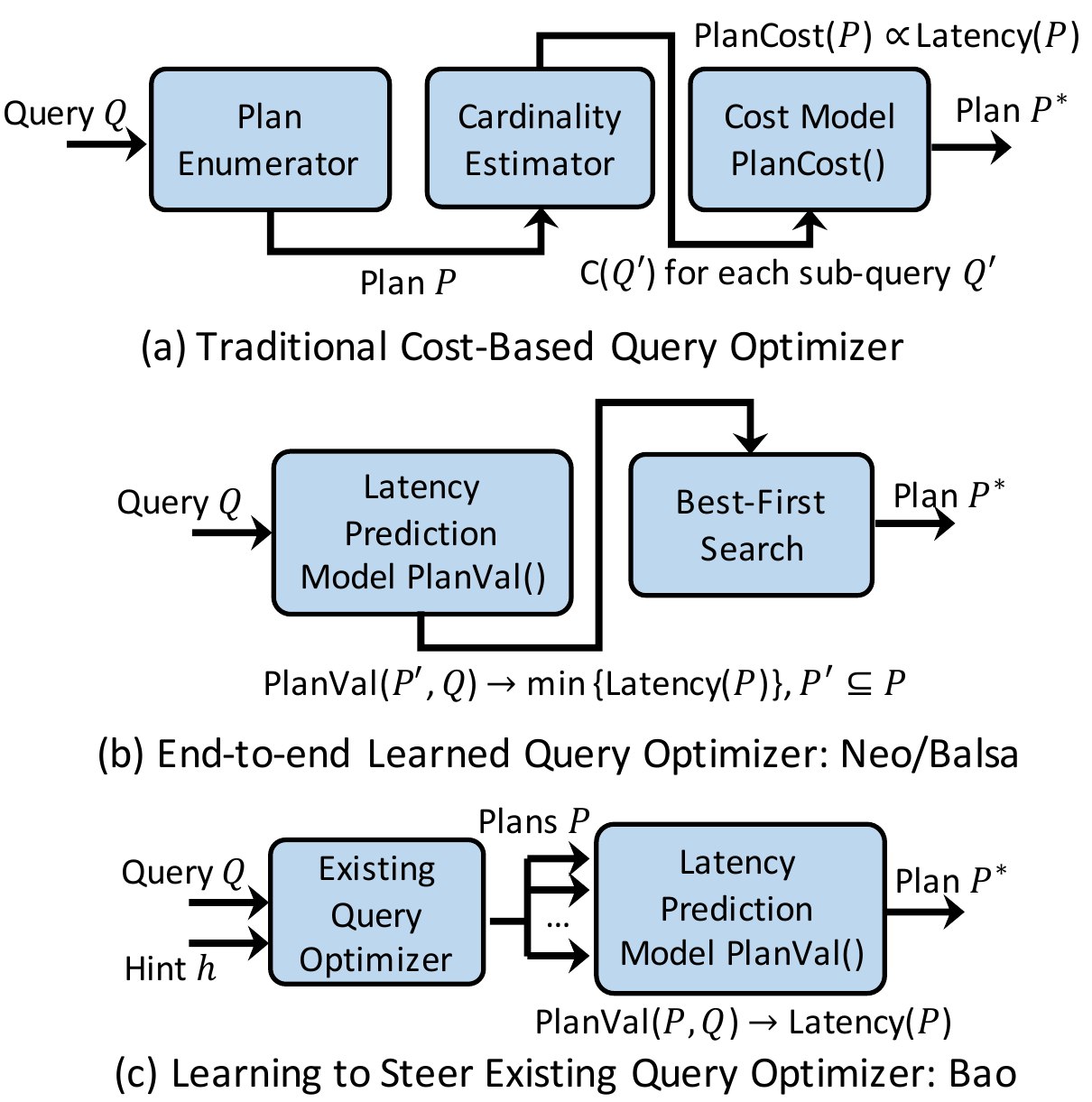}
	\vspace{-1em}
	\caption{System architecture of traditional and existing learned query optimizer.}
	\label{fig: CostArch}
	\vspace{-1.3em}
\end{figure}
}

It is a natural idea to develop machine learning models to replace traditional cardinality estimators, heuristic-based cost models, and plan enumerators.
For example, there are works on learning cardinality estimators (refer to \cite{han2021CEbenchmark, zhu2022learned} for a survey).
For learning cost models, different model parameters, instead of {\em fixed} magic constant numbers in traditional cost models, can be learned from different datasets and workloads to enable finer-grained characterization of various data distributions and system configurations, thus providing instance-level optimization of each query.
% 
% However, learning to replace one or more components, such as a learned cardinality estimator (refer to \cite{han2021CEbenchmark} for a survey), only partially solves the problem of query optimization.
% 
A recent line of works on {\em learned query optimizers} are built upon the above idea and demonstrate some promising results~\cite{marcus2019neo,yang2022balsa,marcus2021bao}.

% \sstitle{Learned Query Optimizer.}
% Recently, ML techniques are broadly applied to aid the query optimizer. ML models enable more fine-grained characterization of the data distribution, thus providing instance-level optimization of each query. The representative learned query optimizers adopt one of the following two technique routines:
% 
{Neo}~\cite{marcus2019neo} and {Balsa}~\cite{yang2022balsa} provide end-to-end {\em learned} solutions for query optimization.
% Their paradigm is shown in Figure~\ref{fig: CostArch}(b).
A {\em plan value function} $\PV()$, inspired by the value networks in deep reinforcement learning, is learned to replace the traditional cost model $\PC$.
For a partial plan $P'$ of the query $Q$, the machine learning model $\PV(P', Q)$ predicts the minimum latency of a complete plan $P$ that contains $P'$ as a sub-plan, with statistics and patterns about the tables, predicates, and joins involved in $P'$ and $Q$ as {input features}.
With $\PV$, Neo and Balsa use their best-first search strategies to find the best plan with the minimum estimated latency.
% for execution.

% \ssstitle{End-to-end learned query optimizers.}
% \textbf{Neo}~\cite{marcus2019neo} and \textbf{Balsa}~\cite{yang2022balsa} provide end-to-end solutions. As shown in Figure~\ref{fig: CostArch}(b), a {\em plan value function} $\PV()$, inspired by the value networks in deep reinforcement learning, is learned to replace the traditional cost model $\PC()$. As an ML model, for a partial plan $P'$ of the query $Q$, $\PV(P', Q)$ predicts the minimum possible $\LAT(P)$ ({\em label}) of a complete plan $P$ that contains $P'$ as a sub-plan, and takes the tables, predicates, and joins in $P'$ and $Q$ as {\em input features}. Based on $\PV()$, Neo and Balsa apply different best-first search strategies to find the plan $P^{*}$ with the minimum estimated latency for execution.

% \ssstitle{Learning to steer existing query optimizers.}
{Bao}~\cite{marcus2021bao} learns to steer a native query optimizer\eat{instead of replacing all the components with machine learning models}. 
%
% As shown in Figure~\ref{fig: CostArch}(c), 
It tunes the native query optimizer with different sets of {\em hints} to generate a number of different candidate plans for the query $Q$. Each hint set forces/disables some operations, e.g., {index scan} or {hash join}, \eat{in the plan search space,} so the query optimizer may output a different, and possibly better, execution plan. 
% 
% For all the candidate plans of $Q$ generated with different hint sets, 
% 
Bao uses a machine learning model $\PV(P,Q)$ to estimate the quality (e.g., latency) of each candidate plan $P$ and select the best candidate for execution. Meanwhile, Bao periodically updates its model parameters based on execution statistics using Thompson sampling~\cite{thompson1933likelihood}, as solving a contextual multi-armed bandits problem~\cite{zhou2015survey} to minimize the overall performance regret.
% 
% Bao uses an ML model $\PV(P,Q)$, which estimates the quality (e.g., latency) of a complete plan $P$, to choose the best plan $P^{*}$ from those candidates for execution. Meanwhile, Bao periodically updates the model parameters of $\PV(P,Q)$ based on execution experience using Thompson sampling~\cite{thompson1933likelihood}, as solving a contextual multi-armed bandits problem~\cite{zhou2015survey} to minimize the overall performance regret.

Although learned query optimizers exhibit superiority than traditional ones in some applications~\cite{negi2021steering}, their performance is far from satisfactory. They still suffer from three major deficiencies:

\squishlist
% \begin{itemize}
    \item {\bf Unstable performance.} These learned models are easily to produce inaccurate latency estimates, which leads to sub-optimal plans. Sometimes, the performance regression is very significant. As shown in Section~\ref{sec: eval-gain}, their performance could even be worse than PostgreSQL's native query optimizer on TPC-H benchmark.
    \item {\bf High learning cost.} The cost of learning a query optimizer includes both the cost of exploring and executing different query plans and the cost of model training.
    % Existing learned query optimizers may need a very long time for model training: 
    Some models~\cite{marcus2019neo,yang2022balsa} require tens to hundreds of training iterations to coverage, and {execute all newly explored plans in each iteration}.
    % \sout{each iteration to execute {all} newly explored plans}.
    For complex datasets, this would consume several days to even weeks. 
    \item {\bf Slow model updating.} Existing learned optimizers need to update their latency prediction models to fit dynamic data, which is very challenging especially considering that the latency of the same plan may vary on dynamic data. Thus, model updating in this case requires huge efforts (training data and cost), and easily leads to performance regression, as shown in Section~\ref{sec: eval-dynamic}.
    % hey can Some learned query optimizers~\cite{marcus2019neo,yang2022balsa} can not be incrementally maintained. Even if some can be updated~\cite{marcus2021bao}, they require large size of training data and long time to reflect the changes in data and query workload. This 
    % \highlight{
    % \item {\bf Slow model updating.} Some learned query optimizers~\cite{marcus2019neo,yang2022balsa} can not be incrementally maintained. Even if some can be updated~\cite{marcus2021bao}, they require large size of training data and long time to reflect the changes in data and query workload. This easily leads to performance regression, as shown in Section~\ref{sec: eval-dynamic}. 
    % }
    % 
    % \item {\bf Slow model updating.} Such learned query optimizers cannot quickly update their models to fit the latest data. Some~\cite{marcus2019neo,yang2022balsa}\eat{Both {Neo} and {Balsa}} require to retrain the model when data changes. Others~\cite{marcus2021bao}\eat{Bao's model} can be incrementally updated, but not fast enough to keep up with the updated datasets. As shown in Section~\ref{sec: eval-dynamic}, it could cause significant performance regression on changing data. 
% \end{itemize}
\squishend

\eat{
\squishlist
% \begin{itemize}
    \item {\bf Unstable performance.} These learned models are easily to produce inaccurate latency estimates, which leads to sub-optimal plans. Sometimes, the performance regression is very significant. As shown in Section~\ref{sec: eval-gain}, their performance could even be worse than PostgreSQL's native query optimizer on TPC-H benchmark.

    \item {\bf High learning cost.} The cost of training a learned query optimizer includes both the cost of exploring and executing different query plans and the cost of model training itself.
    % Existing learned query optimizers may need a very long time for model training: 
    Some~\cite{marcus2019neo,yang2022balsa} require tens to hundreds of iterations for the trained models to coverage, and each iteration requires to execute the newly explored plans. For complex datasets, this would consume several days to even weeks.

    \item {\bf Slow model updating.} Such learned query optimizers cannot quickly update their models to fit the latest data. Some~\cite{marcus2019neo,yang2022balsa}\eat{Both {Neo} and {Balsa}} require to retrain the model when data changes. Others~\cite{marcus2021bao}\eat{Bao's model} can be incrementally updated, but not fast enough to keep up with the updated datasets. As shown in Section~\ref{sec: eval-dynamic}, it could cause significant performance regression on changing data. 
% \end{itemize}
\squishend
}

{We recognize that, the deficiencies of both traditional and newly proposed learned query optimizers stem from the notoriously difficult problem of predicting the execution latency or cost of a plan.} The exact execution latency depends on numerous factors~\cite{ReddyH05Analyzing, marcus2019plan, leis2015good, wu2013predicting, li2012robust}, e.g., underlying data distribution, workload patterns, and system environment.
Training such a prediction model is a costly operation, which requires collecting a large volume of training data, by executing {query plans} and measuring latency statistics, and a lengthy training process to explore the huge hypothesis space.
% The resulting hypothesis space is very large and too complex. 
% 

% There are also works on learning to replace other components, such as learned cardinality estimators (refer to \cite{han2021CEbenchmark, zhu2022learned} for a survey).
% only partially solves the problem of query optimization.
{
Moreover, cardinality and cost estimations are only partial factors for query optimization.
No matter whether with the traditional calibrated estimators/models or with the learned models trained on previous statistics, improving the estimations' accuracy does not necessarily lead to improvement in query optimization.
}

% No matter whether with the traditional calibrated cost models or with the learned models trained on previous statistics, it is difficult to characterize how inaccurate estimation/prediction affects the quality of the plan selected for execution (except some very loose worst-case upper bound \cite{MoerkotteNS09preventing}).
% 
% the models easily produce inaccurate results and may be very costly to train/update.

\sstitle{Is it really necessary to predict the latency?} 
We ask such a fundamental research question: {\em for the purpose of query optimization, do we really need to estimate/predict the execution latency (or any other performance-related metric) of every possible query plan?}
With the goal of {finding} the best execution plan, training a machine learning model to predict the exact latency (cost) is an overkill.

\vspace{-1em}
\subsection{A Learning-to-Rank Query Optimizer}
\label{sec: intro-solution}

In this paper, we introduce a {\em \underline{le}arning-to-\underline{r}ank} query \underline{o}ptimizer, called \rao. In essence, what we need for query optimization is a {\em learned} oracle that is able to \emph{rank} a set of candidate query plans with respect to their execution efficiency. 

Looking back, traditional cost models are essentially ``human learning'' models, whose parameters, i.e., magic constant numbers, have been tuned with decades of engineering efforts. Under the learning-to-rank paradigm, traditional and learned cost models can be regarded as {\em pointwise} approaches \cite{liu2009ltr}, which outputs an ordinal score (i.e., estimated cost) for each plan to rank them.
\rao adopts an effective pairwise learning-to-rank approach without discarding human wisdom in developing traditional cost models and query optimizers.
We summarize our contributions as follows.

\squishlist
% \begin{itemize}
    \item \rao applies \ltr paradigm to query optimization. Compared to previous proposals to learn cost models or cardinalities which are partial factors for query optimization, \rao directly ranks and learns to improve plans’ quality. \rao is more effective in selecting plans with good quality.

    \item {\rao adopts a {\em pairwise} approach which learns to compare two plans and predict the better one. Compared to other {\em pointwise} approaches, training such a binary classifier is often easier\cite{JohannesF2011Preference}. \rao has significantly lower training costs, requiring fewer training samples, and much less time to train. 
    A pairwise comparison model is also used for index tuning in \cite{ding2019ai}. \rao implements a novel deep learning model which consists of a classifier over two plan embeddings with shared parameters. The model is able to capture the detailed structural properties for given plans and is proven more effective in the context of query optimization.}
    
    \eat{
    % \item \eat{\rao is the first to apply \ltr paradigm to query optimization.}
    % 
    \item \revise{\rao applies \ltr paradigm to query optimization, which learns to select the best plan for a query in two steps, plan {\em exploration} and {\em pairwise comparison}. Compared to other learned optimizers, \rao has significantly lower training costs, requiring fewer training samples, and much less time to train.}

    \item \revise{\rao learns to compare two plans and predict the better one.}
    \revise{As is demonstrated in preference learning literature \cite{JohannesF2011Preference} and database tuning tasks, e.g., index tuning \cite{ding2019ai}, training a binary classifier for pairwise comparisons is often easier than training regression models from the perspective of machine learning. Compared to the previous pairwise comparison model for index tuning in \cite{ding2019ai}, we need more structural features and a different model architecture for plan selection, which are more effective in comparing two plans under the same index configuration.}
    }
    
    % 
    % \rao adopts a {\em pairwise} approach which learns to compare two plans and predict the better one. 
    % 
    % Compared to previous proposals to {\em learn} cost models or cardinalities which are partial factors for query optimization, \rao directly ranks and learns to improve plans' quality. \rao is more effective in selecting plans with good quality.
    % 
    % Compared to other learned query optimizers, \rao has significantly lower training costs, requiring fewer training samples, and much less time to train.
    % 
    % \revise{Motivated by~\cite{ding2019ai}, training such a binary classifier is often easier than solving the regression problems in DBMS tasks. Compared to other learned query optimizers, \rao has significantly lower training costs, requiring fewer training samples, and much less time to train.}

    \item \rao is able to {\em explore} and {\em learn} new plan space (and new optimization opportunities) more effectively. Instead of using query-level hints as previous proposals~\cite{marcus2021bao}, \rao can be equipped with various strategies to adjust cardinality estimates at expression level to effectively consider diversified plans and prioritize exploring more promising ones.

    \item \rao \ {\em leverages} decades of wisdom of databases and query optimization, rather than building a brand new optimizer from scratch. \rao starts with the default behavior of the native query optimizer and gradually {\em learns} to improve. Therefore, \rao is guaranteed with a good initial quality and does not need a lengthy cold-start training process.

    \item \rao employs a {\em non-intrusive} design. It can be implemented on top of any existing DBMS and integrated seamlessly with the native query optimizer. To be specific, \rao takes advantage of public interfaces provided by most DBMSs and jointly works with the native query optimizer to improve optimization quality.
% \end{itemize}
\squishend

We implement \rao on PostgreSQL~\cite{LeroImp} and conduct extensive experiments 
% on several benchmarks 
to evaluate its efficiency and effectiveness. \rao reduces the execution time of the native query optimizer in PostgreSQL by up to $70\%$ and other learned query optimizers by up to $37\%$. Experiments show that \rao adapts faster with stable performance to various query workloads and dynamically changing data. 
% 
% Meanwhile, it is able to explore better query plans and eliminate the cold-start problem.
 
%\jrzhou{exp numbers and claims need to be refreshed.}

% The rest of this paper is organized as follows. 
\sstitle{Organization.} Section~\ref{sec: sys} outlines \rao's {\em pairwise} \ltr system architecture and key components. 
Section~\ref{sec: train} presents the design of \rao's comparator model and describes how to train and infer using the model.
Section~\ref{sec: policy} describes how \rao explores different plans and new optimization opportunities, for the purposes of both plan selection and model training.
% 
%Section~\ref{sec: impl} presents the implementation details on PostgreSQL. 
Extensions are discussed in Section~\ref{sec: discussion}.
We report detailed evaluation results in Section~\ref{sec: eval}. We cover related work in Section~\ref{sec: related} and conclude in Section~\ref{sec: con}.
% and Section~\ref{sec: con} concludes this paper.

% \note{\textbf{Part VIII: paper organization.} \\
% Section 2 reviews the background.
% Section 3 outlines the rank-preserving system.
% Section 4 introduces the key techniques.
% Section 5 describe the implementation details.
% Section 6 reports the evaluation results.
% Section 7 concludes this paper.
% }

\eat{
% the old version

% 
In essence, what we essentially need is just a {\em learned} oracle that \emph{ranks} a set of candidate plans for the query with respect to execution efficiency. In addition, instead of fully ranking all the possible plans, the relative order of plans suffices for query optimizers to select the best plan. To further improve training efficiency, we employ a \emph{pairwise} approach which learns to compare two plans and predict the better one. From the perspective of machine learning, training such a comparator is a binary classification problem which is often easier than regression problems~\cite{JohannesF2011Preference}. Together with a co-designed plan exploration strategy,
\rao demonstrates outstanding performance and efficiency with such a \emph{pairwise} model and requires minimal modification to the native query optimizer. 

\eat{
With the goal of searching the best execution plan, training a machine learning model to predict the exact latency (cost) is an overkill. What we essentially need is just an oracle that is able to compare two plans and thus rank a set of candidate plans for the query. From the perspective of machine learning, training such a comparator is a binary classification problem which is often easier than regression problems~\cite{JohannesF2011Preference}. Our solution \rao is built on such a \ltr paradigm with a co-designed plan exploration strategy and minimal modification to the native query optimizer. 
}

\eat{
\rzhu{Another version to replace after overskill:}

\highlight{Essentially, the {\em relative order}, or {\em rank}, of plans suffices or query optimizers to select the best plan. Our solution \rao is built on such a \ltr paradigm with a co-designed plan exploration strategy and minimal modification to the native query optimizer.}
}

We can replace the cost model used in traditional query optimizers with a {\em plan comparator}, namely, an oracle that compares two input plans and outputs which one is better (with lower cost or latency): when the costs of two plans need to be compared during plan enumeration (in dynamic programming or heuristic algorithms, e.g., the ones in \cite{MoerkotteN08dynamic}), the plan comparator can be invoked.
Conceptually, for $n$ candidate plans of a query, we can compare each plan with the rest $n-1$ using the plan comparator and the one that {\em wins} (asserted by the oracle to be better) the most times is the best plan.

% The learning-to-rank paradigm of \rao is based on the fact that the relative order of plans suffices for query optimizers to select the best plan. With very minor modification to most existing dynamic programming or heuristic algorithms (e.g., the ones in \cite{MoerkotteN08dynamic}), we can replace the cost model used in these algorithms with a {\em plan comparator}, namely, an oracle that takes two plans as the input and output which one is better (with lower cost or latency).

In \rao, we train a machine learning model $\CP$, called {\em comparator model} or {\em comparator} for short, to approximate the above oracle. At a very high level, $\CP(P_1, P_2)$ maps two plans of a query, $P_1$ and $P_2$, from the original feature space (with information such as tables/attributes involved in the query, join orders, and predicates) to a low-dimensional space, and compare the corresponding two low-dimensional vectors to decide whether $P_1$ or $P_2$ is better. The comparator $\CP$ is trained to fit a training dataset, with pairs of plans and labels indicating, for each pair, which plan is better (in terms of latency or other user-specified metric).

% A fully trained comparator $\CP$ may still make mistakes, and thus may not preserve the {\em transitivity}\footnote{$P_1$ is better than $P_2$ and $P_2$ is better than $P_3$ $\Rightarrow$ $P_1$ is better than $P_3$.}. In general, without transitivity, for a list of $n$ candidate plans of a query, we can compare each plan with the rest $n-1$ using $\CP$ and choose the one that {\em wins} (asserted by $\CP$ to be better in a comparison) the most times as the best plan for execution. In Section~\ref{sec: train}, We will introduce how to design $\CP$ so that the best plan can be picked in a more efficient and equivalent (in expectation) way.

In comparison to other possible approaches in \ltr paradigm, e.g., pointwise and listwise approaches~\cite{liu2009ltr}, our pairwise approach makes the best trade-off between model accuracy and learning efficiency. While it is too expensive to execute the complete list of valid plans for a given query and rank them by latency to obtain training data with labels, $\CP$ can be trained with only pairs of plans in a partial list.
A fully trained comparator $\CP$ may still make mistakes, and thus may not preserve the {\em transitivity}, i.e., $P_1$ is better than $P_2$ and $P_2$ is better than $P_3$ imply that $P_1$ is better than $P_3$. However, we can also use the number of wins (asserted by $\CP$) to choose the best plan from a candidate list. In Section~\ref{sec: train}, we will introduce how to design $\CP$ so that the best plan can be picked in a more efficient but equivalent way, and the model's output preserves transitivity.

\rao centers around the comparator model and another important component called {\em plan explorer}, as in Figure~\ref{fig: SysArch}. 
For a query, the plan explorer generates a list of $n$ candidate plans $P_1, \ldots, P_n$. While the best one chosen by the comparator is executed, the rest (or as many as possible) candidate plans are run on idle workers under in parallel to collect training data. In real-world distributed systems, idle workers are available for numerous reasons~\cite{kanet2000scheduling, chen2016revisiting, chaiken2008scope, negi2021steering}. 
% To satisfy both optimizing and training requirements, 
Our plan explorer, introduced in Section~\ref{sec: policy}, uses the cardinality estimator as the tuning knob: the output cardinality is purposely scaled up/down (or unchanged) before being fed into the cost model for the native query optimizer to generate different candidate plans. 

\eat{For a query, the plan explorer generates a list of $n$ candidate plans $P_1, \ldots, P_n$. While the best one chosen by the comparator is executed, the rest (or as many as possible) candidate plans are run on idle workers under the same system configuration in parallel to collect training data, in the form of ($P_i$, $P_j$; label--which one is better) for all pairs of executed plans. In real-world distributed systems, idle workers are available for numerous reasons~\cite{kanet2000scheduling, chen2016revisiting, chaiken2008scope, negi2021steering}. For cold start, in the initial stage, we can pre-train the comparator using the estimated costs of candidate plans (without executing them) generated from a sample workload of queries.
Our design turns waste into treasure: the comparator can be trained and continuously improved as more and more information is collected about the possibly dynamic data distribution and workloads by executing candidate plans, while normal database services are not disturbed.}

\eat{
Candidate plans generated by our plan explorer serve the purposes of both plan selection and model training, and thus they should: i) contain some truly good candidates (although we do not have to know who they are) and ii) be 
% \rzhu{not too long but}
sufficiently diversified so that the model could learn to distinguish between good and bad plans. To satisfy these two requirements, our plan explorer uses the cardinality estimator as the tuning knob: the output cardinality is purposely scaled up/down (or unchanged) before being fed into the cost model for the native query optimizer to generate different candidate plans.
Details about our plan exploration strategy will be introduced in Section~\ref{sec: policy}. One obvious advantage of our plan explorer is that the cardinality estimator is an essential component in almost all query optimizers, and thus the strategy and the implementation of our plan explorer can be easily migrated to different databases. Unlikely, {Bao}~\cite{marcus2021bao} relies on query hints as the tuning knob which may vary from one database to another, and requires significant manual efforts to select proper candidate hint sets for each DBMS.}

The design of \rao is system-agnostic. We implement it as a middleware on PostgreSQL. 
% by modifying only two hook functions in its native query optimizer. 
It is easy to deploy \rao in other DBMS, as the major modification to DBMS is on the cardinality estimator, which has system provided interfaces in most databases.

We conduct extensive experiments on several benchmarks to evaluate \rao. \rao largely outperforms the traditional query optimizer in PostgreSQL by up to $3.3\times$, a commercial DBMS by up to $2.9\times$, and existing learned query optimizers by up to $1.6\times$. It is demonstrated in our experiments that model training in \rao is more lightweight, and \rao adapts faster and performs more stably in dynamic data and query workloads. It is also shown that our plan explorer is more effective than alternate approaches.

\sstitle{Paper Organization.}
% 
% Our contributions can be summarized as follows:
% 
% 1) We propose a novel rank-based paradigm, which paves a new way for learned QO system.
% 
% 2) We design \rao, a new rank-based learned QO system with better accuracy, less training cost and faster update speed.
% 
% 3) We provide an implementation of \rao on top of PostgreSQL, which serves as a blueprint to deploy \rao into any DBMS.
% 
% 4) We comprehensively evaluate \rao on multiple benchmarks to demonstrate its superiority and effectiveness.
% 
%\smallskip
% \sstitle{Organization.}
% The rest of this paper is organized as follows.
% Section~\ref{sec:bkg} introduces the background on QO systems. 
Section~\ref{sec: sys} outlines \rao's system architecture under our \ltr paradigm. 
Section~\ref{sec: train} introduces the design of our comparator model, and how to train and infer using the model.
Section~\ref{sec: policy} introduces our plan explorer, for the purposes of both plan selection and model training.
Section~\ref{sec: impl} presents the implementation details on PostgreSQL. 
We report detailed evaluation results in Section~\ref{sec: eval}. Other related works are reviewed in Section~\ref{sec: related}. Section~\ref{sec: con} concludes this paper.
% and Section~\ref{sec: con} concludes this paper.

% \note{\textbf{Part VIII: paper organization.} \\
% Section 2 reviews the background.
% Section 3 outlines the rank-preserving system.
% Section 4 introduces the key techniques.
% Section 5 describe the implementation details.
% Section 6 reports the evaluation results.
% Section 7 concludes this paper.
% }
}
%\input{1-intro-new-rzhu.tex}
% \input{1-intro.tex}

% !TeX spellcheck = en_US
% \vspace{-1em}
\section{System Overview}
\label{sec: sys}
% 
% We introduce the paradigm and system architecture of 
% 
\rao is a \ltr query optimizer which continuously explores different query plans, observes their performance, and {\em learns} to {\em rank} them more accurately. \rao adopts a \emph{pairwise} approach whose objective is to predict which of two plans is more efficient. Compared with other \ltr approaches, e.g., pointwise and listwise~\cite{liu2009ltr}, \rao's pairwise approach makes the best trade-off between model accuracy and learning efficiency for our task. 

% \highlight{
% \rao is a \ltr query optimizer which continuously explores different query plans, observes their performance, and {\em learns} to {\em rank} them more accurately. Among all \ltr techniques, the \emph{pairwise} approach makes the the best trade-off between model accuracy and learning efficiency than e.g., pointwise and listwise approaches.~\cite{liu2009ltr}. As a result, \rao also adopts a \emph{pairwise} approach whose objective is to predict which of two plans is more efficient.
% }

\begin{figure}[t]
    \includegraphics[width=0.9\linewidth]{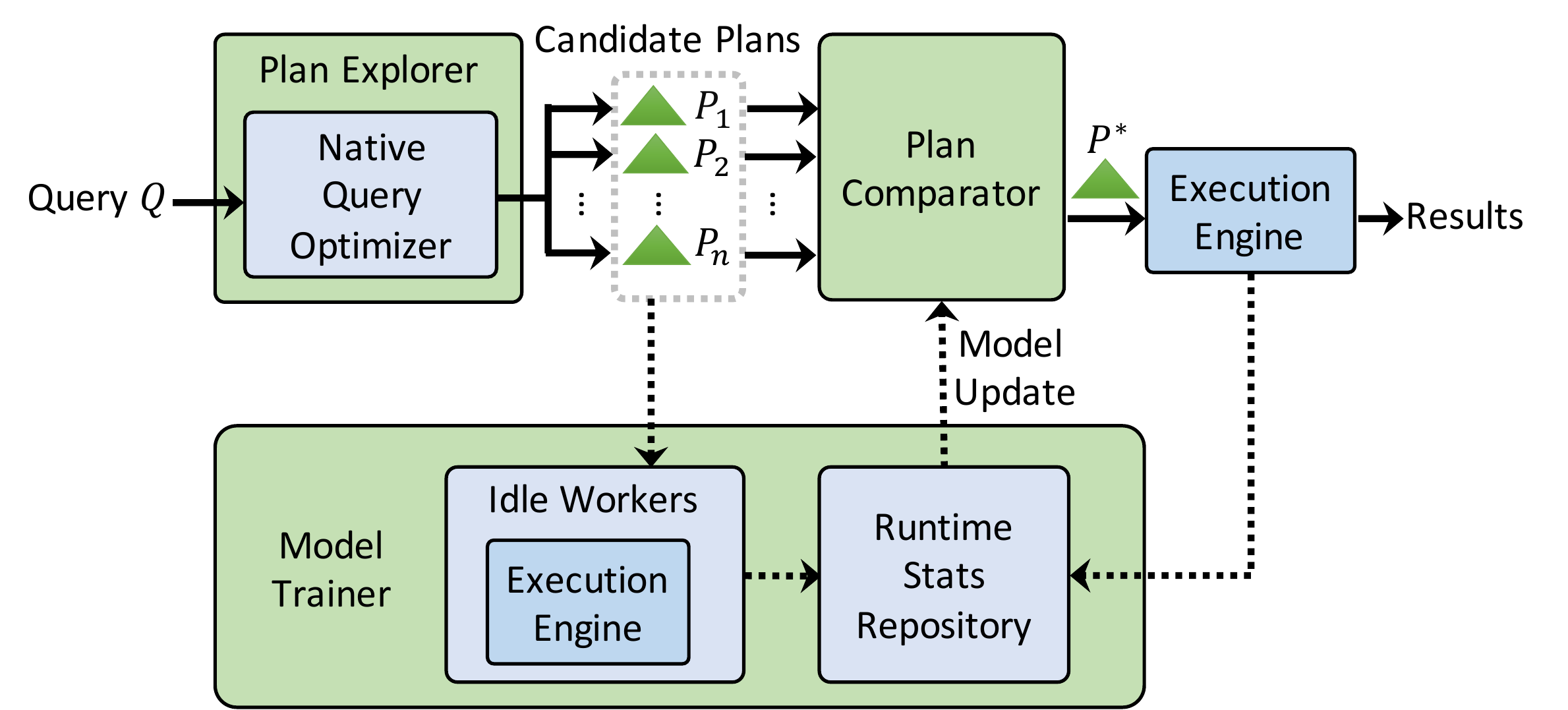}
	\vspace{-1em}
	\caption{System architecture of \rao.}
	\label{fig: SysArch}
	\vspace{-2em}
\end{figure}

Figure~\ref{fig: SysArch} shows the overall architecture of \rao. For each input query $Q$, the \textit{\textbf{plan explorer}} works with the native query optimizer to generate a number of potentially good and diversified {\em candidate plans} $P_1, P_2, \dots, P_n$. The \emph{pairwise} \textit{\textbf{plan comparator model}} $\CP$ is then invoked to select the best plan $P^{*}$ from the candidates for answering $Q$. In system background, the \textit{\textbf{model trainer}} executes other candidate plans\eat{ in queue} using idle workers whenever system resources become available, collects the latency information into the {\em runtime stats repository}, and continuously trains the comparator model $\CP$ and updates it periodically. Such design enables the comparator model to be continuously trained and become better over time without affecting normal database services.
\eat{
Our design turns waste into treasure: the comparator can be trained and continuously updated as more and more information is collected about the possibly dynamic data distribution and workloads by executing candidate plans, while normal database services are not disturbed.}
The three components are briefly introduced as follows.

% In this section, we present our new {\em learning-to-rank} paradigm, together with the architecture of \rao under this paradigm. 
% 
% \rzhu{Remove this subsection to merge into Section 1.2.}
% \section{Learning-to-Rank: A New Paradigm}
% \label{sec: sys-paradigm}

\sstitle{Plan Comparator Model.}
\eat{
A traditional query optimizer enumerates different plans of a query based on equivalence and transformation of relational algebra in certain order, and chooses the best plan based on their estimated costs. The cost model induces a full order for all plans; however, a binary relation induced by a comparator suffices for the purpose of query optimization. 
}
Formally, let $\CP(P_1, P_2)$ be an oracle comparing any two plans $P_1$ and $P_2$ of a query: 
%\vspace{-0.25em}
\begin{equation}\label{equ: cmp}
{\CP}(P_1, P_2) = 
  \begin{cases} 
   0   & \text{if } \LAT(P_1) < \LAT(P_2) \\
   1   & \text{if } \LAT(P_1) > \LAT(P_2)
  \end{cases},
%\vspace{-0.25em}
\end{equation}
with ties broken arbitrarily. 
% For ease of presentation, 
We use execution latency $\LAT(P)$ as the performance metric, but it can be easily generalized.
% in this paper, but our solution can be easily generalized to other metrics. 

Our comparator model is to learn the above oracle $\CP$. To be more specific, we organize the training datasets (e.g. from runtime stats repository) in the form of ($P_i$, $P_j$, $label$) for all pairs of executed plans of each query where $label$ indicates which plan in a pair is better. The learning goal is to fit the output of the oracle in Eq.~\eqref{equ: cmp}. We also use $\CP$ to denote the {\em learned plan comparator model} (or {\em comparator} for short).
% 
% \eat{The learned comparator can be noisy.}
% 
% By~\cite{2015Simple, JohannesF2011Preference, 2018Approximate}, the optimal plan $P^{*}$ also maximizes $\WP(P_i)$ in terms of the learned model $\CP$ in expectation. 
% 
% With $\CP$ trained sufficiently, we choose the plan $P_i$ among all the candidate plans which maximizes $\WP(P_i)$ (with $\CP$ in Eq.~\eqref{equ: cmp} as the learned comparator) to execute. The design details of comparator model $\CP$, including training and inference techniques, are described in Section~\ref{sec: train}.
% 
% With $\CP$ trained sufficiently, we can choose the best execution plan via pairwise comparisons.
% 
% $P_i$ among all the candidate plans which maximizes $\WP(P_i)$ (with $\CP$ in Eq.~\eqref{equ: cmp} as the learned comparator) to execute. 
% 
The design of $\CP$, including training and inference techniques, is described in Section~\ref{sec: train}.

\eat{
To leverage decades of wisdom of cost models developed in DBMS, we pre-train the comparator model using the estimated costs of candidate plans (without executing them) generated from a sample workload of queries in an offline training stage. By doing so, the model performs similarly to the native query optimizer at the beginning. As more and more queries are executed, $\CP$ can be continuously improved and tuned to adapt itself to possibly dynamic data distribution and workloads that drifts over time.
}

\sstitle{Plan Explorer.}
For a query, the plan explorer generates a variety of $n$ candidate plans $P_1, P_2, \ldots, P_n$. {While the best one chosen by the comparator is returned {by the optimizer} for query execution, the rest candidate plans are used by the model trainer to refine the model.}
% , in the form of ($P_i$, $P_j$; label--which one is better) for all pairs of executed plans. 
% In real-world distributed systems, idle workers are available for numerous reasons~\cite{kanet2000scheduling, chen2016revisiting, chaiken2008scope, negi2021steering}.
% 
Therefore, the candidate plans serve the purposes of both plan selection and model training, and they should: i) contain some truly good candidates ({\em although we do not have to know who they are}) and ii) be 
% \rzhu{not too long but}
sufficiently diversified so that the model could learn to distinguish between good and bad plans. 

To satisfy the two requirements, our plan explorer uses the cardinality estimator as the tuning knob to generate more plans for each query: the cardinality estimates of sub-queries are purposely scaled up/down before being fed into the cost model for the native query optimizer to generate different candidate plans. 
In particular, it is shown that diversity can be introduced in the generated plans by tuning selectivities (cardinalities) of predicates in a query \cite{DDH08identifying, DeyBDH08approximating}.
Our plan explorer is built on the intuition that the plan from the native query optimizer is usually not too bad; by tuning cardinality estimates, it may either generate some better plans in the neighboring plan space (if tuning towards the true cost of a sub-query which was\eat{ unknown and} incorrectly estimated), or some worse plans (tuning in the opposite direction).
Plans with various quality also increase the diversity among candidates to train the comparator, and it is the comparator's job to identify the best among them.

% It is shown to be able to introduce diversity in the \cite{DDH08identifying, DeyBDH08approximating} by tuning the estimated cardinalities. 
% \rzhu{Feeling of tuning cardinality to decrease q-error.} 
% \highlight{Moreover, the qualities of plans generated by traditional query optimizers at least loosely depend on the q-errors of cardinality estimations \cite{MoerkotteNS09preventing}. However, it is important to note that the purpose of tuning cardinality estimates is not to bound the estimation error. We just want to ensure that, during iterations of guesses of the ``true cardinality'', some good alternative plans would be generated for the model to consider and the candidate list is sufficiently diversified.}
% 
% \jrzhou{We need to argue for the benefit of cardinality tuning, or defer it to a later section: 1) potentially able to generate more plans. Need to cite the papers "Efficiently Approximating Query Optimizer Plan Diagrams" and "Identifying Robust Plans through Plan Diagram Reduction" by Jayant Haritsa; 2) able to purposely generate good alternative plans for the model to consider, pruning out some obviously bad plans.}

Details about our plan exploration strategy are described in Section~\ref{sec: policy}. One obvious advantage of our plan explorer is that the cardinality estimator is an essential component in almost all query optimizers, and thus the strategy and the implementation of our plan explorer can be easily migrated to different databases. 
% \rzhu{Is it better to not mention Bao here?} 
% \highlight{Unlikely, {Bao}~\cite{marcus2021bao} relies on query hints as the tuning knob which may vary from one database to another, and requires significant manual efforts to select proper candidate hint sets for each DBMS~\cite{negi2021steering}.}

\sstitle{Model Trainer.} 
% \eat{It utilizes idle workers to collect training data for the comparator model.}
{For each query, the trainer executes other candidate plans generated by the plan explorer, whenever system resources become available, and adds their plan information and execution statistics into the runtime stats repository. Such information is further used for training and updating the comparator model $\CP$. By doing so, \rao is able to \emph{explore} new plan space as much as possible and \emph{learn} from its potential mistakes.} 
In real-world distributed systems, idle workers and computation resources commonly exist due to scheduling~\cite{kanet2000scheduling} or synchronization~\cite{chen2016revisiting}. Some platforms~\cite{chaiken2008scope, negi2021steering} provide an individual environment for performance testing, which can be also used as a resource for executing candidate plans and collecting runtime stats. 

% \revise{We assume that the resource budget allocated to execute each plan is stable. When a dramatic change of the resource status significantly affects the performance of subsequent plans, the model trainer can re-train the plan comparator with fresh runtime stats.}

% Therefore, under the same resource budget (e.g. work memory),

\eat{
In this paper, we assume that there is no resource contention and the amount of resources available for executing the plans is properly configured and stable. \rao is also able to handle dramatic changes in the available resources, due to, e.g., updates of the resource allocation strategy or resource contention in highly concurrent workloads.
A naive method is to retrain the plan comparator model when the amount of available resources has significantly changed.
Our experiments demonstrate \rao’s resilience to highly concurrent workloads without retraining the model.
}

% \eat{This design turns waste into treasure: the performance of \rao could be enhanced without disturbing normal services.}

% \revise{
% \sstitle{Remarks to Runtime Environment.} 
% In this paper, we design \rao without taking consideration of the runtime environment factors, e.g., concurrency, locking, resource and etc. This is consistent to existing work, including both traditional~\cite{1994Volcano} and learned~\cite{marcus2019neo,marcus2021bao,yang2022balsa} query optimizer, which also isolate runtime factors from the query optimization process. The reasons are two fold. First, runtime factors often change frequently. Considering them would greatly increase the difficulty to design query optimizer. For example, in traditional query optimizer, if we tune the cost model to specify magic numbers to fit the cost for each runtime setting, it would be too costly. Second, it is easy to resolve the conflict between generated plans and runtime environment. In practice, if the generated plan is found to be not adaptive with the runtime environment, e.g., without enough memory for a merge sort, the query optimizer could be re-triggered with some constraints, e.g., disable the merge sort join, to generate a new plan to execute. We will consider how to model these runtime factors in \rao in the future work.  
% }
% \input{2-system.tex}

% !TeX spellcheck = en_US

\section{A Learned Plan Comparator}
\label{sec: train}
The comparator $\CP$ is trained/updated to fit a training dataset from the runtime stats repository which continuously monitors query execution and collects execution information. 
% 
% For each incoming query, the plan explorer generates a list of candidate plans from which the best is picked for execution using $\CP$.

\begin{figure*}[t]
\includegraphics[width=0.85\linewidth]{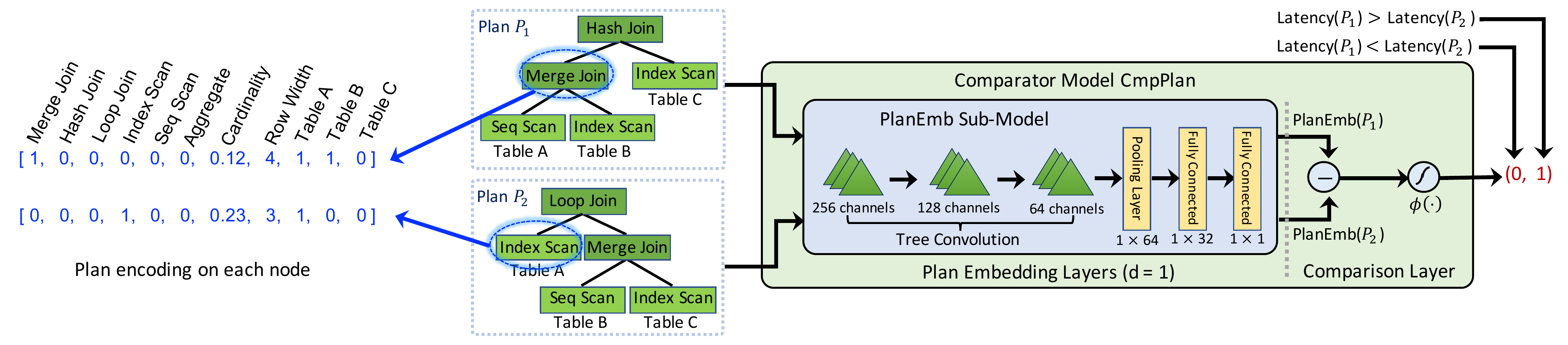}
	\vspace{-1.6em}
	\caption{Plan encoding and the structure of our plan comparator model ($d=1$).}
	\label{fig: PairLearn}
	\vspace{-1.2em}
\end{figure*}

\eat{\subsection{A General Design}}
\eat{The architecture of our comparator $\CP(P_1,P_2)$ is in Figure~\ref{fig: PairLearn}. It takes two featurized plans $P_1$ and $P_2$ of a query as the input.
Information about each plan, e.g., tables and operators involved in each sub-plan and the join order, is encoded as a feature vector.}

\subsection{Model Design}
\label{sec: train: general}
The overall model architecture of our comparator $\CP(P_1,P_2)$ is shown in Figure~\ref{fig: PairLearn}. It consists of \emph{plan embedding layers} followed by a \emph{comparison layer}. The plan embedding layer is carefully designed to effectively capture all the crucial information about a query plan, including tables, operators, and plan structural properties, etc., and the comparison layer compares {\em embeddings} ({$1$-$dim$ or $d$-$dim$} vectors output by the embedding layers) from the two plans and calculates their difference in terms of plan quality for the comparison purpose.

\sstitle{Plan Embedding Layers.}
The {\em plan embedding layers} of $\CP$ map $P_1$ and $P_2$ from the original feature space to a {\em one-dimensional (1-$dim$) embedding space}, in order to learn {differences} between plans. A sub-model $\PRR$ takes features from each plan and generate its {\em plan embedding}, $\PRR(P_i) \in \mathbb{R}$ for each plan $P_i$ ($i=1,2$). We use the {\em parameter sharing} technique in machine learning: the two plan embeddings, $\PRR(P_1)$ and $\PRR(P_2)$, are generated by two copies of $\PRR$, which are two components in $\CP$ sharing the same model structure and learnable parameters.

% , to learn the oracle in Eq.~\eqref{equ: cmp}.
% in order to approximate \highlight{learn} the oracle defined in .

% Parameters in both the plan representation layers and the comparison layer are learnable. 

% In order for the comparator $\CP$ to learn {\em differences} between plans, 

Judicious design of embedding model to capture critical information in the tree-structured plan is crucial for the overall model's effectiveness and efficiency. \rao builds on top of the tree convolution model, similar to~\cite{mou2016convolutional,marcus2019neo,marcus2021bao} but with significant improvements. As in Figure~\ref{fig: PairLearn}, a plan $P$ is featurized as a tree structure of vectors. The vector for each sub-plan $P_j$ that answers sub-query $Q_j$ (corresponding to a tree node) concatenates: a one-hot encoding of the last operation on $P_j$, the cardinality estimate, the row width of $Q_j$'s output, and 0/1 encoding of tables touched by $Q_j$. As the cardinality $\Card(Q_j)$ spans in a wide range, we use a min-max normalization over $\log(\Card(Q_j))$ in the feature vector. Unlike previous approaches~\cite{marcus2021bao,marcus2019neo}, the vector does not include the estimated cost, as it is strongly correlated with the estimated cardinality and row width and it may introduce additional inaccuracy of the cost model.
% 
% \eat{Unlike with Neo \cite{},  we do not encode the columns and predicates information specified by query $Q_i$ on each node, which simplifies the space of features.}

\eat{
Any embedding model can be used as $\PRR$ in the above framework. In this work, we use the tree convolution model proposed in~\cite{mou2016convolutional}, which is also used in other learned query optimizers, e.g., Neo \cite{marcus2019neo} and Bao \cite{marcus2021bao}, for latency prediction. As shown in Figure~\ref{fig: PairLearn}, each plan $P$ is featurized as a tree structure of vectors. The vector for each sub-plan $P_i$ that answers sub-query $Q_i$ (corresponding to a node of the tree) concatenates: a one-hot encoding of the last operation on $P_i$, the (normalized) cardinality, the row width of $Q_i$'s output, and 0/1 encoding of tables touched by $Q_i$. As the cardinality $\Card(Q_i)$ spans in a wide range, we use a min-max normalization over $\log(\Card(Q_i))$ in the feature vector. 
}

The tree convolution operation slides multiple triangle shaped filters over each node and its two children to transform the plan into another tree. Finally, vectors on the tree nodes are flattened to be fed into a neural network to generate the plan embedding.

% \jrzhou{For the completeness, shall we describe tree convolution model a bit in details? We also cite Neo and Bao too many times.}

% It is worth noting that, although similar tree convolution model structures are used in Neo \cite{marcus2019neo} and Bao \cite{marcus2021bao}, 
% 
% in comparison to how it is previously used, 

% We can extend the design to a $d$-dimensional embedding space, which is introduced and compared with the $d=1$ version in Appendix~\ref{app:multidim}.

It is worth noting that previous works use machine learning models to predict latency. Our embedding model $\PRR$ and comparator $\CP$ are trained and used differently. Instead of predicting plan latency, $\PRR$ tries to extract key information from plans and enables the {\em comparison layer} of $\CP$ to compare two plans.
% 
% and to induce a ranking order. 
% 
The learned 1-$dim$ plan embedding can be interpreted as a ranking criteria, and all pairs of plans for a query are comparable based on $\PRR(\cdot)$.
Thus, the scale of $\PRR(P)$ for a plan $P$ is not restricted, and its value does {\em not} have to be proportional to (or approximate) $P$'s performance metric, e.g., latency. Such flexibility allows the overall model to be more effective and easier to train.

% The one-dimensional embedding $\PRR(P)$ of plan $P$ could be interpreted as a ranking criteria, and all pairs of valid plans of a query are comparable based on $\PRR(P)$.
% 
% \jrzhou{With $d = 1$, $\PRR(P)$ becomes comparable.}

% the sub-model $\PRR$ in $\CP$ is trained and used differently. Instead of approximating the latency of plans (as in Neo and Bao), $\PRR$'s outputs enable the {comparison layer} of $\CP$ to compare two plans more easily in a low-dimensional space. We will make it more clear in the rest of this section.

\eat{
\vspace{-0.5em}
\subsection{Comparator with 1D Plan Embeddings}
\label{sec: train: 1D}

A special case of the general design is to set $d=1$, i.e., the plan embedding $\PRR(P)$ of a plan $P$ is a 1D embedding, or, a singleton value.
This choice of $d$ simplifies the comparison layer as well as how the best plan is selected after $\CP$ is fully trained. It also allows \rao to seamlessly inherit the wisdom of the native query optimizer during a pre-training stage.
{In our experiments, the performance of query processing using the plan selected by $\CP$ with $d=1$ is even better than that with $d>1$, while the training process with $d=1$ converges faster.} 
{
Specifically, on our evaluation benchmarks (described in Section~\ref{sec: eval-set-benchmark}), the execution time of \rao with a larger $d \in \{2, 4, 8, 16\}$ is $1.4$ times to $2.2$ times as long as the case with $d = 1$.
}
{The reason could be that, with $d > 1$, the embedding model $\PRR$ tries to summarize more sophisticated information in plan embeddings, but requires larger training dataset and more training time to obtain more accurate embeddings.}
In practice, we find $d = 1$ suffices. We now introduce and analyze the design of $\CP$ with $d=1$.
}

% 
% as well as how to train and model inference with $\CP$.

% \jrzhou{The reason is that, with $d > 1$, the model is more sophisticated so that it could capture more information but require more training dataset and more time to train. In practice, we find $d = 1$ suffices.}

% \backup{Instead, it is only required that the value could preserve the relative order of different plans for the same query. We will make it more clear by introducing and analyzing the loss function of our comparator $\CP()$ to match its learning goal.}

\sstitle{Comparison Layer.}
The plan embedding model $\PRR(\cdot)$ is learned together with $\CP$ via pairwise comparisons of plans, with binary labels indicating which one is better for each pair. 
% 
% \revise{The one-dimensional embedding $\PRR(P)$ of plan $P$ could be interpreted as a ranking criteria, and all pairs of valid plans of a query are comparable based on $\PRR(P)$.}
% \jrzhou{With $d = 1$, $\PRR(P)$ becomes comparable.}
% 
% It is learned together with $\CP$ from pairwise comparisons of plans and binary labels indicating which one is better for each pair. 
% 
% Therefore, the scale of $\PRR(P)$ is not restricted and its value does {\em not} have to be proportional to (or approximate) $P$'s performance metric, e.g., latency.
% 
Thus, in the {\em comparison layer} of $\CP$, we feed the difference $x = \PRR(P_1) - \PRR(P_2)$ of embeddings of two plans into a {logistic activation function} $\phi(x) = (1+\exp(-x))^{-1}$ to generate the {\em model's final output} (indicating whether $P_1$ or $P_2$ is better):
\vspace{-0.2em}
\begin{equation} \label{equ: modeloutput}
\CP(P_1, P_2) = \phi(\PRR(P_1) - \PRR(P_2)),
\vspace{-0.2em}
\end{equation}
which is within $(0,1)$ and can be interpreted as how likely $P_2$ is more preferable than $P_1$. 
{This is consistent with the learning goal of our comparator model as in Eq.~\eqref{equ: cmp}: $P_1$ is more preferable if 
% 
% \rzhu{$\CP(P_1, P_2)$ approaches to $0$ and vice versa.}
% 
{$\CP(P_2, P_1) < 0.5$ or approaches to $0$, and $P_2$ is more preferable if $\CP(P_2, P_1) > 0.5$ or approaches to $1$.}
Equivalently, a smaller 1-$dim$ plan embedding is more preferable.}
More formally,
\vspace{-0.2em}
\[\small
\CP(P_1, P_2) \rightarrow
\begin{cases}
0 & \text{if } \PRR(P_1, Q) \ll \PRR(P_2, Q) \\
1 & \text{if } \PRR(P_1, Q) \gg \PRR(P_2, Q)
\end{cases}.
\vspace{-0.2em}
\]
% which matches the learning goal of the comparator model in Eq.~\eqref{equ: cmp}. 

{From how the model's output is derived in \eqref{equ: modeloutput} and the fact that $\phi(a-b) + \phi(b-a) = 1$, our comparator model preserves two nice properties: i) ({\em commutativity}) $\CP(P_1, P_2) = 1 - \CP(P_2, P_1)$, that is, exchanging the order of input plans does not affect the comparison result; ii) ({\em transitivity}): $\CP(P_1, P_2) < 0.5$ and $\CP(P_2, P_3) < 0.5$ $\Rightarrow$ $\CP(P_1, P_3) < 0.5$, that is, $P_1$ is better than $P_2$ and $P_2$ is better than $P_3$ imply that $P_1$ is better than $P_3$.}
% 
% Meanwhile, the transitivity of the comparator model naturally holds, because the 1D plan embedding induces a total order of all plans.
% 
{Thus, the comparator $\CP(\cdot, \cdot)$ induces a total order of all plans, and we could select $P^{*} = \argmin_{P_1 \ldots P_n} \PRR(P_i)$ with the minimum value of $\PRR(\cdot)$ as the best plan for execution.}

\sstitle{Loss Function.} 
\eat{We now consider how to design the loss function for model training.}
% 
% For a list of candidate plans of a query, we want to recover their ranking order via $\CP()$ (and $\PRR()$), and choose the best one for query execution. 
% 
The goal of $\CP$ is to maximize the likelihood of outputting the right order between any two plans, so that the best plan can be selected. 
Thus, the loss function of $\CP$ is designed towards this goal.
Conceptually, let $\mathcal{A}$ be a (randomized) algorithm that decides which one of $P_1$ and $P_2$ is more preferable based on the model's output $\CP(P_1, P_2) \in (0,1)$:
\begin{equation} \label{equ: cmp_algo}
\mathcal{A}(P_1, P_2) \rightarrow
\begin{cases}
P_1 & \text{with probability } 1 - \CP(P_1, P_2)\\
P_2 & \text{with probability } \CP(P_1, P_2)
\end{cases}.
\end{equation}

We have $\CP(P_1, P_2) + \CP(P_2, P_1) = 1$ by the commutativity.
% our model construction (this is from how the model's output is derived in \eqref{equ: modeloutput}, and the fact that, for the logistic activation function $\phi(x)$, $\phi(a-b) + \phi(b-a) = 1$). 
Thus, the algorithm $\mathcal{A}$ is well-defined: $\mathcal{A}(P_1, P_2)$ and $\mathcal{A}(P_2, P_1)$ output $P_1$ with the same probability.
Let $P_1 \prec P_2$ denote the event that $\LAT(P_1) < \LAT(P_2)$ and $P_1 \succ P_2$ vice versa. 
% 
% The probability that $\mathcal{A}$ makes the right decision (its accuracy) is:
% \[
% \small
% {\rm ACC}(\mathcal{A}) \! = \!\prt[\substack{Q \sim \mathcal{Q}\\(P_1,P_2)\sim\mathcal{P}(Q)\times\mathcal{P}(Q)}]{\indicator{\mathcal{A}(P_1, P_2) = P_1} \!\!\cdot\! \indicator{P_1 \prec P_2} + \indicator{\mathcal{A}(P_1, P_2) = P_2} \!\!\cdot\! \indicator{P_1 \succ P_2} = 1},
% \]
% %
% where $\mathcal{Q}$ is the distribution of queries, $\mathcal{P}(Q)$ is the distribution of plans for a given query $Q$, and the indicator function $\indicator{\rm{x}}$ returns $1$ if the condition $\rm{x}$ holds and $0$ otherwise. 
% 
The goal is to train $\CP$ such that $\mathcal{A}$ is as accurate as possible.
Suppose we have a workload of queries in $\mathcal{W} = \{Q\}$ and a set of candidate plans ${P}(Q) = \{P_1, \ldots, P_n\}$ for each query, as the training data. 
We use $\mathcal{A}$ to compare all pairs of candidate plans for each $Q \in \mathcal{W}$;
the probability that $\mathcal{A}$ makes no mistake is:
\vspace{-0.7em}
\begin{align}
\small
{\rm ACC}(\mathcal{A}, \mathcal{W}) = \prod_{Q \in \mathcal{W}} & \left(\prod_{P_i \prec P_j \in {P}(Q)} (1 - \CP(P_i, P_j))\right. \\ \nonumber
& \left.\cdot \prod_{{P_i \succ P_j \in {P}(Q)}} \CP(P_i, P_j) \right).
\vspace{-0.5em}
\end{align}
We train $\CP$ to fit the observed orders of plans for queries in $\mathcal{W}$, so the loss function is chosen to be $L = -\log{\rm ACC}(\mathcal{A}, \mathcal{W})$:
\vspace{-0.5em}
\begin{align}
\small
\label{equ: lossfunc}
L = -\sum_{Q \in \mathcal{W}}\sum_{P_i, P_j \in {P}(Q)} & \left(\indicator{P_i \prec P_j} \cdot \log(1 - \CP(P_i, P_j))\right.\\ \nonumber
& \left.+ \indicator{P_i \succ P_j} \cdot \log(\CP(P_i, P_j)) \right),
\vspace{-0.8em}
\end{align}
where the indicator function $\indicator{\rm{x}}$ returns $1$ if the condition $\rm{x}$ holds and $0$ otherwise. $L$ coincides with the {\em cross-entropy loss} in classification.

\subsection{Model Training}
\label{sec: train: train}
% \label{sec: train-train}
% 
Instead of training from scratch, we pre-train our comparator model $\CP$ offline first on synthetic workloads to inherit the wisdom of the native query optimizer and its cost model.
% 
% align its performance with the native query optimizer to avoid cold-start. 
% 
It is then continuously trained and updated online on real workloads with training data collected during actual executions of query plans. 

\sstitle{Model Pre-training: Starting from Traditional Wisdom.}
% Jingren: copied from the previous section; needs to be cleaned up
To\\ leverage decades of wisdom of cost models developed in DBMSs, we pre-train the comparator model using the estimated costs of candidate plans (without executing them) generated from a sample workload of queries in an offline training stage.
{After pre-training, the 1-$dim$ embedding $\PRR(P)$ is expected to be an approximation to the native estimated cost $\PC(P)$. Thus, the model is bootstrapped to perform similarly to the native query optimizer at the beginning.}
As more and more queries are executed, $\CP$ is continuously improved and tuned to adapt to possibly dynamic data distribution and workloads that drift over time.

%\revise{To be more specific, the sub-model $\PRR(P)$ is pre-trained to fit the native query optimizer's cost model $\PC(P)$ without executing the plans. After pre-training, the 1D embedding $\PRR(P)$ is expected to be an approximation to the native estimated cost $\PC(P)$. Thus, we expect that, before pairwise training in the online stage, the comparator $\CP$ and the native cost model would agree on which candidate plan is the best.}

% 
% That is, \rao could seamlessly inherit the wisdom of the native query optimizer after a pre-training stage.

% The hope is that, with proper pre-training, the initial performance of \rao is comparable to that of the native query optimizer.

% We aim at bootstrapping the comparator model $\CP()$ with simulated data offline so that it could \rao could inherent the wisdom of the native query optimizer. 
% 

% Recall that, the cost of each sub-query is computed by some experience-driven rules. 

In the native cost model, the cost of a plan is usually a piecewise linear or quadratic  function of estimated cardinalities of sub-queries with magic constants as co-coefficients for different operators.
The goal of pre-training $\PRR(P)$ is to learn such functions in a data-agnostic way, so as to handle any unseen query.
% 
% The plan cost $\PC(P, Q)$ simply adds the cost of each sub-query occuring in $P$. If our representation model $\PRR(P, Q)$ could capture these data-agnostic computation rules on sub-query and learn to add them together, it is able to output the cost of any unseen query.
% 
Thus, we could pre-train $\PRR(P)$ purely using synthetic workloads. Namely, we randomly generate a number of plans with different join orders and predicates on different tables, and featurize them as the training data; we also randomly set the cardinality for each sub-plan of each plan and feed them into the native cost model (without executing the plans) to derive the estimated plan costs as labels. 
% 
% These synthetic plans and target cost value are fed into the model $\PRR()$ for training. 
% 
As the cost models are often in a class of functions with simple structures, the model pre-training converges very fast. 
% Meanwhile, the pre-trained model $\PRR()$ generalizes across data.

% The literature work Balsa~\cite{yang2022balsa} also proposes a pre-training method to warm up its learned model $\textsf{PlanVal}(P', Q)$. It tries to approach the minimum latency of the plan $P$ for $Q$ expanded from the partial plan $P'$. In the offline phase, it obtains the minimum cost value for plan $P$ instead of its actual latency for pre-training. Therefore, the model needs to learn to recognize the optimal cost for each partial plan, which is data specific and more difficult. In comparison to it, our pre-training method is more efficient, general and practical. 

% \bd{I will try to shorten the following ``Model Training'' part.}

\sstitle{Pairwise Training.}
In the online stage, we train and update the comparator $\CP$ in the pairwise comparison framework. For each query $Q$, its candidate plans $\{P_1, \ldots, P_n\}$ are generated by the plan explorer to be introduced in Section~\ref{sec: policy}; these plans are executed by idle workers, whenever system resources become available,  with execution statistics collected in the runtime stats repository. According to the loss function in Eq.~\eqref{equ: lossfunc}, we construct $n(n-1)$ training data points for 
each query: for each pair $(i,j)$ with $1 \leq i \neq j \leq n$, {we construct a data point with {\em features} $(P_i, P_j)$ and {\em label} $1$ if $\LAT(P_i) > \LAT(P_j)$, and $0$ otherwise}.

Periodically, we use an SGD optimizer to update $\CP$ together with $\PRR$ by backward propagation with the training dataset constructed above. For parameters in the sub-model $\PRR$, as they are shared in two copies of $\PRR$ that output $\PRR(P_i)$ and $\PRR(P_j)$, respectively, we add up the gradients from the two copies for each shared parameter together to update this parameter in the model $\PRR$. 

Our comparator, by nature of its design, adapts fast to dynamically changing data during the processing of model training and updating. When tuples are inserted or deleted into the database, the relative orders of two plans (i.e., labels of $\CP$) are more robust than their execution latencies (i.e., labels of a latency prediction model). The latter would vary for the same plan even when tuples from the same distribution are inserted, which makes the training of a latency prediction model costly especially on dynamic data.
Our experimental study in Section~\ref{sec: eval-dynamic} verifies the above intuition.

%The superiority of training rank-based models using pairwise comparison has been verified by literature from other domains~\cite{furnkranz2010preference, hullermeier2008label, beutel2019fairness}. If we directly train the scoring model $\PR()$ to fit a rank score $r_i$ for each plan $P_i$, it is difficult to specify the proper value of $r_i$. Specifically, if the gap between $r_{i+1} - r_{i}$ is too small, the model is less freedom and degenerates to a regression learning task. Otherwise, if the gap is too large, the model needs to witness lots of value between $r_{i}$ and $r_{i+1}$ to recognize the score interval for each plan. Whereas, the pairwise learning method is automatically adaptive to different data distributions. From our evaluation in Exp-8 in  Section~\ref{sec: eval-detail}, it always outperforms other training methods.

% \sstitle{Selecting the Best Plan using Comparators.}

\eat{
\subsection{Selecting the Best Plan using Comparators}
After a comparator is fully trained, we can use it to pick the best plan among candidates $P_1, \ldots, P_n$ of a possibly unseen query $Q$.

In the general design (Section~\ref{sec: train: general}), we do not require a fully trained $\CP$ to preserve transitivity, i.e., $P_1$ is better than $P_2$ and $P_2$ is better than $P_3$ imply that $P_1$ is better than $P_3$. In order to pick the best plan among candidates $P_1, \ldots, P_n$, 
% generated by the plan explorer for a possibly unseen query $Q$, 
we can invoke $\CP$ $n(n-1)$ times to compare all pairs of candidates. In general the output $\CP(P_i,P_j) \in (0,1)$ gives a soft prediction (i.e., the predicted probability of whether $P_i$ or $P_j$ is better), so the definition of $\WP$ in Eq.~\eqref{equ: wins} does not apply.
We use the algorithm in Eq.~\eqref{equ: cmp_algo} to pick the better one, and define {\em randomized wins} as
\vsshrink
\[
\RWP(P_i) = |\{ P_j \mid \mathcal{A}(P_i, P_j) = P_i, j \neq i \}|.
\vsshrink
\]
% \rzhu{The definition of wins is different and inconsistent with former one.}
Accordingly, we choose the best plan as
\vsshrink
\[
P^* = \argmax_{P_i \in \{P_1, \ldots, P_n\}} \RWP(P_i)
\vsshrink
\]
with the most randomized wins--ties are broken arbitrarily.
% 
% we compare all the $n(n-1)$ pairs of candidates with $\CP$, and choose the one with the max $\WP$ defined in Eq.~\eqref{equ: wins}

In the more practical design of $\CP$ with 1D plan embeddings (Section~\ref{sec: train: 1D}), we do not need to derive $\RWP(P_i)$ as above.
% , i.e., the number of plans less preferable by $\CP$ than $P_i$ (defined in Eq.~\eqref{equ: cmp}) of each plan $P_i$ and then choose the plan maximizing it. 
Instead, the output of the sub-model $\PRR$ defines a {\em total order} on all candidate plans. For any pair of plans $(P_i, P_j)$, if $\PRR(P_i) > \PRR(P_j)$, we always have $\CP(P_1, P_2) > 0.5$, i.e., $P_j$ is more preferable than $P_i$ with higher probability than vice versa.
Therefore, we could select $P^{*} = \argmin_{P_i} \PRR(P_i)$ with the lowest value of $\PRR$ as the best plan for execution. We can show that, $P^*$ chosen in this way is indeed the one with the most randomized wins {\em in expectation} (thus, the choice is equivalent to the one for the general design, in expectation). 
%Due to space limits, we put all proofs in Appendix~B of the full version~\cite{fullversion}.
% 
\begin{proposition}
$P^{*} = \argmin_{P_i} \PRR(P_i)$ chosen with the lowest 1D embedding also wins the most: it satisfies
\vsshrink
\[
P^* = \argmax_{P_i \in \{P_1, \ldots, P_n\}} \ep{\RWP(P_i)}.
\vsshrink
\]
\end{proposition}

\begin{proof}
By the model construction, for each $P_i$, we have
\begin{align}
    & \ep{\RWP(P_i)} = \sum_{j \in [n], j \neq i} \pr{\mathcal{A}(P_i, P_j) = P_i} \nonumber 
    \\
=   & \sum_{j \in [n], j \neq i} \!\!\!\!\!\! (1 - \CP(P_i, P_j)) = \!\!\!\!\!\! \sum_{j \in [n], j \neq i} \!\!\!\!\!\! \CP(P_j, P_i) \nonumber
    \\
=   & \left(\sum_{j \in [n]} \!\! \phi(\PRR(P_j) - \PRR(P_i))\right) - \phi(0). \label{equ: epwins}
\end{align}

From Eq.~\eqref{equ: epwins}, it is straightforward that $P^* \in \{P_1, \ldots, P_n\}$ with the lowest $\PRR(\cdot)$ would maximize $\ep{\RWP(\cdot)}$.
\end{proof}
}
% \input{3-train.tex}

% \input{4-policy-rzhu.tex}
% !TeX spellcheck = en_US

\section{Plan Exploration Strategy}
\label{sec: policy}
The plan explorer of \rao generates a list of candidate plans $P_1,$ $\ldots,$ $P_n$ for a query $Q$, for two purposes. {First, for the purpose of {\em query optimization}, \rao applies the comparator model to identify the best plan among the candidates for execution. Thus the candidate list must include some truly good plans for consideration.}

{Second, for the purpose of {\em plan exploration}, \rao \emph{prioritizes} exploring other promising plans for the query.
% , especially in the neighboring plan space. 
By executing them and comparing their performance pairwise, \rao is able to catch past optimization mistakes and adjust the model timely using the newly observed runtime information. In addition, whenever system resources are available, other candidate plans which are {\em diversified} sufficiently (e.g., with different join orders or in different shapes such as left-deep and bushy trees) are also considered so that \rao learns new plan space and improves the model over time.}

\eat{
The plan explorer of \rao generates a list of candidate plans $P_1,$ $\ldots,$ $P_n$ for a query $Q$, for two purposes. First, for the purpose of {\em query optimization}, we apply the comparator model to identify the best plan among the candidates for execution. Thus the candidate list must include some truly good plans for consideration. Second, for the purpose of {\em plan exploration}, the candidate plans should be {\em diversified} sufficiently (e.g., with different join orders or in different shapes such as left-deep and bushy trees) so that the comparator model learns new plan space by executing them and comparing their performance to refine the model in a pairwise fashion.

The plan space is huge for a complex query. It is costly or impossible to explore and execute every single plan. Obviously bad plans are not worth consideration. Thus the plan explorer should prioritize exploring good and promising plans. The goal of plan exploration is to find a list of {\em good} and {\em diversified} candidate plans. 
}

\eat{
The plan explorer of \rao generates a list of candidate plans $P_1,$ $P_2,$ $\ldots,$ $P_n$ for a query $Q$. 
% 
% We pick the best from these candidates using the comparator $\CP$ ({\em optimization}), and meanwhile, we would use their execution statistics to train and update $\CP$ ({\em training}).
% 
For the purpose of {\em query optimization}, we want the candidate list generated by the plan explorer to include some truly good plans (with low latency when being executed); we would use the comparator to identify the best one among the candidates for execution.
For the purpose of {\em training}, execution statistics of plans in the list would be used to learn and update $\CP$; thus,  the comparator is able to learn enough information to distinguish between better and worse plans.
}

% \rzhu{List not too long? -  to be mentioned later}

For the ease of deployment, \rao makes minimal changes to the native query optimizer. Therefore, the plan explorer should be able to be implemented in a lightweight way, e.g., through re-implementing some system provided interfaces. Meanwhile, we hope that the framework of \rao could be generally used in any DBMS; thus, the strategy should not be system-specific.

% By its role in \rao, we functionally require the candidates are highly possible to include some truly good plans for plan selection and sufficiently diversified, i.e., in different forms including left-deep tree, bushy tree and right-deep tree plans, for model training. Moreover, we also have two non-functional requirements for the plan exploration:

% \squishlist
% \item We do not want the candidate list to be too long ($n$ should be small, e.g., 5 or 10) as each of them needs to be executed by idle workers to provide the ``labels'' in training data. Small size of training data also enables the model to learn fast to converge, which is important in dynamic settings.

% \item For the ease of deployment, \rao makes minimal changes to the native query optimizer. Therefore, the plan generator should be able to be implemented in a lightweight way, e.g., through re-implementing some system provided interfaces. Meanwhile, we hope that the framework of \rao could be generally used in any DBMS; thus, the strategy should not be system-specific.
% \squishend

\eat{
\squishlist
\item For the purpose of optimizing, we want the candidates are highly possible to include some truly good plans (with low latency when being executed), although we do not know which ones are good before applying our comparator model.

\item For the purpose of training, we do not want the list to be too long ($n$ should be small, e.g., 5 or 10) as each of them needs to be executed by idle workers to provide the ``labels'' in training data. Small size of training data also enables the model to learn fast to converge, which is important in dynamic settings. Meanwhile, plans in the list should be diversified sufficiently so that the comparator model is able to learn enough information to distinguish between good and bad plans. 

\item For the ease of deployment, \rao makes minimal changes to the native query optimizer. Therefore, the plan generator should be able to be implemented in a lightweight way, e.g., through re-implementing some system provided interfaces. Meanwhile, we hope that the framework of \rao could be generally used in any DBMS; thus, the policy should not be system-specific.
\squishend
}

\sstitle{Existing Plan Exploration Methods.}
A straightforward strategy is to explore a random sample of valid plans for each query, which is used by reinforcement learning-based approaches such as Neo~\cite{marcus2019neo} and Balsa~\cite{yang2022balsa}.
The obvious drawback is that, with high likelihood, high-quality plans could be missing in a random sample; otherwise, the sample size has to be so large that executing these sample plans would be too costly for model training.
% 
% By these requirements, the reinforcement learning approaches used in Neo~\cite{marcus2019neo} and Balsa~\cite{yang2022balsa} are not applicable for \rao. They randomly explore a large number (hundreds to thousands) of plans for each query for model training, which is very costly and does not focus on generating truly good plans.
% Another nature strategy is to reserve all plans, which are filtered by some optimization rules~\cite{warshaw1999rule} in the native query optimizer, as candidates. However, as these rules can often ensure to improve plan quality, these discarded plans are often worse than the output one.

% Such tuning coui ked be done in different ways. One strategy proposed in 
Bao~\cite{marcus2021bao} explores plans by tuning a set of hints (boolean flags) to disable/force certain types of optimization rules. For example, the native query optimizer initially generates a plan with merge join for a query; with a hint set that disables merge join and forces indexed nested loop join, it would generate a different candidate plan.
However, this optimizer-level tuning strategy has two drawbacks.

First, a hint set is typically applied for the whole query during the entire plan search procedure. 
If different parts of a query have different optimal choices, tuning a flag at a query
level may miss opportunities for finding high-quality plans.
% 
% Some high-quality plans can be easily missed due to the coarse granularity of tuning.
% 
For example, let $Q$ be a query joining two sub-queries $Q_{1}$ and $Q_{2}$ where the best join operations for $Q_{1}$ and $Q_{2}$ are indexed nested loop join and merge join, respectively. The query optimizer may use merge join for both $Q_{1}$ and $Q_{2}$ due to cardinality estimation errors. With a hint set disabling/forcing either merge join or indexed nested loop join, at most one of $Q_1$ and $Q_2$ could select the right operation. In such cases, the optimal plan can never be 
% obtained by hint set tuning and 
included in the candidate list.

Second, the set of available hints is system specific. An optimizer\eat{, such as SCOPE~\cite{chaiken2008scope},} usually contains hundreds of flags to enable/disable certain optimization rules. Enumerating all kinds of combinations is infeasible in practice. Selecting an effective subset of hint sets manually requires a deep understanding on the system and comprehensive analysis on the workload~\cite{negi2021steering}. 

Thus, it motivates us to pursue other routes for designing a plan exploration strategy. We introduce our tuning knob in Section~\ref{sec: policy: knob}, followed with practical heuristic strategies in Section~\ref{sec: policy: algo}.

\subsection{Cardinality as Knob for Plan Explorer}
\label{sec: policy: knob}
% 
% Motivated by the pipeline of cost-based query optimizer (see Section~\ref{sec: intro-challenge} and Figure~\ref{fig: CostArch} (a)) and literature analysis~\cite{leis2015good, han2021CEbenchmark}, 
% 
\rao uses the cardinality estimator as the tuning knob for our plan explorer. In the native cost-based query optimizer, the estimated cardinality for each sub-query of an input query $Q$ is fed into the cost model to guide plan enumeration and selection. Each time with different cardinality estimates on one or more sub-queries, the query optimizer would select a different plan for $Q$. In \rao's plan explorer, we tune (magnify or reduce) the estimated cardinalities multiple times to generate a list of different candidate plans. 

Formally, let $\Card()$ be the native cardinality estimator in DBMS. For each sub-query $Q'$ of a query $Q$, $\Card(Q')$ gives a cardinality estimate. Instead of invoking $\Card()$ in the cost model, we ask the query optimizer to invoke a {\em tuned estimator} $\TCard()$, so it would generate a different plan. Our plan explorer would tune $\Card()$ in different ways; with different tuned estimators fed into the cost model, the query optimizer generates different plans as the candidates $P_1, \ldots, P_n$.

Using the cardinality estimator as a tuning knob has following advantages.
First, in cost-based query optimizers, cardinality estimates decide the estimated costs, and thus determine join orders and physical operations on tables and sub-queries. {Therefore, under the same resource budget (e.g. work memory) for plan execution, tuning cardinality estimates would be highly possible to introduce diversity in candidate plans, possibly with different join orders or different operators (e.g., sort-merge join v.s. nested loop join).} Second,\eat{ it allows \rao to explore various optimization opportunities in the plan space and include some truly good ones in the candidate list; it is the comparator's job to identify the best from them.
Meanwhile,} cardinality tuning is platform-independent. For most DBMSs, there exist system-provided interfaces to modify the estimated cardinality, which is friendly for system deployment.
% 
% \rzhu{Feeling of tuning cardinality to decrease q-error.}
% 
% \highlight{Last but not least, it is shown that the more accurate the cardinality results, the higher the generated plan's quality is~\cite{MoerkotteNS09preventing, leis2015good, han2021CEbenchmark}. Ideally, if all possible cardinality estimates of each sub-query are fed into the cost model to generate candidate plans, the native query optimizer would give at least one near-optimal plan.}

%\sstitle{An Ideal Exploration Strategy.}
\sstitle{A Brute-Force Exploration Strategy.}
% 
% Ideally, we should generate different plans for the learned comparator to consider
% \sstitle{Tuning Cardinality for a Single Sub-query.}
Let $\Card^{*}()$ be the true cardinality. For any query $Q$, the difference between $\Card^{*}(Q)$ and the DBMS's estimate $\Card(Q)$ is unknown. 
However, with a reasonable number of different ways to tune $\Card(Q)$, we can ensure that at least one tuned estimator $\TCard(Q)$ is close to $\Card^*(Q)$, in terms of {\em q-error}, which is defined as ${\rm QE}({\rm estimate}, {\rm true}) = \max\{\frac{\rm estimate}{\rm true}, \frac{\rm true}{\rm estimate}\}$.

% While it is impossible for the tuned estimator $\TCard$

% Literature work~\cite{leis2015good,han2021CEbenchmark,zhu2020flat,yang2019deep,hilprecht2019deepdb} often uses {\em q-error}, defined as ${\rm QE}(\Card(Q), \Card^*(Q)) = \max\{\Card^{*}(Q)/\Card(Q), \Card(Q)/\Card^{*}(Q)\}$, to measure the accuracy of estimated cardinality. We first focus on one sub-query $Q'$ and how to make guesses of its cardinality, starting from the default estimator $\Card(Q')$, e.g., histograms, in databases.

% In our approach, we tune $\Card(Q')$ multiple times: for the purpose of optimizing, it suffices to ensure that at least one tuned estimation is close to $\Card^{*}(Q')$ so that some truly good plan is in the candidate list; for the purpose of training, the purposely tuned cardinality easily lead to different join orders and operations for the diversity of plans. 

We tune the cardinality estimator $\Card()$ with exponentially varying step sizes.
Suppose we know an upper bound of the q-error of $\Card()$, namely, ${\rm QE}(\Card(Q'), 
\Card^*(Q')) \leq \Delta$ for any sub-query $Q'$ of $Q$.
Let 
\vsshrink
\begin{equation}
F_\alpha^\Delta = \{\alpha^{t} \mid \lfloor -\log_\alpha\Delta \rfloor \leq t \leq \lceil \log_\alpha\Delta \rceil, t \in \mathbb{Z}\}
\vsshrink
\end{equation}
be the set of {\em guesses of scaling factors}. For each $f = \alpha^t \in F_\alpha^\Delta$, we tune $\Card(Q')$ as $\TCard(Q') = f \cdot \Card(Q')$.
Then there is at least one $f \in F_\alpha^\Delta$ such that ${\rm QE}(f \cdot \Card(Q'), \Card^*(Q')) \leq \alpha$. 
% 
% We only need $2\log_{\alpha}\Delta$ guesses such that one of them has q-error no larger than $\alpha \ll \Delta$. Specifically, we scale up or down each guess by a factor of $\alpha$. 
% 
% \begin{proposition}\label{prop: qeguess}
% Starting from a cardinality estimation $C(Q')$ with ${\rm QE}(\Card(Q'), \Card^*(Q') \leq \Delta$, for a specific constant $1 < \alpha \ll \Delta$, there is at least one $f \in F_\alpha^\Delta$ such that ${\rm QE}(f \cdot C(Q'), C^*(Q')) \leq \alpha$. 
% \end{proposition}
% 
The estimation of $\Delta$ does not need to be tight, since the number of guesses $|F_\alpha^\Delta| = \Theta(\log_\alpha\Delta)$ depends logarithmically on $\Delta$. In practice, it can be set based on users' experience or q-error distribution on historical queries.
% , or derived using analytical tools, e.g., Chernoff-Hoeffding bound~\cite{schmidt1995chernoff} on sampling-based cardinality estimation algorithms.

% We now consider how to tune the estimator $\Card()$ for all sub-queries of the query $Q$ together before feeding it into the cost model in order to derive candidate plans.
% Suppose we start from a default estimator $C()$ with q-error bounded by $\Delta$ for any sub-query of $Q$. 
In order to explore plans for the query $Q$, we repeat the above tuning process recursively for all of its sub-queries: for each sub-query $Q'$ of $Q$, we pick a scaling factor $f_{Q'} \in F_\alpha^\Delta$ and set $\TCard(Q') = f_{Q'} \cdot \Card(Q')$. For each combination of scaling factors $\langle f_{Q'} \rangle_{Q' \subseteq Q}$, we construct an estimator $\TCard()$ defined for all the sub-queries of $Q$, and we can feed it into the cost model to generate a candidate plan.

From the way how tuned estimators are constructed above, there is at least one combination of scaling factors such that the resulting $\TCard()$ satisfies ${\rm QE}(\TCard(Q'), \Card^*(Q')) \leq \alpha$ for all sub-queries of $Q$. Theoretically, a near-optimal candidate plan under a specific cost model can be generated by the optimizer using an estimator $\TCard()$ with q-error bounded by $\alpha$; more formally, from \cite{MoerkotteNS09preventing}, with some mild assumptions about the cost model, the optimal plan under $\TCard()$ is no worse than the optimal plan under $\Card^*()$ by a factor of $\alpha^4$.

Ensuring that at least one candidate plan is near-optimal suffices for the purpose of optimization, as it is the comparator's job to identify it among all candidates. However, the overhead of the above method is too high: the number of different sub-queries $Q'$ is at least $2^q$ where $q$ is the number of tables in $Q$, and $|F_\alpha^\Delta| = \Theta(\log_\alpha\Delta)$; thus, the total number of combinations of scaling factors is $\Theta(\log_\alpha^{2^q}\Delta)$. 
From the above discussion, we can prove the following results.
\begin{proposition}
In the above brute-force exploration strategy, the number of candidate plans generated is at most ${\rm O}(\log_\alpha^{2^q}\Delta)$; with some mild assumptions about the cost model as in \cite{MoerkotteNS09preventing}, at least one of them is no worse than the optimal plan by a factor of $\alpha^4$.
\end{proposition}
The list of candidate plans generated in the above way is too long, which is unbearably costly for model training. 
% 
% Since this method is only applicable for queries joining a very small number of tables,
% 
% For queries joining larger number of tables, 
% 
We propose more effective heuristic methods in the following subsection. 

% \bd{Not sure whether we want to highlight the second part of the above proposition... to handle Reviewer \#2 D3-D4. We do have a brute-force strategy with guarantees but it is too expensive in terms of the number of candidates it generates; so we propose an heuristic below. The first part of Proposition 1 and Proposition 2 are also for Reviewer \#2 D1.}

\eat{
The key idea of reducing the overhead of plan exploration and the number of candidates is to priofocus on where the native optimizer makes mistakes, instead of exploring all possibilities and tuning cardinality estimates for all sub-queries simultaneously as in the brute-force strategy. We introduce a heuristic based on this idea, which focuses on mistakes in estimating cardinalities for size-$k$ sub-queries (for a different $k$ at one time).
}

\subsection{Priority-Based Heuristic Methods}
\label{sec: policy: algo}
% % 
Plan exploration is conducted in background whenever system resources become available. The key idea of our heuristics is to \emph{prioritize} exploring where the native optimizer is likely to make mistakes. In addition, instead of exploring all possibilities and tuning cardinality estimates for all sub-queries simultaneously as in the brute-force strategy, we introduce a heuristic which focuses on mistakes in estimating cardinalities for size-$k$ sub-queries on $k$ tables (for each different $k \geq 1$ at one time).
~\footnote{We also consider heuristics based on plan diagram~\cite{DDH08identifying, DeyBDH08approximating}, which can be found in Appendix~\ref{app:diagram}.}
%~\footnote{We also consider heuristic based on plan diagram\cite{DDH08identifying, DeyBDH08approximating}, which can be found in the Appendix~\ref{app:diagram}.}

\eat{We introduce two heuristics based on this idea.
% 
% \rzhu{The two methods are progressive, but not equal, the first one is not appliable.}
% 
% \highlight{
The first one focuses on mistakes in estimating predicate selectivities, and the second one focuses on mistakes in estimating cardinalities for size-$k$ sub-queries (for a different $k$ at one time).}
% }

% Since the plan generated by the native query optimizer cannot be too far away from the optimal plan, we should 

\eat{
\subsubsection{Heuristic based on Plan Diagram}
Errors in estimating predicate selectivities can be propagated from bottom to top during the plan search, and incur a wrong join order and sub-optimal choices of join types.
A {\em plan diagram} \cite{DDH08identifying, DeyBDH08approximating} can be constructed by varying values of parameters in one or more predicates of the query $Q$, so that the optimizer generates a set of different plans.
% 
% (e.g., replacing $20 \leq {\sf Age} \leq 30$ with $10 \leq {\sf Age} \leq 40$)
% 
For example, for a query with a parameterized predicate ``${\sf @lower}$ $\leq$ ${\sf Age}$ $\leq$ ${\sf @upper}$'', the optimizer may generated different plans by setting $({\sf @lower}, {\sf @upper})$ as $(20,30)$ or $(10,40)$.
These plans form a diagram of different regions on the 2D space $({\sf @lower}, {\sf @upper})$: within each region, the optimizer outputs the same plan.

It is shown in \cite{DDH08identifying, DeyBDH08approximating} that replacing selectivity error-sensitive plan choices with alternatives in the plan diagram provides potentially better performance. Therefore, we can use plans in the plan diagram as the list of candidates to explore the uncertainty from estimating predicates' selectivities in the native optimizer.

The drawbacks of this strategy are obvious. 
It is not affordable to generate a plan diagram by varying parameters on more than two tables, as the number of candidate plans would be too large~\cite{DeyBDH08approximating}. If we tune parameters on two tables, the candidates may not be diversified enough to include plans with quality sufficiently higher than the plan generated by the native optimizer.
In our evaluation on the STATS benchmark~\cite{han2021CEbenchmark}, even the best candidate in a plan diagram does not have significant performance improvement. Specifically, the average performance improvement of the best candidate is less than $3\%$, and the best candidates of only less than $5\%$ of queries are faster than those generated by PostgreSQL's optimizer.
% However, if we tune cardinality on more tables, the number of candidate plans would be too large and not affordable for execution.
}

\begin{figure}[t]
	\small
	\rule{\linewidth}{1pt}
	\leftline{~~~~\textbf{Algorithm} ${\sf plan\_explorer}(Q, \alpha, \Delta)$}
	\vspace{-1.5em}
	\begin{algorithmic}[1]
        \STATE Priority queue ${\sf candidate\_plans} \gets \varnothing$
		\FOR{each $f \in F_\alpha^\Delta$ in the increasing order of $|\log f|$}
    		\FOR{$k \gets 1$ to $q$ (the number of tables in $Q$)}
			 %   \STATE Send parameter $w = (k,f)$ to query optimizer
			    \STATE \underline{\bf Inside query optimizer:}
       let $\Card()$ be the default cardinality estimator in native query optimizer
			    \STATE \hspace{1em} $\TCard(Q') \gets f \cdot \Card(Q')$ for size-$k$ sub-queries $Q' \in {\sf sub}_k(Q)$
			    \STATE \hspace{1em} $\TCard(Q') \gets \Card(Q')$ for sub-queries $Q' \in {\sf sub}(Q) - {\sf sub}_k(Q)$
				\STATE \hspace{1em} feed cardinality $\TCard()$ into the cost model to generate a plan $P$
				\STATE ${\sf candidate\_plans} \gets {\sf candidate\_plans} \cup \{P\}$
			\ENDFOR
		\ENDFOR
		\RETURN ${\sf candidate\_plans}$
	\end{algorithmic}
	\vspace{-0.5em}
	\rule{\linewidth}{1pt}
	\vspace{-2.5em}
	\caption{Plan exploration strategy in \rao.}
	\label{alg: plangen}
	\vspace{-1.7em}
\end{figure}

\eat{\subsubsection{Heuristic based on Sub-query Grouping}
The plan diagram-based strategy can be interpreted as a processing of tuning cardinality estimates for sub-queries on single tables.

More specifically, to introduce more diversity among candidates, we can tune cardinality estimates for size-$k$ sub-queries (on $k$ tables) for each different $k \geq 1$ at one time.
}
The necessity of considering $k > 1$ is because the difficulty (as well as the error) of estimating cardinality of size-$k$ sub-queries increases significantly as $k$ increases (due to cross-table correlation among columns). Thus, the native optimizer tends to make mistakes in generating partial plans for sub-queries on a larger number $k > 1$ of tables.
% 
% Second, sub-queries with large cardinalties dominate the total cost of a plan; thus, either overestimating or underestimating a 
% 
As a trade-off between efficiency and effectiveness, the following strategy tries to guess where the biggest mistake is (by enumerating all possible values of $k$), and tune cardinality estimates for all size-$k$ sub-queries together.

The plan explorer in \rao is illustrated in Figure~\ref{alg: plangen}. In the outer loop (line 2), we enumerate a scaling factor $f \in F_\alpha^\Delta$, in the increasing order of $|\log f|$ with ties broken arbitrarily. That is, we prioritize exploring the neighboring plan space surrounding the native optimizer's choice (with $f$ closer to $1$), where plan quality is not too bad and the optimizer is likely to make mistakes, if any. 
Let ${\sf sub}_{k}(Q)$ be the set of size-$k$ sub-queries of a query $Q$ on $q$ tables.
An extensive experimental study in \cite{leis2015good} shows that sub-queries with the same size are often underestimated together with similar q-errors.
Thus, in the inner loop, we enumerate $k$ and tune the cardinality estimates for all sub-queries $Q' \in {\sf sub}_{k}(Q)$ together, resulting in a tuned estimator $\TCard$ (lines~5-6) to be fed into the optimizer.
\eat{This heuristic explores the neighboring plan space surrounding the native optimizer's choice, which is usually not too bad. Both better candidates and worse ones can be generated and it is the comparator's job to identify the best from them.} 
The resulting candidate plans are maintained in a priority queue for evaluation and execution whenever system resources allow. Considering the sizes of the two loops in lines~2-3, the following result is obvious.

\eat{
The plan explorer in \rao is illustrated in Figure~\ref{alg: plangen}.
Let ${\sf sub}_{k}(Q)$ be the set of size-$k$ sub-queries of a query $Q$ on $q$ tables.
An extensive experimental study in \cite{leis2015good} shows that sub-queries with the same size are often underestimated together with similar q-errors.
Thus, in the outer loop, we enumerate $k$, i.e., where the biggest mistake made by the native optimizer in estimating cardinality and plan choice is. In the inner loop, we choose a scaling factor $f \in F_\alpha^\Delta$ to tune the cardinality estimates for all sub-queries $Q' \in {\sf sub}_{k}(Q)$ together, resulting in a tuned estimator $\TCard$ (lines~5-6) to be fed into the optimizer.
% \rzhu{Feeling of tuning cardinality to decrease q-error.}
% 
% Our plan explorer is built on the intuition that the plan from a native query optimizer is usually not too bad; by tuning cardinality estimates, it may either generate some better plans in the neighboring plan space (if tuning towards the true cost of a sub-query which was unknown and incorrectly estimated), or some worse plans (tuning in the opposite direction).
% 
% Plans with various quality also increase the diversity among candidates to train the comparator, and it is the comparator's job to identify the best from them.
% 
This heuristic explores the neighboring plan space surrounding the native optimizer's choice, which is usually not too bad. Both better candidates and worse ones can be generated and it is the comparator's job to identify the best from them.
% 
% \highlight{The hope is that, with some choice of $f$, the max q-error of their cardinality and cost estimation can be reduced, and the optimizer may generate a better plan in the candidate list.}
}

\begin{proposition}
In the heuristic strategy in Figure~\ref{alg: plangen}, the number of candidate plans generated is at most ${\rm O}(q \cdot \log_\alpha\Delta)$.
\end{proposition}

\begin{figure}[t]
	\includegraphics[width=0.95\linewidth]{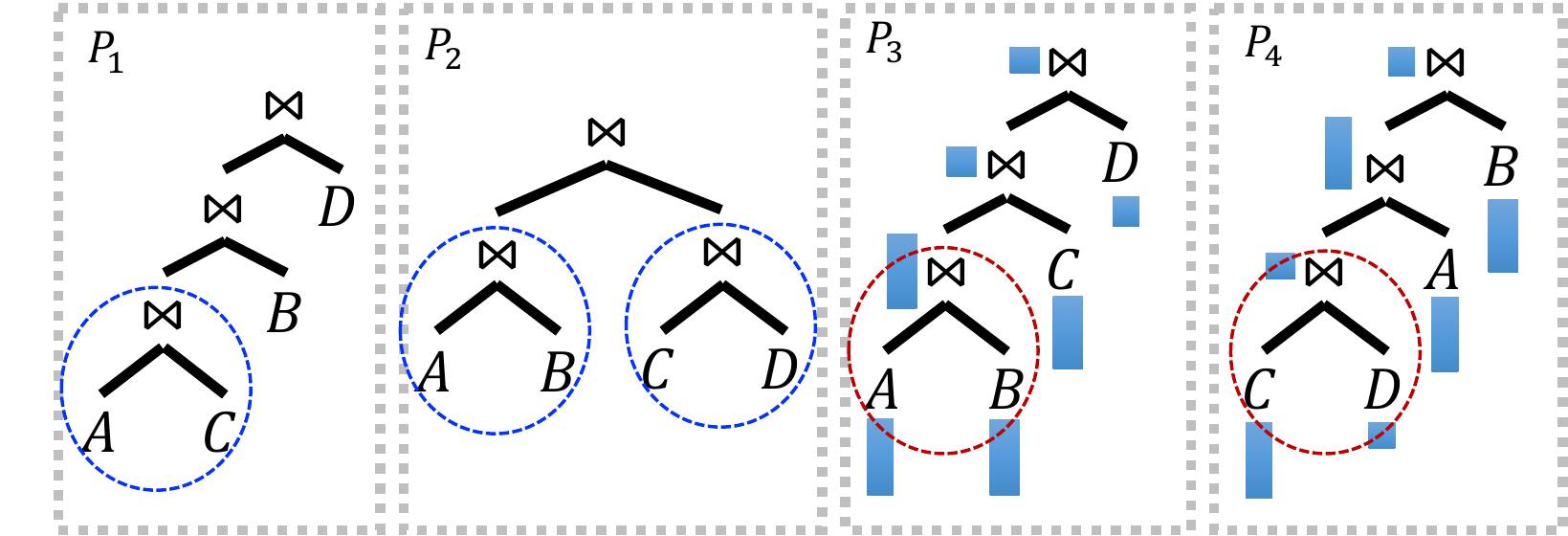}
	\vlshrink
	\caption{Intuitions on why our plan explorer based on sub-query grouping encourages diversity.}
	\label{fig: PlanEample}
	\vlshrink
\end{figure}

\sstitle{Encouraging Diversity in Candidates.} An important goal of our plan explorer is to generate diversified candidates. In Figure~\ref{fig: PlanEample}, we give conceptual examples of typical cases when our plan explorer encourages diversity in the candidates (with different plan shapes and join orders), for a query joining four tables, $A \bowtie B \bowtie C \bowtie D$:

({\em Diversity of plan shapes}) Two valid plans $P_1$ and $P_2$ have different shapes (left-deep tree and bushy tree, respectively). When we tune cardinality estimates for size-2 sub-queries ($k=2$ in line~2 of Figure~\ref{alg: plangen}), different values of $f$ would encourage different plan shapes. Namely, when $f \ll 1$, two sub-queries $A \bowtie B$ and $C \bowtie D$ in $P_2$ have reduced cardinality estimates in $\TCard()$, while only one sub-query $A \bowtie C$ in $P_1$ has a reduced cardinality estimate. As a result, the estimated cost of $P_2$ is reduced more than that of $P_1$, and thus a bushy tree is more likely to be generated by our plan explorer and included in the candidate list. When $f \gg 1$, with a similar logic, a left-deep tree is more likely to appear in the candidate list.

({\em Diversity of join orders}) Plans $P_3$ and $P_4$ are both left-deep trees but have different join orders. {In Figure~\ref{fig: PlanEample}, the length of bar on each single and joined table represents its size.} $P_3$ first executes $A \bowtie B$ which has a large estimated cardinality, while $P_4$ first executes $C \bowtie D$ which is estimated to be much smaller. When we tune cardinality estimates for size-2 sub-queries, we multiply both estimates with a scaling factor $f$. Suppose $P_3$ and $P_4$ have similar total estimated cost before tuning. When $f \ll 1$, the {\em decrease} of the tuned cardinality estimate $\TCard(A \bowtie B) - \Card(A \bowtie B)$ for $A \bowtie B$ is more significant than $\TCard(C \bowtie D) - \Card(C \bowtie D)$, and thus $P_3$ is more likely to be in the candidate list; when $f \gg 1$, the {\em increase} $\TCard(A \bowtie B) - \Card(A \bowtie B)$ is more significant,
% than $\TCard(C \bowtie D) - \Card(C \bowtie D)$, 
and thus $P_4$ is more preferable.

\sstitle{Some Implementation Details.} 
The tuned estimator $\TCard()$ in our plan explorer (in Figure~\ref{alg: plangen}) is not constructed explicitly by setting the cardinality estimate for every possible sub-query $Q'$ of $Q$ (as in lines~5-6). Instead, we only need a ``hook'' to the native cardinality estimator $\Card()$: if the size of a sub-query $Q'$ is $k$, we invoke $\Card(Q')$ and return $f \cdot \Card(Q')$ as $\TCard(Q')$; otherwise, we just return $\Card(Q')$.
We implement \rao on PostgreSQL.
% \rzhu{in a non-intrusive and straightforward way}.
% , which has a cost-based query optimizer based on dynamic programming.
%
Thanks to \rao's non-intrusive design, the implementation is straightforward. The only major modification to PostgreSQL is a hook function \textsf{pg\_hint\_plan} that implements the tuned cardinality estimator $\TCard()$ in the above way.
Our implementation on PostgreSQL can be easily ported to other databases that offer similar interfaces for cardinality estimators.

\eat{
\sstitle{Some Implementation Details.} The tuned estimator $\TCard()$ in our plan explorer (in Figure~\ref{alg: plangen}) is not constructed explicitly by setting the cardinality estimate for every possible sub-query $Q'$ of $Q$ (as in lines~5-6). Instead, we only need a ``hook'' to the native cardinality estimator $\Card()$: if the size of a sub-query $Q'$ is $k$, we invoke $\Card(Q')$ and return $f \cdot \Card(Q')$ as $\TCard(Q')$; otherwise, we just return $\Card(Q')$.}

\eat{
Let $|Q|$ denote the number of tables joined in query $Q$. We have two explicit observations for the sub-queries $Q'$ of $Q$: 1) the q-error of $\Card(Q')$ grows exponentially w.r.t.~$|Q'|$; and 2) the execution time of $Q'$ also grows exponentially w.r.t.~$|Q'|$. This implies that the native query optimizer is much easier to make mistakes in generating the partial plans for sub-queries joining a large number of tables~\cite{leis2015good, han2021CEbenchmark}. But at the same time, these plans may dominate the execution time of query $Q$. Following our design principle, our tuning strategy should pay more attention to such sub-queries joining larger number of tables.

Before introducing our heuristic approach, we review a bit on dynamic programming, the exact and widely used approach for plan generation in most DBMS. Let ${\sf sub}_{k}(Q)$ be the set of size-$k$ sub-queries of the query $Q$ and $|Q| = n$. On the last round to generate the plan for query $Q$, for each $1 \leq k \leq n - 1$, we pick a combination of one sub-query $Q_{1}^{k} \in {\sf sub}_{k}(Q)$ and another sub-query $Q_{2}^{k} \in {\sf sub}_{n - k}(Q)$ with the minimum estimated cost. The combination $(Q_{1}^{k^{*}}, Q_{2}^{k^{*}})$ minimizes the cost across all possible $k$ is identified as the plan for $Q$. When $n$ is large, the estimated cardinality of the sub-queries ($Q_{1}^{k}$ and $Q_{2}^{k}$) may be not accurate, so the selected combination $(Q_{1}^{k^{*}}, Q_{2}^{k^{*}})$ may be not the truly optimal choice. 

Then, we discuss how to intervene the plan generation process of $Q$ using tuned cardinality. We apply a heuristic strategy to guide the query optimizer to explore plans in different join forms. We present the method details in Figure~\ref{alg: plangen}.
Specifically, for each $1 \leq k \leq n - 1$, we simultaneously modify the cardinality of all queries $Q' \in {\sf sub}_{k}(Q)$ with each scaling factor $f \in F_\alpha^\Delta$. Obviously, when $f$ is very small, the cardinality, as well as the cost of the selected sub-query $Q_{1}^{k} \in {\sf sub}_{k}(Q)$ would also become very small. As a result, the native query optimizer would prefer to pick up the combination from ${\sf sub}_{k}(Q)$ and ${\sf sub}_{n - k}(Q)$. By changing $k$, the native query optimizer explores plans in different sub-spaces with different forms, e.g., $k = 1$, $1 < k < n-1$ and $k = n - 1$ correspond to the left-deep tree, busky-tree and right-deep tree, respectively. This could increase the plan diversity, as well as the opportunity to witness better plans. 

On the contrary, if the scaling factor $f$ is enough large, the native query optimizer tends to pick up the combination from ${\sf sub}_{k'}(Q)$ and ${\sf sub}_{n - k'}(Q)$ where $k' \neq k$. In original, if the combination of $(Q_{1}^{k}, Q_{2}^{k})$ with large estimation errors is wrongly selected as the resulting plan, it is implicitly excluded in this setting and better plans could be found. By~\cite{leis2015good}, the sub-queries are often likely to be underestimated together. Therefore, for proper scaling factor $f$, the maximum q-error across all sub-queries could be reduced. By~\cite{MoerkotteNS09preventing}, this could also increase the plan quality.

Afterwards, we obtain a list of candidate plans and we identify the best one from the combination $(Q_{1}^{k^{+}}, Q_{2}^{k^{+}})$ using our comparator model. We does not recursively apply the above heuristic method to tune the sub-plans of $Q_{1}^{k^{+}}$ and $Q_{2}^{k^{+}}$ due to three reasons: 1) The execution time of $Q_{1}^{k^{+}}$ and $Q_{2}^{k^{+}}$ is much less in comparison to $Q$; 2) The number of candidate plans would increase by recursion; and 3) The wisdom in the native query optimizer could generate  good enough plans if the size of $Q_{1}^{k^{+}}$ and $Q_{2}^{k^{+}}$ is not too large.
}

% \input{4-policy.tex}

%\input{5-implement.tex}

%\input{6-eval.tex}
% !TeX spellcheck = en_US

\section{Extensions and Discussion}
\label{sec: discussion}
\sstitle{Comparator with $d$-$dim$ Embedding.}
Our comparator $\CP$ could be extended to a $d$-dimensional ($d$-$dim$) embedding space. We only need to leave the dimensionality of the last plan embedding layer (in Figure~\ref{fig: PairLearn}) as $d>1$, with the hope to summarize more sophisticated statistical and structural information about each plan in plan embeddings. After that, the {comparison layer} is a learnable linear layer that compares the two $d$-$dim$ embeddings $\PRR(P_1) \in \mathbb{R}^d$ and $\PRR(P_2) \in \mathbb{R}^d$, and outputs $0$ ($P_1$ is better) or $1$ ($P_2$ is better). We defer detailed analysis to Appendix~\ref{app:multidim}.
%We defer detailed analysis to the Appendix~\ref{app:multidim}.

\sstitle{How to Handle Varying Resource Budget.}
Traditionally, each query is assigned a resource budget (e.g. work memory) under which the query is optimized and later executed. It is the database engine's responsibility to guarantee each query receives its assigned budget at runtime and ensure performance isolation among different queries to avoid any resource contention. This is an orthogonal task to query optimization. When resource condition in the database engine changes dramatically and the assigned budget for a query can no longer be guaranteed at runtime, re-optimization for the query under a different budget might be triggered. 

% \eat{In \rao, the comparator model keeps being updated with recent runtime stats to adapt to gradual changes of resource status; dramatic changes, however, can only be handled by re-training the comparator model.}

We assume constant resource budget per plan execution in this paper.
In principle, \rao can include resource budget information into the feature set of our comparator model, namely, as inputs to the plan embedding layers. In this way, the resulting plan embeddings encode such information about available resource, and plans can be compared in the comparison layer under different resource budgets at runtime. This augmented design will bring more challenges in model training, as the comparator now needs to observe runtime stats under various resource budgets to be fully trained, which we leave for future work.
% 
% For simplicity of presentation, 
% `reserve it as a future work' is not a right sentence

\section{Experimental Evaluation}
\label{sec: eval}
% 
% The design of \rao is system-agnostic. We implement it as a middleware on PostgreSQL. 
% by modifying only two hook functions in its native query optimizer. 
% It is easy to deploy \rao in other DBMS, as the major modification to DBMS is on the cardinality estimator, which has system provided interfaces in most databases.
% 
We make our implementation of \rao open-source \cite{LeroImp}. We also implement recently proposed learned query optimizers, Neo~\cite{marcus2019neo} and Bao~\cite{marcus2021bao}, and use Balsa's open-source implementation in~\cite{yang2022balsa, BalsaImp} on PostgreSQL.
% 
% We implement \rao on PostgreSQL, which has a cost-based query optimizer based on dynamic programming.
% 
% taking advantages of publicly available functions/interfaces.
% 
% \revise{
% Thanks to \rao's non-intrusive design, the implementation is straightforward. The only major modification to PostgreSQL is a hook function
% 
% \textsf{pg\_hint\_plan} 
% 
% that implements the tuned cardinality estimator $\TCard()$ introduced in Section~\ref{sec: policy}.
% }
% 
% To make minimal impact on the DBMS's kernel, we take advantage of publicly available functions/interfaces. 
% We make our implementation open-source \cite{LeroImp}.
% ~\footnote{\tiny \url{https://github.com/AlibabaIncubator/Learned-Query-Optimizer-by-Ranking-Plans}}. 
% Our implementation on PostgreSQL can be easily ported to other DBMSs that offer similar interfaces for cardinality estimators.
% 
We describe our experimental setup in Section~\ref{sec: eval-set}. 
% To extensively evaluate \rao, 
We first answer the most crucial questions about \rao's performance:

\squishlist
\item How much improvement on query execution performance could \rao achieve in comparison with PostgreSQL's native and other learned query optimizers? (Sections~\ref{sec: eval-gain} and \ref{sec: eval-balsa})
\item How much is \rao's query optimization cost? (Section~\ref{sec: eval-opt})
\item Could \rao adapt to workloads on dynamic data? (Section~\ref{sec: eval-dynamic})
\squishend

We then examine the design choices and settings in \rao and understand how they affect the performance of \rao:

\squishlist
\item What are the benefits of using pre-training  in \rao? (Section~\ref{sec: eval-pretrain})
\item Is the proposed plan exploration strategy efficient and effective? How is it compared with alternative strategies? (Section~\ref{sec: eval-policy})
% 
% \item How does idle resource usage affect \rao's performance? 
\item With less or limited idle resource, could \rao still achieve similar performance gains? (Section~\ref{sec: eval-idle})
% 
% \delete{\item What is the impact of concurrent query execution? Could \rao bring benefits under this setting? (Section~\ref{sec: eval-conc})}
\squishend

% and then report the evaluation results in a top-down manner. Specifically, Section~\ref{sec: eval-overall} presents the overall performance of our \rao and other learned QO systems in different settings. Section~\ref{sec: eval-policy} and  
%Section~\ref{sec: eval-train} examine the detailed performance of the plan generation policy and training methods in \rao, respectively.

\vspace{-0.7em}
\subsection{Experimental Setup}
\label{sec: eval-set}
% 
% \eat{\highlight{The idle computation resource in this machine is used for executing candidate plans in background.}}
% 
% \eat{We disable parallel computing to minimize the impact of other factors. Meanwhile, we close the GEQO function as it uses a heuristic genetic algorithm but not dynamic programming to search plans for queries joining a large number of tables.}

\sstitle{Benchmarks.}
% \label{sec: eval-set-benchmark}
We evaluate all the query optimizers on three widely used benchmarks. \eat{We summarize their properties in Table~\ref{tab:exp-dataset}.}

\squishlist
\item The IMDB dataset has 21 tables on movies and actors, and its JOB workload~\cite{leis2015good} has 113 realistic queries. 
%We denote the original JOB workload as IMDB-S in our experiments.
% 
For large-scale evaluation in most of our experiments, we generate a workload of 1,000 queries from JOB (similar to the one in \cite{marcus2021bao}):
% of 1,000 queries. Specifically, 
each time we randomly sample a query template from JOB, fetch its join template and attach some randomly generated predicates to it. 

\item The STATS dataset and STATS-CEB workload~\cite{stats2021benchmark, han2021CEbenchmark} are recently proposed to evaluate the end-to-end performance of query optimizer. STATS contains $8$ tables of user-contributed content on the Stats Stack Exchange network. Its data distribution is more complex than IMDB. STATS-CEB contains 146 query templates varying in join sizes and types. We generate a query workload using the same approach as described above.

\item The TPC-H~\cite{tpch2021} has its data synthetically generated under a uniform distribution. We set the scale factor to $10$, and use its query templates \#3, 5, 7, 8, 9, 10 for workload generation.
% Under each template, we generate 45 queries for training and 5 queries for testing. 
Under each template, we generate a number of queries with varying predicates.
% 50 queries. 
{We exclude other templates which are either too simple (on only one or two tables), or with views or nested SQL queries. 
Queries with views or nested SQL queries cannot be fully optimized by \rao due to a limitation of our implementation: to tune cardinality estimations for operators in the plans, we modify the hook function \textsf{pg\_hint\_plan}, which does not support to impose any hints on views and nested SQL queries.
% so we can not change the cardinality of some sub-queries for plan generation. 
We would try to fix this limitation in the future implementation.}

\item The TPC-DS~\cite{tpcds2021} is another benchmark for evaluating database performance. Similar to TPC-H, we also set its scale factor to $10$ and exclude all templates that are either too simple or can not be supported by the \textsf{pg\_hint\_plan} hook function. The remaining {23} templates are used to generate a query workload {with various predicates in a similar way to TPC-H}.
\squishend

\sstitle{Learned Optimizers in Comparison.}
% 
% Among the four learned optimizers, both Neo and Balsa require rebuilding the query optimizer from scratch, while \rao and Bao are designed to work with the native query optimizer. For Neo and Balsa, their training process is very expensive. 
% 
% They 
Neo and Balsa need to find a plan to execute using their latest model in each training epoch, and use the execution statistics to update the model. The generated plans are unknown before each epoch, so the training can only be done sequentially, which leads to a very long training time even with unlimited resource. By their evaluation~\cite{marcus2019neo, yang2022balsa}, the models converge after tens of epochs. On the contrary, both Bao and \rao can simultaneously run the selected plans and collect the training data in background, which greatly expedite the training process. 

{In our experiments, neither Neo nor Balsa could match the performance of PostgreSQL's native query optimizer after training for 72 hours on all of our datasets, except the original JOB workload (with 113 queries) on the IMDB dataset. On the contrary, Bao often outperforms PostgreSQL after training for several hours. Thus, we only report performance of Bao and \rao in most of our experiments in Sections~\ref{sec: eval-gain}-\ref{sec: eval-idle}. We compare \rao with Balsa on IMDB with the original JOB workload in Section~\ref{sec: eval-balsa}.
Bao and Balsa have demonstrated their superiority over Neo in~\cite{marcus2021bao, yang2022balsa}, so we do not further compare with it in the rest experiments.}
%
% We compare \rao with Balsa on the original JOB workload (IMDB-S).}

As described in~\cite{marcus2021bao}, Bao selects one candidate plan by {\em Thompson sampling} to execute and collects its execution time to update its model periodically. To have a fair comparison with \rao, we also implement an extended version of Bao, called Bao+, which witnesses more plans for model training. For each training query, it runs all candidate plans generated by its hint set tuning strategy using idle computation resource and collects their execution time to update the model periodically, in a similar way to \rao.

\sstitle{Evaluation Scenarios}
\label{sec: eval-set-scenarios}
We compare \rao with PostgreSQL's native query optimizer and other learned optimizers in different settings. 
PostgreSQL's native query optimizer does not need a separate training phase. For fair comparisons with the native and learned optimizers, we use the ``time series split'' strategy \cite{marcus2021bao} for training and evaluating Bao, Bao+, and \rao. 
Namely, unique queries in a workload are randomly shuffled as $Q_1, Q_2, \ldots$. The learned optimizers are always evaluated on queries that have {\em never} seen during the procedure of model training and updating.
% \eat{in the model training and updating phase, under two realistic scenarios:}. 
% 
In the experiments, we evaluate their performance under two realistic scenarios:
\squishlist
\item {\em Performance curve since\eat{ the first second of} deployment.} The learned optimizers are continuously updated since deployed for each workload. Bao+ and \rao may execute multiple candidate plans for a query on idle workers. When $Q_{t+1}$ is evaluated, all learned optimizers are only trained with information from earlier queries $Q_1, \ldots, Q_{t}$.
{Their models are updated in the background every 100 queries on IMDB and STATS, every 30 queries on TPC-H and every 50 queries on TPC-DS.}
By reporting the (accumulated) latency on each $Q_{t+1}$, we meter the performance of different optimizers since the deployment and how quickly they adapt to a new workload.
\item {\em Performance with stable models.} The learned optimizers tend to be stable (that is, the model training process converges) after seeing a sufficient number of queries in the workload, i.e., $Q_1, \ldots, Q_T$ ({\em training queries}), then we would use the optimizers to evaluate queries $Q_{T+1},$ $Q_{T+2},$ $\ldots$ ({\em test queries}) without further updating the models. {The numbers of test queries on IMDB, STATS, TPC-H and TPC-DS are 113, 146, 30 and 115, respectively.}
By reporting their (average) performance on $Q_{T+1},$ $Q_{T+2},$ $\ldots$, we compare the performance of learned optimizers after they are deployed on a workload and stabilized for a while.
\squishend
\eat{Table~\ref{tab:exp-dataset} summarizes properties of benchmarks and workloads. 
{Note that, in the default setting, we test \rao and other learned query optimizers in the single thread environment where each training or testing query is executed in sequential. We would evaluate the impact of concurrent query execution in Section~\ref{sec: eval-conc}.}}

\sstitle{Setup.}
We deploy learned optimizers on a Linux machine with an Intel(R) Xeon(R) Platinum 8163 CPU running at 2.5 GHz, 96 cores, 512GB DDR4 RAM and 1TB SSD. It is also equipped with one NVIDIA RTX-2080TI GPU for model training and inference. PostgreSQL 13.1 is installed and configured with 4GB shared buffers.

\eat{
%\smallskip
\sstitle{Baselines.}
%We implement \rao on top of the native query optimizer of PostgreSQL. We inject our plan exploration strategy described in Section~\ref{sec: policy} via {\sf planner\_hook} function in PostgreSQL. Our implementation is publicly available~\cite{LeroImp}. 
We compare \rao with both traditional and learned query optimizers:

1) \underline{\bf PostgreSQL}: . 

2) \underline{\bf Bao} and 3) \underline{\bf Bao+}:
Bao is the learned query optimizer proposed in ~\cite{marcus2021bao}. We adopt its implementation~\cite{bao2020impl} developed on PostgreSQL. Bao's latency prediction model is exploited and trained together. Each time, Bao selects one candidate plan by Thompson sampling to execute and collects its execution time to update its model periodically. Bao+ is an extended version of Bao to be fairer to compare with our \rao. It still applies the latency prediction model for plan selection but witnesses more plans for model training. For each training query, it runs all candidate plans generated by hint set tuning on idle workers and collects their execution time to update the model periodically. Therefore, Bao+ follows the similar framework of \rao, but its plan selection model and plan exploration strategy are different.

We do not compare with two other learned query optimizers Neo~\cite{marcus2019neo} and Balsa~\cite{yang2022balsa} due to two reasons:

1) Their design choices are totally different from Bao and our \rao. Neo and Balsa learn to rebuild the query optimizer from scratch, while our \rao and Bao are built on top of the native query optimizer. In~\cite{marcus2021bao}, Bao has exhibited its superiority over Neo in its evaluation results. 

2) Their training process is very time costly and not comparative. Specifically, in Bao+ and our \rao, the generated plans of each training query are all known before their execution, so we could easily run them in parallel on idle workers. However, Neo and Balsa need to find a plan to execute using their latest model in each training epoch. The generated plans are unknown before each epoch, so they could only be executed sequentially. This leads to a long training time. By their testing~\cite{marcus2019neo, yang2022balsa}, the models converge after tens of epochs. Whereas, Bao+ and our \rao could simultaneously collect the training data and run the selected plans in background. 

By our testing, neither Neo nor Balsa could match the performance of PostgreSQL, as well as Bao, after training for 72 hours. However, Bao/Bao+ and our \rao often outperform PostgreSQL after training for several hours.

\begin{table}[t]
    \centering
    \caption{Properties of the benchmarks.}
    \vspace{-1em}
    \scalebox{0.9}
    {
    \begin{tabular}{|c|ccc|} 
    \hline
    \rowcolor{mygrey}
    \sf Item & \sf IMDB & \sf STATS & \sf TPC-H \\ \hline
    Data Type & \multicolumn{2}{c}{Real-World} & Synthetic \\ \hline 
    \# of tables & 21 & 8 & 8 \\ \hline
    \# of test queries & 113 & 146 & 30\\ \hline
    \# of training queries & 1,000 & 1,000 & 270 \\ \hline
    \# of join table size & 4--17 & 2--8 & 3--8 \\ \hline
    \end{tabular}    }
    \label{tab:exp-dataset}
    \vspace{-1.5em}
\end{table}

%\smallskip
\sstitle{Datasets and Query Workloads.}
We evaluate all query optimizers on three widely used benchmarks. We summarize their statistical properties in Table~\ref{tab:exp-dataset} and introduce their details as follows:

1) The IMDB dataset with its JOB workload~\cite{leis2015good} is a well-known benchmark for query optimization. It has 21 tables on movie and actor data. JOB has
113 realistic queries with varied join sizes. We use JOB as the test workload and generate a training workload having 1,000 queries. Specifically, each time we randomly sample a query from JOB, fetch its join template, attach some randomly generated predicates and add it into the training workload. 

2) The STATS dataset and related STATS-CEB workload~\cite{stats2021benchmark, han2021CEbenchmark} are recently proposed to evaluate the end-to-end performance of query optimizer. STATS contains $8$ tables of user-contributed content on the Stats Stack Exchange network. Its data distribution is more complex than IMDB. STATS-CEB contains 146 queries varying in join size and types. We also use it as the test workload and generate a training workload having $1,000$ queries using the same above method.

3) The TPC-H~\cite{tpch2021} is regarded as a standard benchmark for database performance testing in the literature. Its data and queries are synthetically generated under a uniform distribution. We set the scale factor to $10$ and use its query templates 3, 5, 7, 8, 9, 10 for workload generation. Under each template, we generate 45 queries for training and 5 queries for testing. We exclude other templates which are either too simple (joining only two tables) or containing complex features such as views or sub-queries.

%\smallskip
\sstitle{Environments.}
We deploy our \rao and other learned query optimizers on a cluster having 9 Linux machines: one for providing the query optimizer services and the others serving as idle workers for executing candidate plans. Each machine has an Intel(R) Xeon(R) Platinum 8163 CPU running at 2.5 GHz having 96 cores, 512GB DDR4 RAM and 1TB SSD. One of them is equipped with 8 NVIDIA RTX-2080TI GPUs for model training. We install PostgreSQL 13.1 on each machine and build the learned query optimizers on top of it. Each PostgreSQL instance is configured with 4GB shared buffers. We disable parallel computing to minimize the impact of other factors. Meanwhile, we close the GEQO function as it uses a heuristic genetic algorithm but not dynamic programming to search plans for queries joining a large number of tables.
}

\subsection{Query Performance}
\label{sec: eval-gain}
We compare \rao with other optimizers for the two scenarios in Section~\ref{sec: eval-set-scenarios}.
% a realistic setting in terms of query performance. 
For the hyper-parameters of \rao, we set the cardinality tuning factor $\alpha = 10$ and q-error upper bound $\Delta = {10}^2$, so the set of scaling factors is $F_\alpha^\Delta = \{{10}^{-2}, {10}^{-1}, 1, 10, {10}^2\}$. 
% 
% Tuning in full version? We find that this could make the best performance for \rao. 
% 
% We warm up the comparator model in \rao by our pre-training method. 
% 
For Bao and Bao+, we use the same family of 48 hint sets~\cite{bao2020hintset} for plan generation. 
% 
% For fairness, 
% 
As in Bao's original implementation~\cite{bao2020impl}, Bao and Bao+ select the same plan as PostgreSQL's optimizer for the first 100 queries while their machine learning models are trained and warmed up.

\eat{
\begin{table}[t]
    \centering
    \scalebox{0.92}{\delete{
    \begin{tabular}{|l|cccc|} 
    \hline
    \rowcolor{mygrey}
    \sf Comparison Item & \sf IMDB & \sf STATS & \sf TPC-H & TPC-DS \\ \hline
    \# of tables & 21 & 8 & 8 & 25\\ \hline
    \# of training queries \eat{($Q_{1}, \ldots, Q_{T}$)} & 1,000 & 1,000 & 270 & 460 \\ \hline
    \# of test queries \eat{($Q_{T+1}, Q_{T+2}, \ldots$)} & 113 & 146 & 30 & 115 \\ \hline
    \# of tables joined in queries & 4--17 & 2--8 & 3--8 & 3--11 \\ \hline
    \end{tabular}}}
    \vspace{0.5em}
    \caption{\delete{Properties of benchmarks and workloads.}}
    \label{tab:exp-dataset}
    \vspace{-3em}
\end{table}}

\eat{
PostgreSQL's native query optimizer does not need a separate training phase. For fair comparisons with learned optimizers, we use the ``time series split'' strategy \cite{marcus2021bao} for training and evaluating Bao, Bao+, and \rao. 
Unique queries in the workload are randomly shuffled as $Q_1, Q_2, \ldots$. The learned optimizers are always evaluated on queries that have {\em never} seen before\eat{in the model training and updating phase, under two realistic scenarios:}. We consider query performance under two realistic scenarios:
\squishlist
\item {\em Performance curve since\eat{ the first second of} deployment.} The learned optimizers are continuously updated since deployed for each workload. Bao+ and our \rao may execute multiple candidate plans for a query on idle workers. But when $Q_{t+1}$ is evaluated, all learned optimizers are only trained with information from earlier queries $Q_1, \ldots, Q_t$.
\item {\em Performance with stable models.} The learned optimizers tend to be stable (that is, the model training process converges) after seeing a sufficient number of queries in the workload, e.g., $Q_1, \ldots, Q_T$, then we would use the optimizers to evaluate queries $Q_{T+1},$ $Q_{T+2},$ $\ldots$ without further updating the models.
\squishend
}
% In this subsection, we regard \rao as an overall system to answer the key question that whether \rao could bring more performance gains than traditional and existing learned query optimizers. First we introduce the configurations for running \rao and other query optimizers. Then we evaluate them on different settings of query workload and data.

% \sstitle{Configurations.}
% PostgreSQL could directly execute training and test queries on datasets. For Bao, Bao+ and our \rao, their models could be consistently updated w.r.t.~incoming queries. We randomly shuffle the training workload and then sequentially feed each query to the query optimizer. Bao+ and our \rao select one plan to execute but run other plans on idle workers in parallel to collect training data. 

\eat{
\begin{figure*}[t]
    \centering
    \subfigure{\includegraphics[width=0.27\linewidth, height=0.12\linewidth]{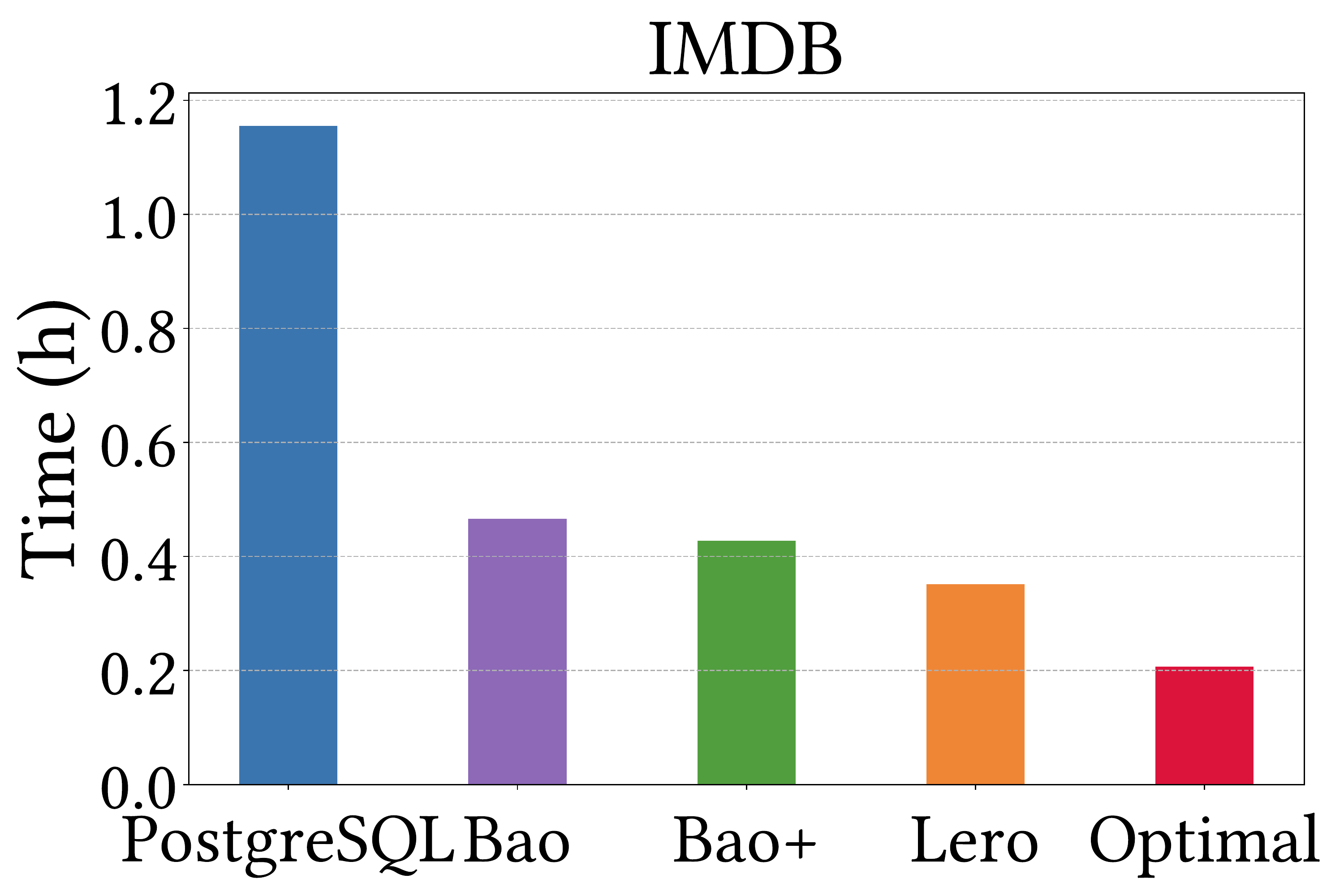}}
    \hspace{0.05\linewidth}
	\subfigure{\includegraphics[width=0.27\linewidth, height=0.12\linewidth]{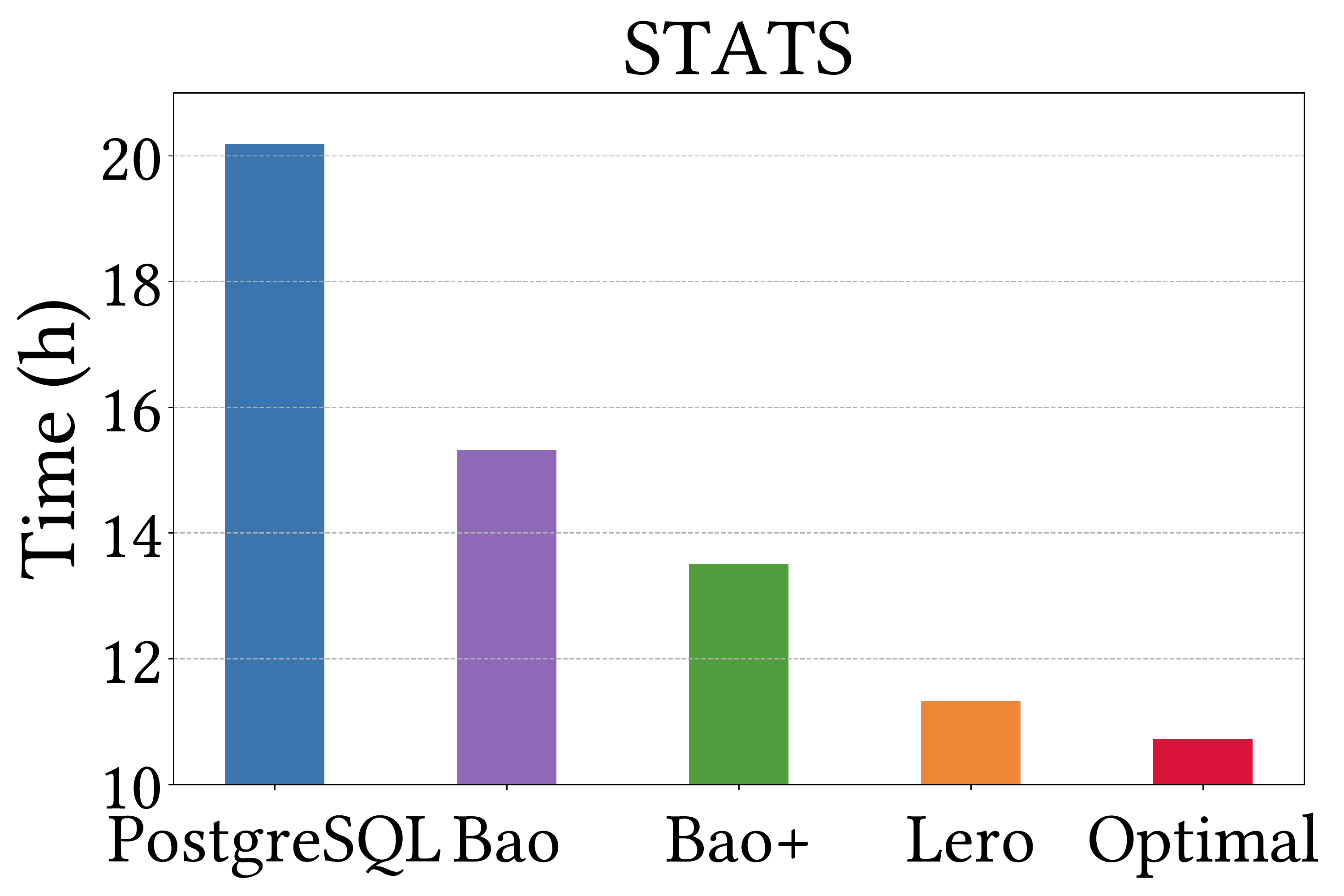}}
	\hspace{0.05\linewidth}
	\subfigure{\includegraphics[width=0.27\linewidth, height=0.12\linewidth]{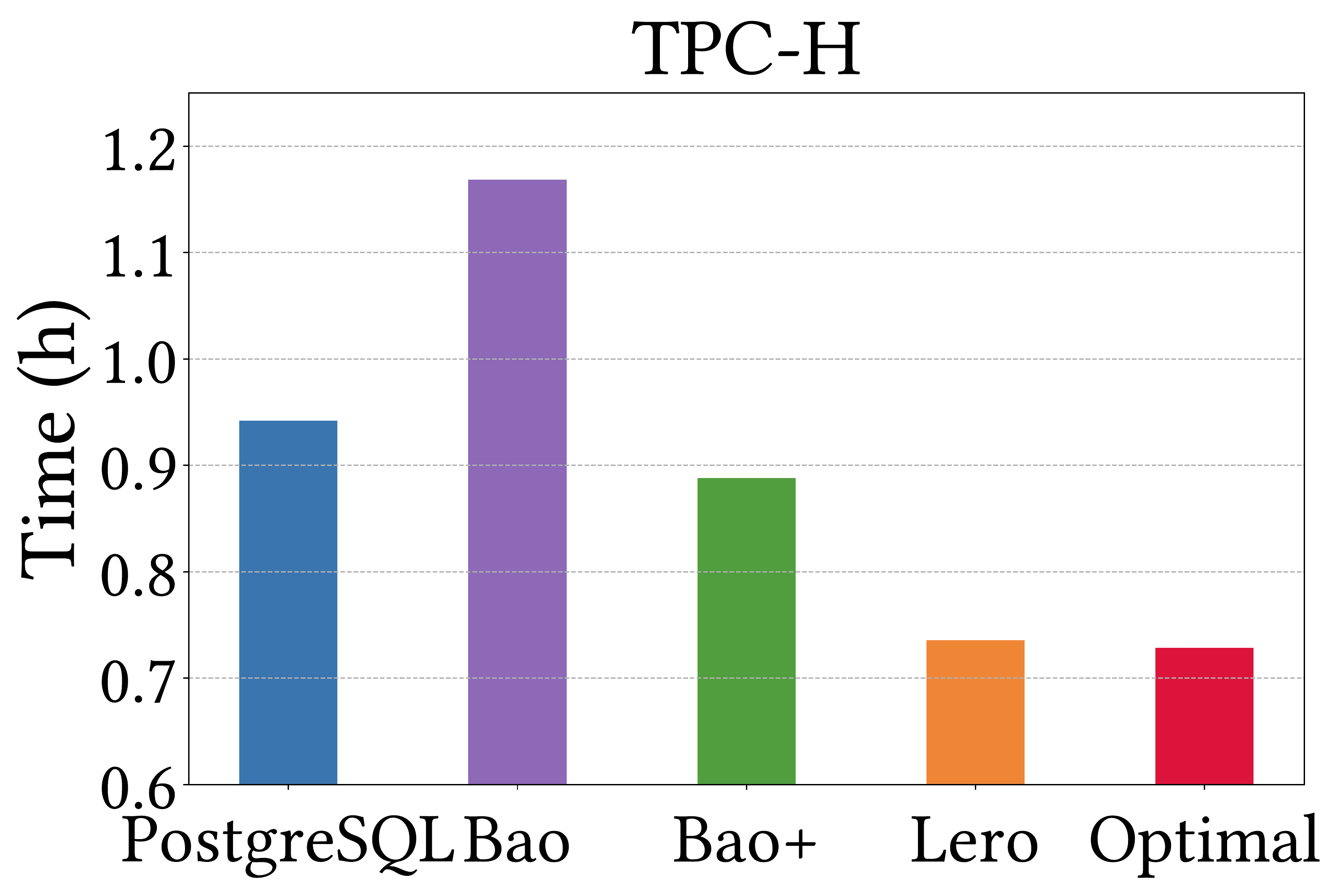}}
	\vspace{-1.5em}
	\caption{\delete{Performance of different query optimizers with stable models on three benchmarks.}}
	\label{fig: exp-overall-static}
	\vspace{-1.5em}
\end{figure*}
}

\begin{table}[t]
    \centering
    \scalebox{0.95}
    {
    \begin{tabular}{|c|cccc|}
    \hline
    \rowcolor{mygrey}
     & \multicolumn{4}{c|}{\text{\textsf{Execution Time (in hour)}}} \\ %\cline{3-6}
     \rowcolor{mygrey}
     \multirow{-2}{*}{\text{\textsf{Query Optimizer}}} & \text{\textsf{STATS}} & \text{\textsf{IMDB}} & \text{\textsf{TPC-H}} & \text{\textsf{TPC-DS}}\\ \hline
     {PostgreSQL} & {20.19} & {1.15} & {0.94} & {1.68}\\     \hline
     {Bao} & {15.32} & {0.47} & {1.17} & {1.55}\\ \hline
     {Bao+} & {13.85} & {0.41} & {0.89} & {1.57} \\ \hline
     {\rao} & {\bf 11.32} & {\bf 0.35} & {\bf 0.74} & {\bf 1.47}\\ \hline 
     {\bf Fastest Found Plan} & \large \bf 10.73 & \large \bf 0.19 & \large \bf 0.72 & \large \bf 1.39\\ \hline
    \end{tabular}
    }
    \caption{{Overall performance of different query optimizers.}}
    \vspace{-3em}
    \label{tab:exp-first}
 \end{table}

\begin{figure*}[t]
    \centering
    \subfigure{\includegraphics[width=0.27\linewidth]{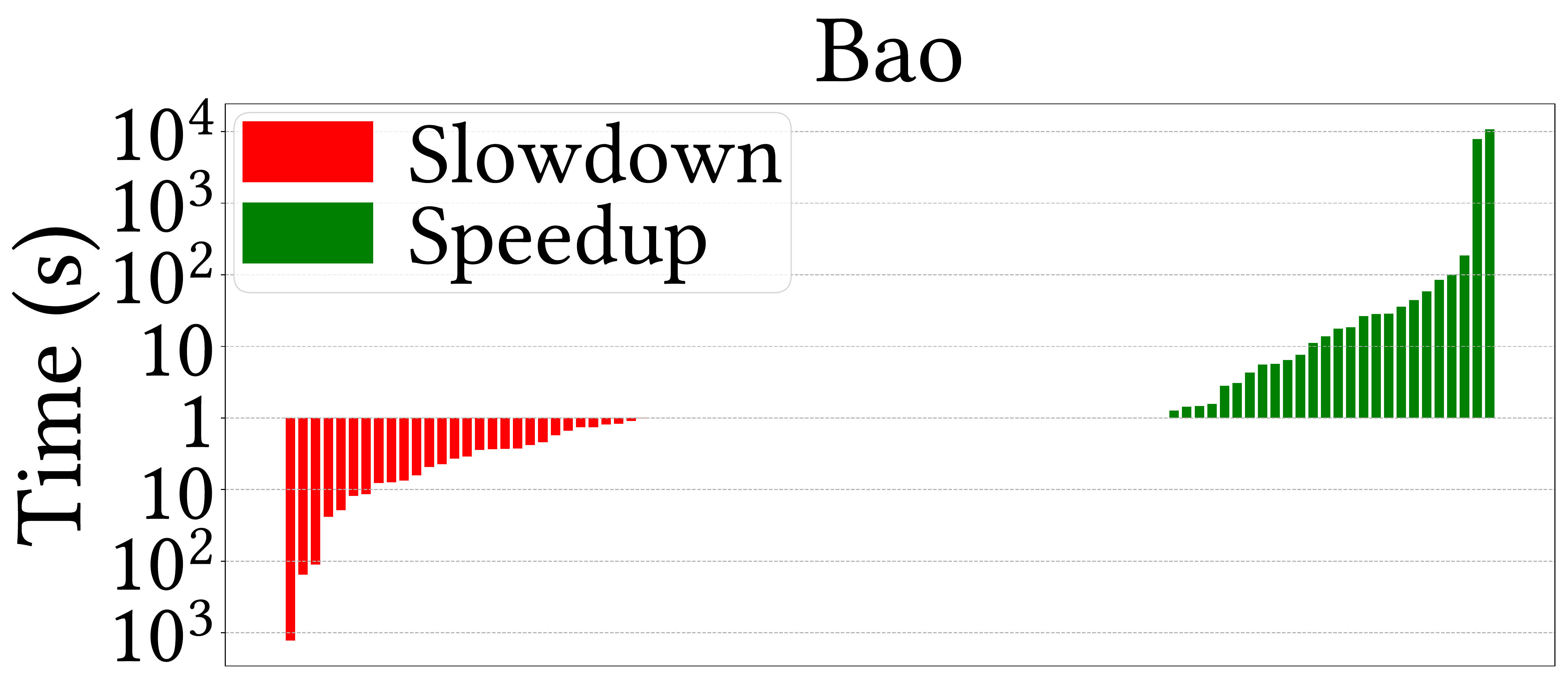}}
    \hspace{0.05\linewidth}
    \subfigure{\includegraphics[width=0.27\linewidth]{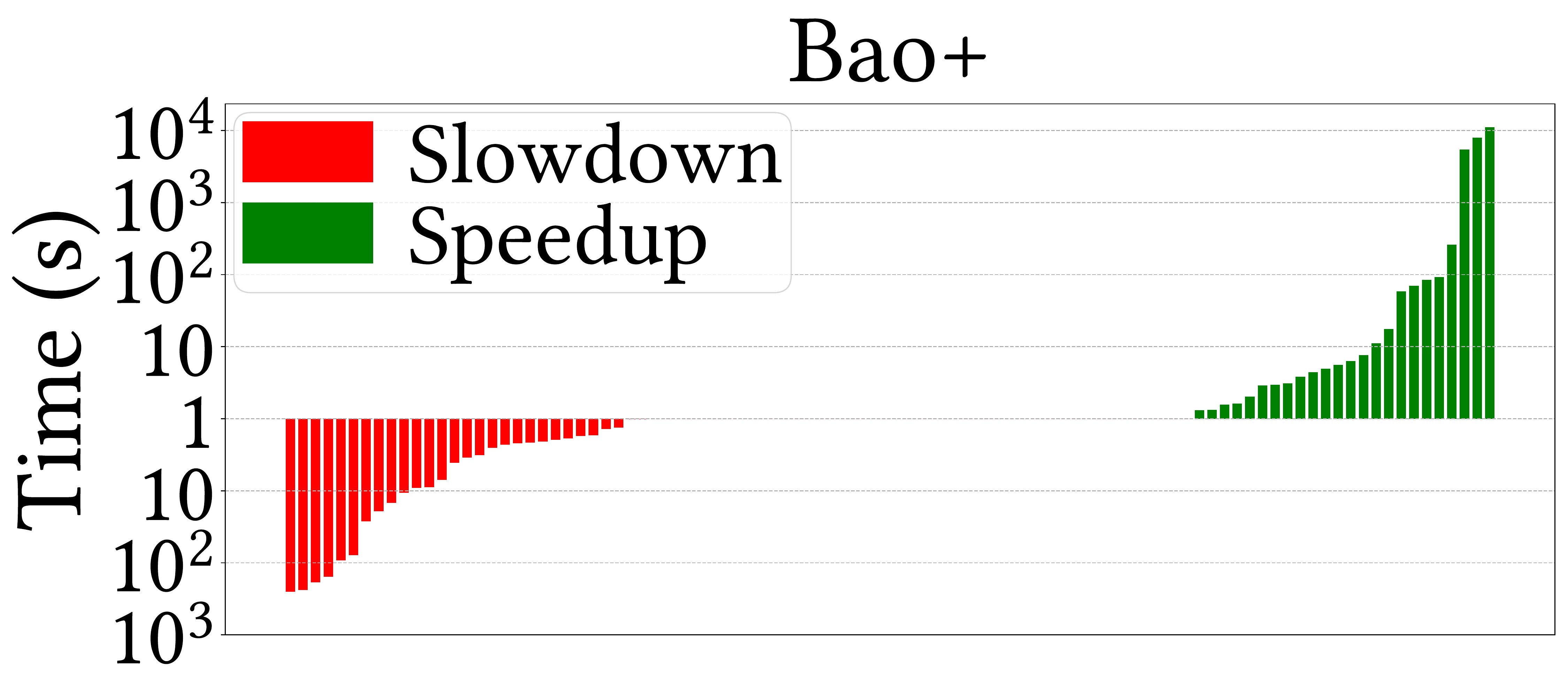}}
    \hspace{0.05\linewidth}
    \subfigure{\includegraphics[width=0.27\linewidth]{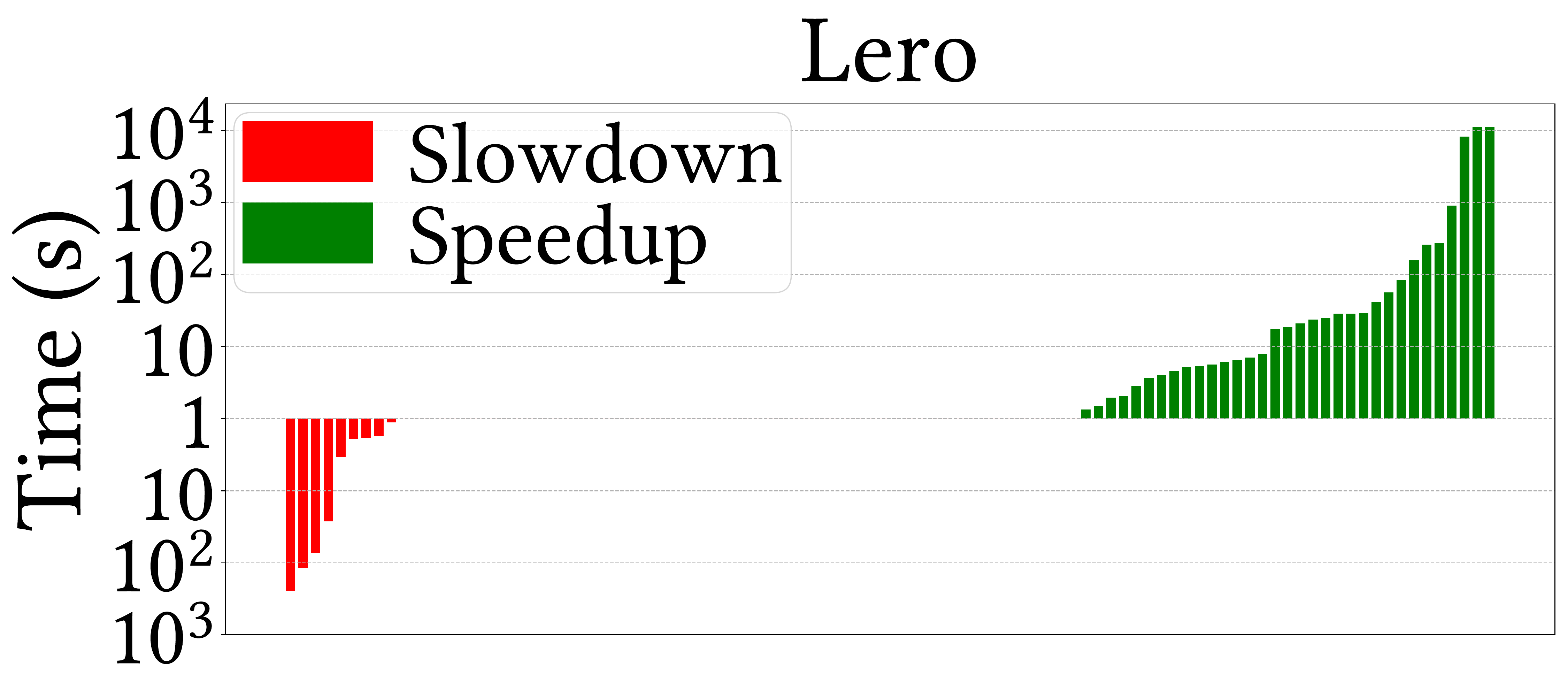}}
    \vspace{-2em}
	\caption{Per-query execution time of different query optimizers in comparison with PostgreSQL on the STATS benchmark.}
	\label{fig: exp-overall-detail}
	\vspace{-1.5em}
\end{figure*}

\subsubsection{Performance with Stable Models}
\label{sec: eval-gain: stable}
We first compare the performance of different optimizers, after Bao, Bao+, and \rao have been deployed for a while and the model training converges after seeing all training queries.
\eat{
In IMDB and STATS, we stop updating the learned optimizers after seeing 1000 training queries; in TPC-H, we stop updating them after 270 seeing queries.}
We will then use the learned optimizers to process the unseen test queries.
% 
% \jrzhou{The test workloads are never defined.}
% 
% First of all, we present the first glance performance of all query optimizers on the test workload of each benchmark after they finish the training workload. At this time, we do not further update the trained models in Bao/Bao+ and \rao but just use them directly to select plans to execute. 
% 
{Table~\ref{tab:exp-first} reports their performance on finishing all test queries on all benchmarks.}
% 
% the execution time on three benchmarks. 
% 
{{\sf ``Fastest Found Plan''} refers to the fastest plan generated by exhaustively search for each query.}
% 
% The optimal bar refers to the execution time of the optimal plans (found by exhaustively search) for all queries. 
% 
Overall, \rao achieves the best performance, compared with Bao, Bao+, and PostgreSQL's optimizer. {Its performance is close to the fastest found plans on STATS, TPC-H and TPC-DS}. 

\squishlist
\item {\rao's execution time is $70\%$, $44\%$, $21\%$ and {$13\%$} less than PostgreSQL's native query optimizer on IMDB, STATS, TPC-H, and {TPC-DS}, respectively.} This demonstrates \rao's practical value and advantages over the traditional optimizer in this scenario.

\item {\rao's execution time is $26\%$, $25\%$, $37\%$ and {$5\%$} less than Bao on IMDB, STATS, TPC-H and {TPC-DS}, respectively.} 
This verifies the effectiveness of our {\em learning-to-rank} paradigm and the {\em pairwise} trained comparator model (rather than a model predicting the exact latency).\eat{ Moreover, \rao takes full advantage of idle computation resources for executing more alternative plans of historical queries, so as to provide more accurate and fine-grained information for training the comparator.} Moreover, compared with the hint set tuning approach in Bao, the plan explorer in \rao is able to generate better and more diversified plans, as described in Section~\ref{sec: policy: algo}, for the comparator model to learn and achieve superior query performance. 
Detailed comparison and experimental analysis of different plan exploration strategies will be given in Section~\ref{sec: eval-policy}.
% 
% First, for each query, \rao generates multiple candidate plans for model training but Bao executes only one plan for each query. Second, \rao uses pairwise learning for plan ranking rather than predicting exact latency, which is much easier to train to attain higher accuracy.
% 
\item Bao+ learns from all the candidate plans generated by hint set tuning and achieves a decent performance improvement over Bao. Nevertheless, \rao's execution time is still around $20\%$ less than Bao+
on the three benchmarks, due to \rao's more effective pairwise learning-to-rank model and plan exploration strategy.
\squishend

\eat{
\squishlist
\item \rao outperforms PostgreSQL's native query optimizer by $1.9$ times, $3.3$ times, and $1.3$ times on the IMDB, STATS, and TPC-H, respectively. This demonstrates \rao's practical value and advantage over the traditional optimizer in this scenario.

\item \rao outperforms Bao by $1.4$ times, $1.3$ times and $1.6$ times on STATS, IMDB and TPC-H, respectively. 
%The improvement arises from two aspects. 
%  
This verifies the effectiveness of our {\em learning-to-rank} paradigm and the {\em pairwise} trained comparator model (rather than a model predicting the exact latency). Moreover, \rao takes full advantage of idle computation resources for executing more alternative plans of historical queries, so as to provide more accurate and fine-grained information for training the comparator.
% 
% First, for each query, \rao generates multiple candidate plans for model training but Bao executes only one plan for each query. Second, \rao uses pairwise learning for plan ranking rather than predicting exact latency, which is much easier to train to attain higher accuracy.
% 
\item Bao+ takes advantage of idle resources in a similar way as \rao to collect more information for its prediction model, with a significant performance improvement over Bao. However, \rao is still $1.2$ times faster than it on the three benchmarks. This is because of our learning-to-rank paradigm, as well as the plan explorer in \rao that encourages more truly good plans and diversity in the candidates (see Section~\ref{sec: policy: algo}) than the hint set tuning strategy in Bao+.
Detailed comparison and experimental analysis of different plan exploration strategies will be given in Section~\ref{sec: eval-policy}.
\squishend
}
% because Bao+ still applies the latency prediction model, which is not as accurate and robust as our pairwise comparator model. Meanwhile, the plan exploration strategy in \rao could find more truly good plans than the hint set tuning strategy in Bao+. 

\begin{figure*}[t]
    \centering
    \subfigure{\includegraphics[width=0.2\linewidth]{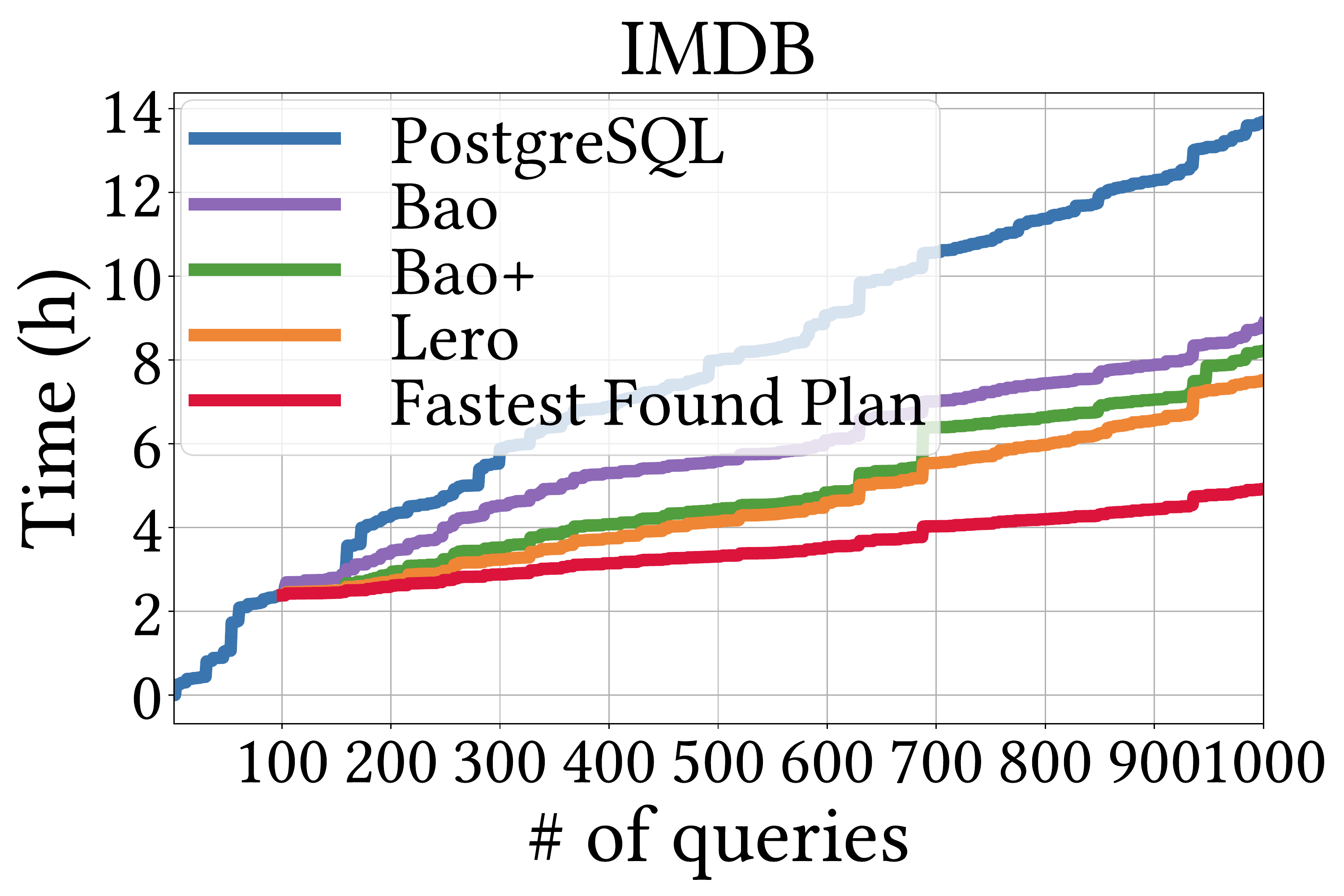}}
    \hspace{0.025\linewidth}
    \subfigure{\includegraphics[width=0.2\linewidth]{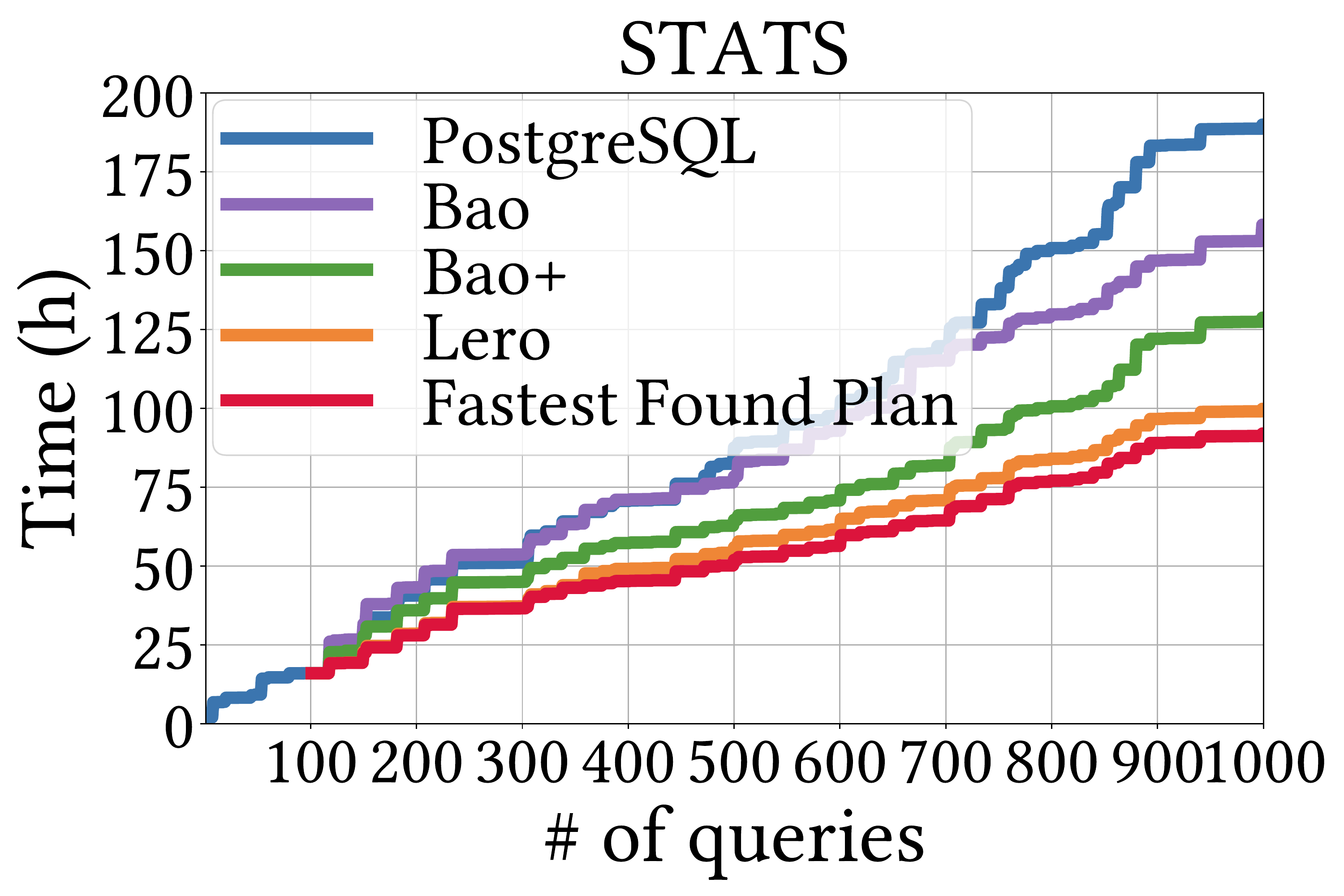}}
    \hspace{0.025\linewidth}
    \subfigure{\includegraphics[width=0.2\linewidth]{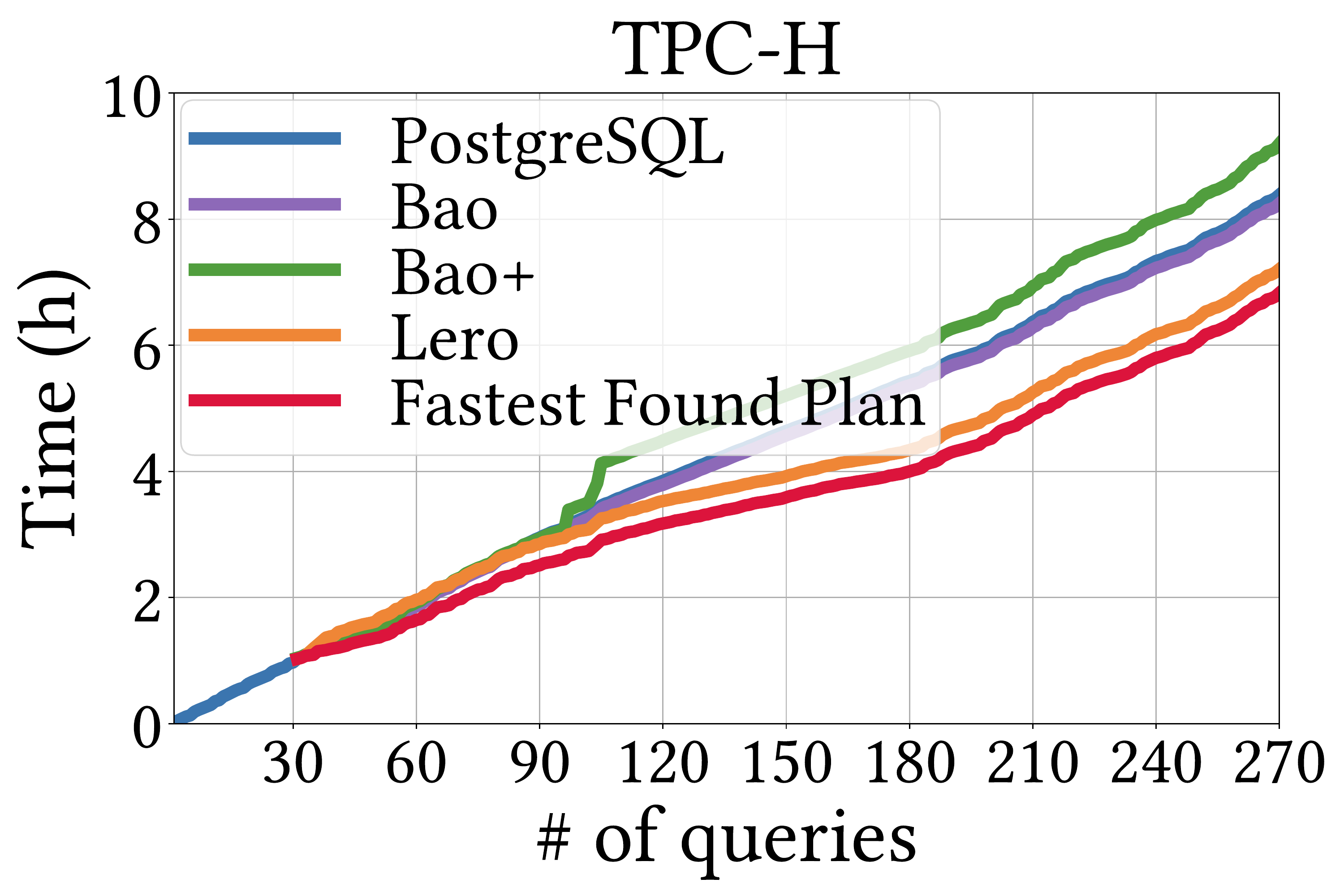}}
    \hspace{0.025\linewidth}
    \subfigure{\includegraphics[width=0.2\linewidth]
        {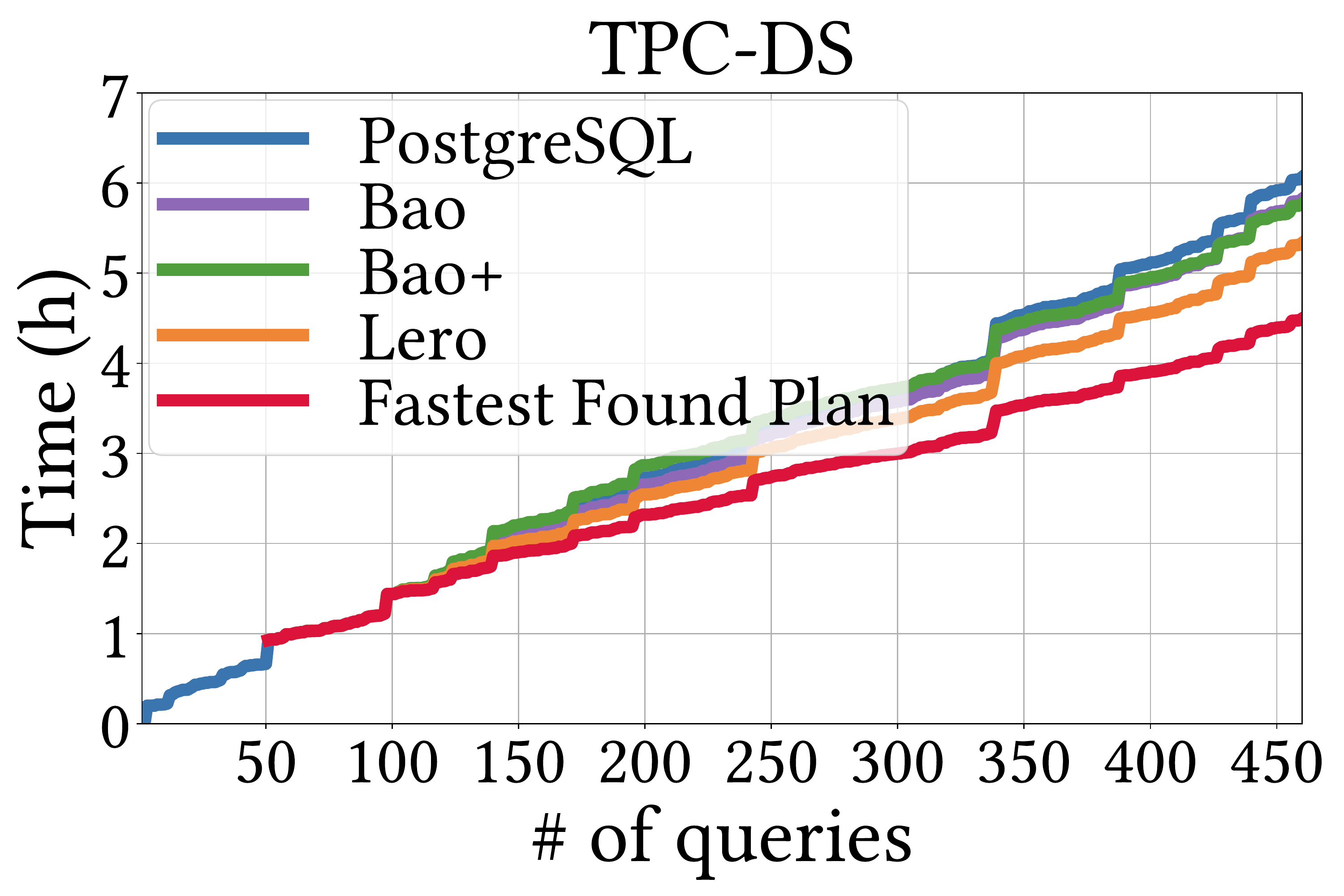}}
	\vspace{-2.0em}
	\caption{{Performance curve of different query optimizers since deployment on benchmarks.}}
	\label{fig: exp-overall-train}
	\vspace{-1.8em}
\end{figure*}

\subsubsection{Analysis of Performance Improvement/Regression.
\label{sec: eval-gain: stable: reg}
~}
Figure~\ref{fig: exp-overall-detail} compares the per-query execution time of each learned optimizer (Bao, Bao+ and \rao) with PostgreSQL on the 146 test queries of STATS.
% \highlight{
We do not plot the 47 queries, for each of which PostgreSQL and all the three learned optimizers choose the same plan for execution. We sort all remaining queries by latency differences between each optimizer and PostgreSQL from slowdown to speedup to visualize performance regression/improvement of \rao and others.
% }
% 
%For each optimizer, we do not plot queries whose performance differs from PostgreSQL's by less than one second (regarded as noise in timing latency), and sort the rest by latency differences from slowdown to speedup with learned optimizers.
% 
% Notice that, as \rao includes the original plan generated by PostgreSQL in the candidates, it could achieve zero regression with a perfectly trained model. The actual regression is due to the model noise and we do not further update model on test workload. 

{In comparison to Bao and Bao+, \rao significantly reduces performance regressions and brings much more performance gains.}
% 
% Only 9 queries are slowed down in \rao (for more than one second) among but accelerates 33 queries. 
% 
Only 9 queries ($6.2\%$) are slowed down in \rao (for more than one second) among 146 queries, while 33 queries are accelerated significantly.
Bao and Bao+ cause regressions for 29 queries each, which are even more than the number of queries they are able to improve (26 in Bao and 24 in Bao+). 
% 
% \rao could optimize lots of queries that were missed by Bao and Bao+. 
% 
When performance regression does happen in \rao, the relative slowdown of execution time, however, is much smaller. For instance, for the STATS workload, the maximum slowdown in \rao is $246s$, which translates to a relative performance regression of $5.2\%$, while the maximum slowdown in Bao is $1,276s$ with a relative performance regression of $46.5\%$. 
% 
%Our technical report~\cite{fullversion}
Appendix~\ref{app:addexp} provides more detailed analysis on how the plans selected by different optimizers rank among the truly best plans, which explains the performance improvement in \rao.

%\jrzhou{what is the percengtage?}
% 
% Thus, \rao provides more robust performance than other learned optimizers do.
% 
% Additionally, \rao can be further improved by considering alternative plans for the small percentage of performance regressions and refining its model gradually.

% The maximum relative slowdown of \rao is only $5.2\%$. These detailed results exhibit that the accuracy of our model is much higher, so \rao could improve more queries with fewer regressions. 
%We have similar observations on other benchmarks. We put them in Appendix~C of the full version~\cite{fullversion} due to space limits.

\eat{
\begin{table}[t]
    \centering
    \scalebox{0.92}
    {
    \revise{
    \begin{tabular}{|c|ccc|ccc|}
    \hline
    \rowcolor{mygrey}
     \text{\textsf{Query}} & \multicolumn{3}{c|}{\text{\textsf{Plans in $\mathcal{P}_{1}$}}} & \multicolumn{3}{c|}{\text{\textsf{Plans in $\mathcal{P}_{5}$}}}\\ %\cline{3-6}
     \rowcolor{mygrey}
     \text{\textsf{Optimizer}} & \text{\textsf{STATS}} & \text{\textsf{IMDB}} & \text{\textsf{TPC-H}} & \text{\textsf{STATS}} & \text{\textsf{IMDB}} & \text{\textsf{TPC-H}} \\ \hline
     
     {PostgreSQL} & 0.137 & 0.000 & 0.167 & 0.459 & 0.336 & 0.867  \\     \hline
     {Bao} & 0.144 & 0.159 & 0.000 & 0.226 & 0.230 & 0.167  \\ \hline
     {Bao+} & 0.096 & 0.221 & \bf 0.833 & 0.171 & 0.319 & \bf 1.000 \\ \hline
     {\rao} & \bf 0.425 & \bf 0.442 & 0.733 & \bf 0.925 & \bf 0.708 & \bf 1.000  \\ \hline 
    \end{tabular}
    }}
    \caption{\revise{{Rank-specific metrics for different query optimizers.}}}
    \vspace{-3em}
    \label{tab:exp-gain-rank}
\end{table}
}

\eat{
\subsubsection{Rank-Specific Analysis.}
\label{sec: eval-gain: stable: rank}
% Next, we analyze the plan quality in terms of the rank-specific metrics. 
\sout{
For a query optimizer, what we care about is whether the selected plan $P^{*}$ is truly the best one; if not, whether it is among the top-$k$ best plans. To this end, let $\mathcal{P}_{k}$ be the set of top-$k$ fastest plans generated by exhaustively search for each query (in terms of execution time). For different benchmarks, we record the ratio of queries for which $P^{*}$ falls into $\mathcal{P}_{k}$. Table~\ref{tab:exp-gain-rank} lists results for each query optimizer for $\mathcal{P}_{1}$ and $\mathcal{P}_{5}$. 
% 
%We find that more than $40\%$ of plans found by \rao is the exact best plan, and more than $70\%$ of plans is in the top-$5$ best plans. 
The percentage of plans selected by \rao falling into $\mathcal{P}_{1}$ and $\mathcal{P}_{5}$ is much higher than that by other query optimizers.
% This verifies the effectiveness of \rao from another perspective. 
}
}

\subsubsection{Performance Curves since Deployment}
\label{sec: eval-gain: curve}
% 
% We set their runtime stats repository to contain at most 2,500 query plans. 
% 
In Figure~\ref{fig: exp-overall-train}, we meter the performance curves of learned optimizers since their deployment for each workload,
% 
% Their models are updated in the background every 100 queries on IMDB and STATS and every 30 queries on TPC-H, respectively.
% 
% With the ``time series split'' evaluation strategy (see Section~\ref{sec: eval-set-scenarios}), we make sure that all the learned optimizers always process new queries, which have not yet been seen before when their models were trained and updated. 
% 
i.e., accumulated execution time of the best plans chosen by different optimizers in each workload (see Section~\ref{sec: eval-set-scenarios} for the definition).
% 
% For Bao/Bao+ and \rao, their models are updated periodically in background and each time they use the latest model for plan selection. 
% 
The ``{\sf Fastest Found Plan}'' curve refers to the conceptual latency lower bound as is defined in Section~\ref{sec: eval-gain: stable}.
\eat{Following are some key findings.}

% execution time of the best possible plan of all queries. We witness that:

\squishlist
\item The training of \rao converges faster. {After seeing 200 queries in IMDB, 100 queries in STATS, TPC-H and {TPC-DS} respectively, \rao's performance is consistently better than PostgreSQL and other learned optimizers.} While Bao and Bao+ eventually outperform PostgreSQL, they do so after seeing much more queries. This confirms that training \rao's learning-to-rank model for query optimization is much more effective than training the latency prediction models in Bao and Bao+.

\item The performance gaps between \rao and Bao/Bao+ gradually enlarge as more and more queries are executed. Eventually, \rao brings much more significant performance gain over PostgreSQL than Bao and Bao+ do. Its total execution time (right-end of each in Figure~\ref{fig: exp-overall-train}) is, e.g., $47\%$, $38\%$ and $23\%$ less than PostgreSQL, Bao and Bao+, respectively, on STATS.
% 
% Once again, this verifies the effectiveness of the \ltr paradigm and the design choices used in \rao. 

\item  \rao's performance is more robust. {It consistently outperforms PostgreSQL on all the benchmarks while the performance of Bao and Bao+ is sometimes worse than PostgreSQL (e.g. on TPC-H and {TPC-DS}).}
This is due to the intrinsic hardness of latency prediction which could require more training data and more time to converge in order to demonstrate any performance gain.
\squishend

% First, the intrinsic hardness of latency prediction affects the learning accuracy of the models, so Bao and Bao+ can not generalize well with varied queries. Second, the limitations of the hint set tuning policy expose more explicitly on TPC-H. 
% We also explain the details in Section~\ref{sec: eval-policy}. 

% As the comparator model in \rao is easier and more lightweight to train, \rao could attain higher learning efficiency to identify better plans earlier.

\subsection{Query Optimization Cost}
\label{sec: eval-opt}
While achieving significant improvement in query execution performance, \rao spends extra query optimization time in generating a list of candidate plans for an input query and applying the comparator model to pick the best candidate. The average query optimization time per-query of different optimizers is reported in Table~\ref{tab:exp-optcost}. We observe that this extra cost in query optimization is very low.
% 
% \highlight{Specifically, the average time \rao spends on generating the candidate plans and selecting the best for each query is a bit longer than the time spent by PostgreSQL's native optimizer on generating one plan.}
% 
% \eat{Specifically, the average time \rao spends on generating the candidate plans and selecting the best for each query is $1,736ms$, $16ms$ and $12.6ms$ on IMDB, STATS and TPC-H, respectively. It is a bit longer than the time spent by PostgreSQL's native optimizer on generating one plan ($842ms$,$7ms$ and $5.3ms$ on IMDB, STATS and TPC-H, respectively).}
% 
{In particular, the total extra cost is only at most $2.4\%$ of the total query execution time on IMDB and less than $0.1\%$ on STATS, TPC-H and {TPC-DS}. Whereas, \rao saves $13\%$ to $70\%$ execution time in comparison with PostgreSQL on these benchmarks.}

\subsection{Adapting to Dynamic Data}
\label{sec: eval-dynamic}
We now examine the performance of learned query optimizers on dynamic data\eat{ since the beginning of deployment}.
% 
% This scenario is more close to real-world settings. 
We use the STATS dataset for the experiment as each tuple is associated with a time stamp. We split the data by time. Initially, the earliest $50\%$ of the data is stored in the database. After the first 200 queries are executed on this database,
% and the learned optimizers trained with execution statistics, 
we add $12.5\%$ of the data into the database in the order of time stamps every 200 following queries. The goal is to evaluate how well different learned optimizers adapt to dynamic data by updating their models.\eat{, as their models are updated with execution statistics}
%
% The difference with the settings considered in Section~\ref{sec: eval-gain: curve} is that the data in tables keep updating as more and more queries come.

\begin{figure}[t]
    \centering
    \subfigure[{\small Performance since deployment}]{\includegraphics[width=0.46\linewidth, height=9em]{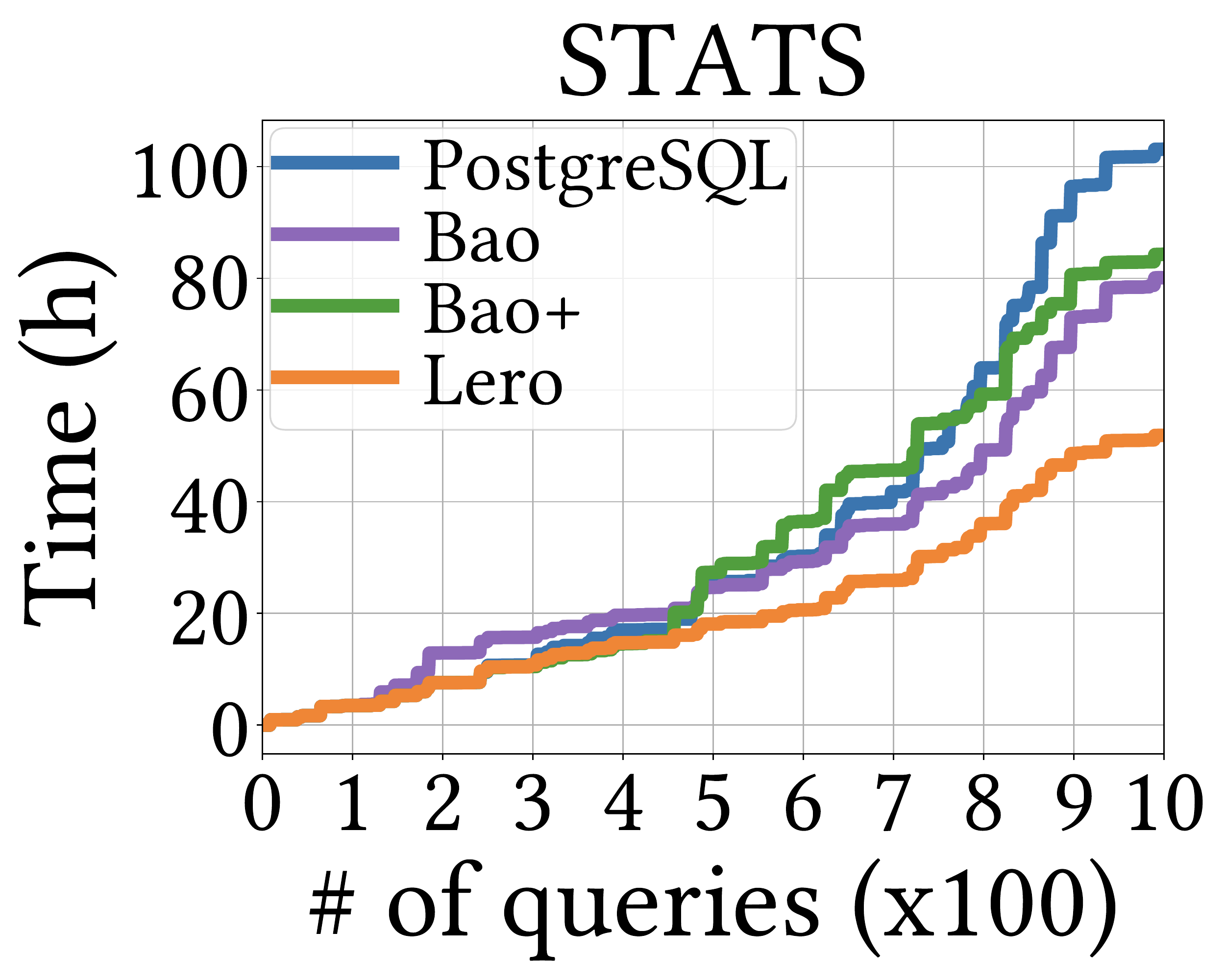}}
    \hspace{0.05\linewidth}
    \subfigure[{\small Performance \! with \! stable \! models}]{\includegraphics[width=0.47\linewidth, height=9em]{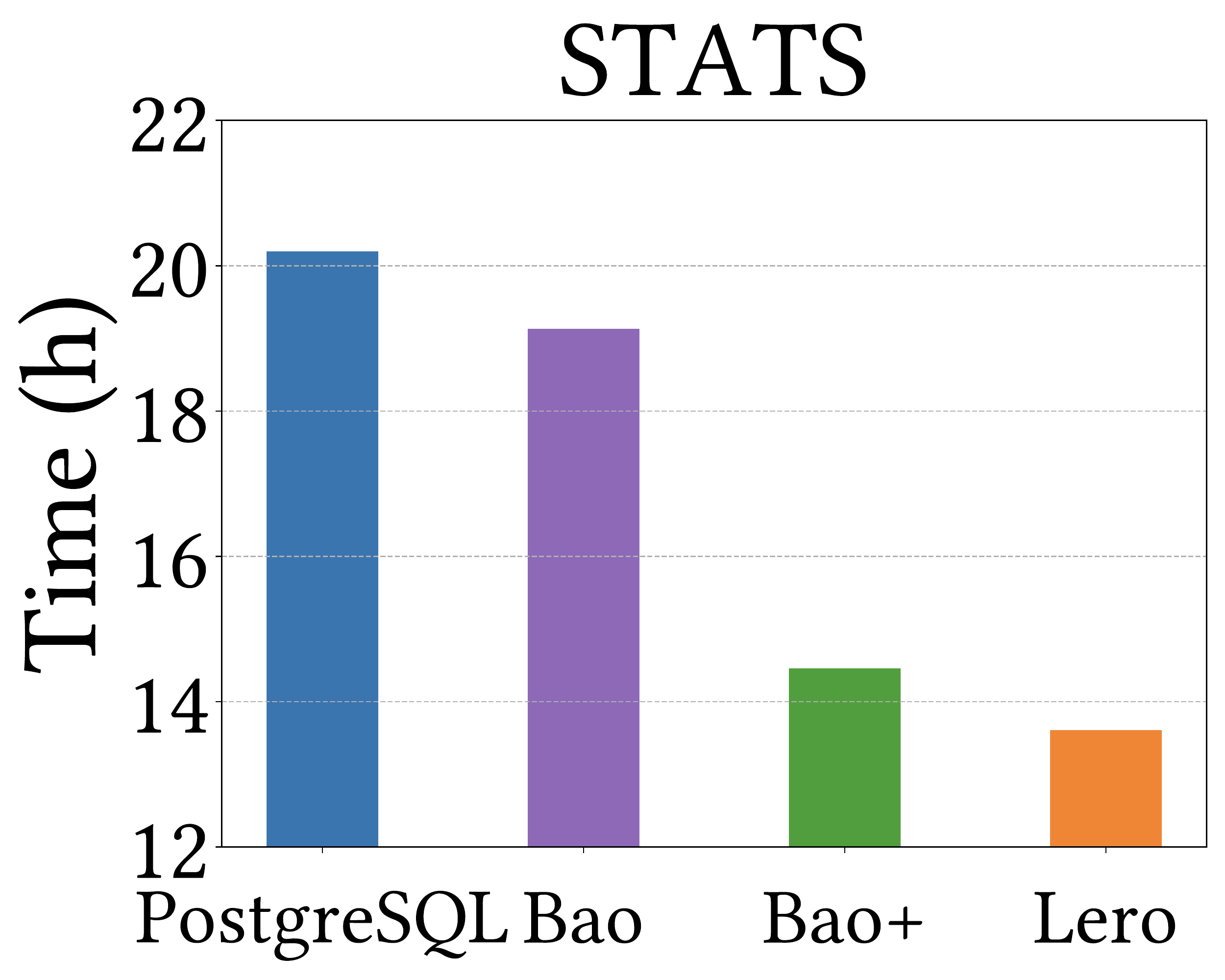}}
    \vspace{-2.0em}
    \caption{Performance of optimizers on dynamic data.}
    \label{fig: exp-overall-dynamic}
    \vspace{-1.8em}
\end{figure}

We report the {\em performance curve since deployment} \eat{(see Section~\ref{sec: eval-set-scenarios})} for each optimizer on dynamic data in Figure~\ref{fig: exp-overall-dynamic}(a), as well as their {\em performance with stable models} in Figure~\ref{fig: exp-overall-dynamic}(b).

% , we observe their performance curves with dynamic data and varied training queries.  reports the execution time of trained models on test workload. We find that: 

\rao outperforms PostgreSQL, Bao, and Bao+ in both settings.
After 1,000 queries, \rao's accumulated query execution time is $50\%$ less than PostgreSQL and $38\%$ less than Bao/Bao+ (right-end of Figure~\ref{fig: exp-overall-dynamic}(a)).
In parallel to query processing, all the learned optimizers continuously refine their models to adapt to the dynamic changes of data. It turns out that \rao adapts to such data changes better than the other two learned optimizers.
Eventually, after all the data changes are done, \rao's query performance (Figure~\ref{fig: exp-overall-dynamic}(b)) is $33\%$, $29\%$ and $10\%$ better than PostgreSQL, Bao and Bao+, respectively. 

% This verifies the effectivenss of \rao on processing dynamic data.

In Figure~\ref{fig: exp-overall-dynamic}(a), we observe \rao's robust performance, for the reason analyzed at the end of Section~\ref{sec: train: train}. Only \rao consistently performs better than PostgreSQL, while Bao is worse than PostgreSQL at the beginning and Bao+ performs worse in the middle. 
\eat{This is mainly because the comparator model in \rao is easier to be trained and updated than the latency prediction model in Bao and Bao+.}

\eat{
\rao outperforms PostgreSQL, Bao, and Bao+ in both settings.
Even with dynamic changes in data, starting from initial deployment to the first 1000 training queries in the workload, \rao's execution time for finishing these queries is $50\%$ less than PostgreSQL and $38\%$ less than Bao/Bao+ (right-end of Figure~\ref{fig: exp-overall-dynamic}(a)). Note that the machine learning models are updated in parallel to the query processing, so that \rao, Bao, and Bao+ keep adapting to the dynamic changes of data. It turns out that \rao adapts to such data changes better than the other two.
After that, \rao's execution time finishing the test queries without updating the model any more (Figure~\ref{fig: exp-overall-dynamic}(b)) is $33\%$, $29\%$ and $10\%$ less than PostgreSQL, Bao and Bao+, respectively. 

% This verifies the effectivenss of \rao on processing dynamic data.

In Figure~\ref{fig: exp-overall-dynamic}(a), we can also observe that only \rao could consistently perform better than PostgreSQL, while Bao is worse than PostgreSQL at the beginning and Bao+ performs worse in the middle of training. 
This is mainly because the comparator model in \rao is easier to be trained and updated than the latency prediction model in Bao and Bao+.
% 
% The reasons are twofold. First, training the comparator model is more lightweight, so \rao attains higher update speed to keep track of data and workload changes. Second, the comparator model is more robust to tolerate errors. It learns a 1-D embeddings but not the exact latency for plan selection. As long as the plan embeddings could keep the relative order of plans, the model could find the best plan. 
}

\begin{table}[t]
    \centering
    \vspace{0.1in}
    \scalebox{0.92}
    {
    {
    \begin{tabular}{|c|ccc|} 
    \hline
    \rowcolor{mygrey}
    \sf Time (in millisecond) & \sf PostgreSQL & \sf Bao/Bao+ & \sf \rao \\ \hline
    IMDB & $842$ & $856$  & $1,736$\\ \hline
    STATS & $7$ & $8.1$  & $16$\\ \hline
    TPC-H & $5.3$ & $6.2$ & $12.6$\\ \hline
    {TPC-DS} & {$6.7$} & {$8.5$} & {$15.8$}\\ \hline
    \end{tabular}
    }}
    %\vspace{0.05in}
    \caption{{Average query optimization time per query.}}
    \label{tab:exp-optcost}
    \vspace{-0.4in}
\end{table}

\subsection{Importance of Pre-Training}
\label{sec: eval-pretrain}
\rao relies on the pre-training procedure to learn from the native query optimizer and better bootstrap its own model. {The pre-training time is only around 5 minutes on each of these benchmarks. It converges fast as the native cost model often consists of functions with simple structures, i.e., piecewise linear or quadratic function of estimated cardinalities of sub-queries with magic constants as coefficients for different operators.}

\begin{figure}[t]
    \subfigure{\includegraphics[width=0.4\linewidth]{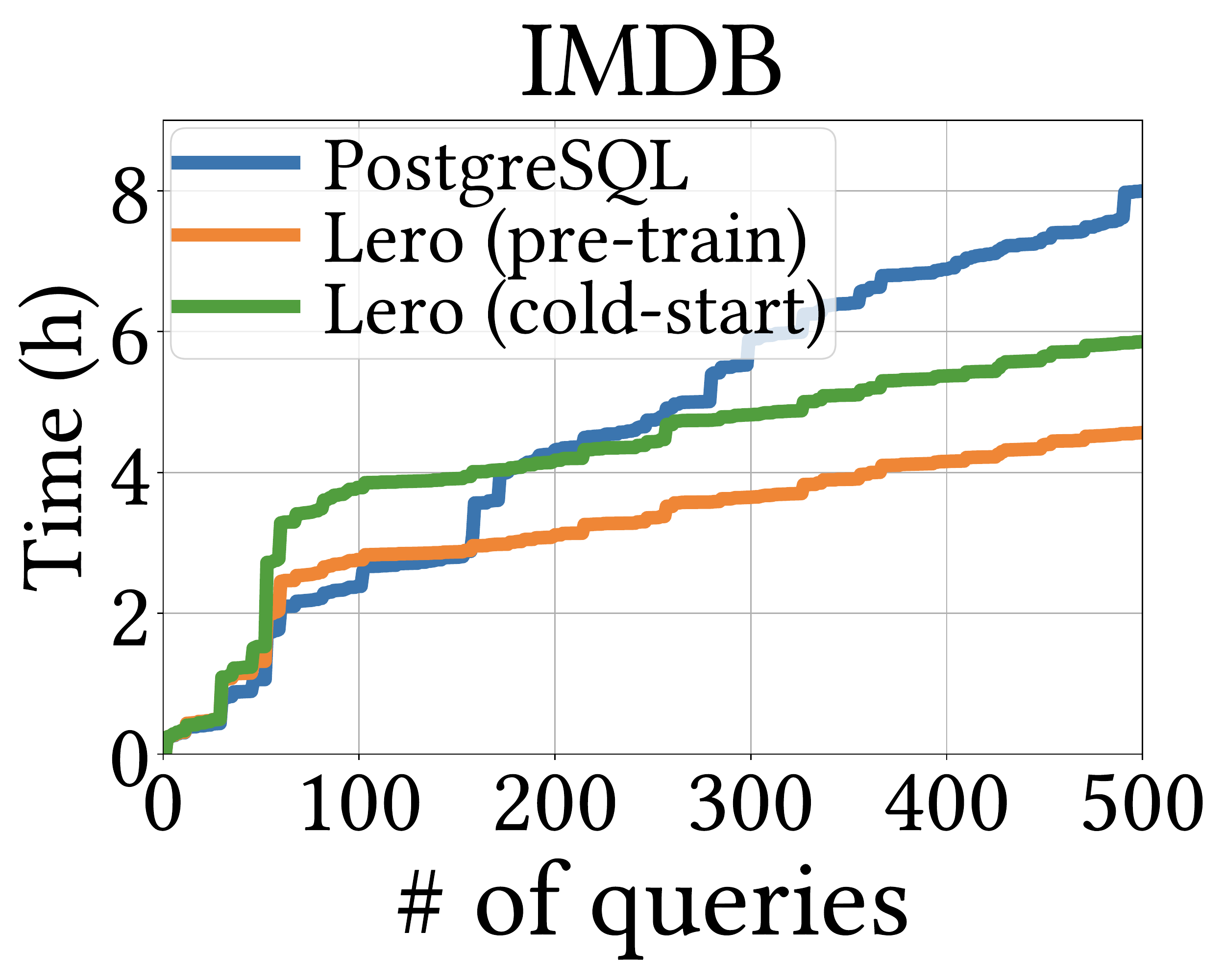}}
    \hspace{0.05\linewidth}
    \subfigure{\includegraphics[width=0.4\linewidth]{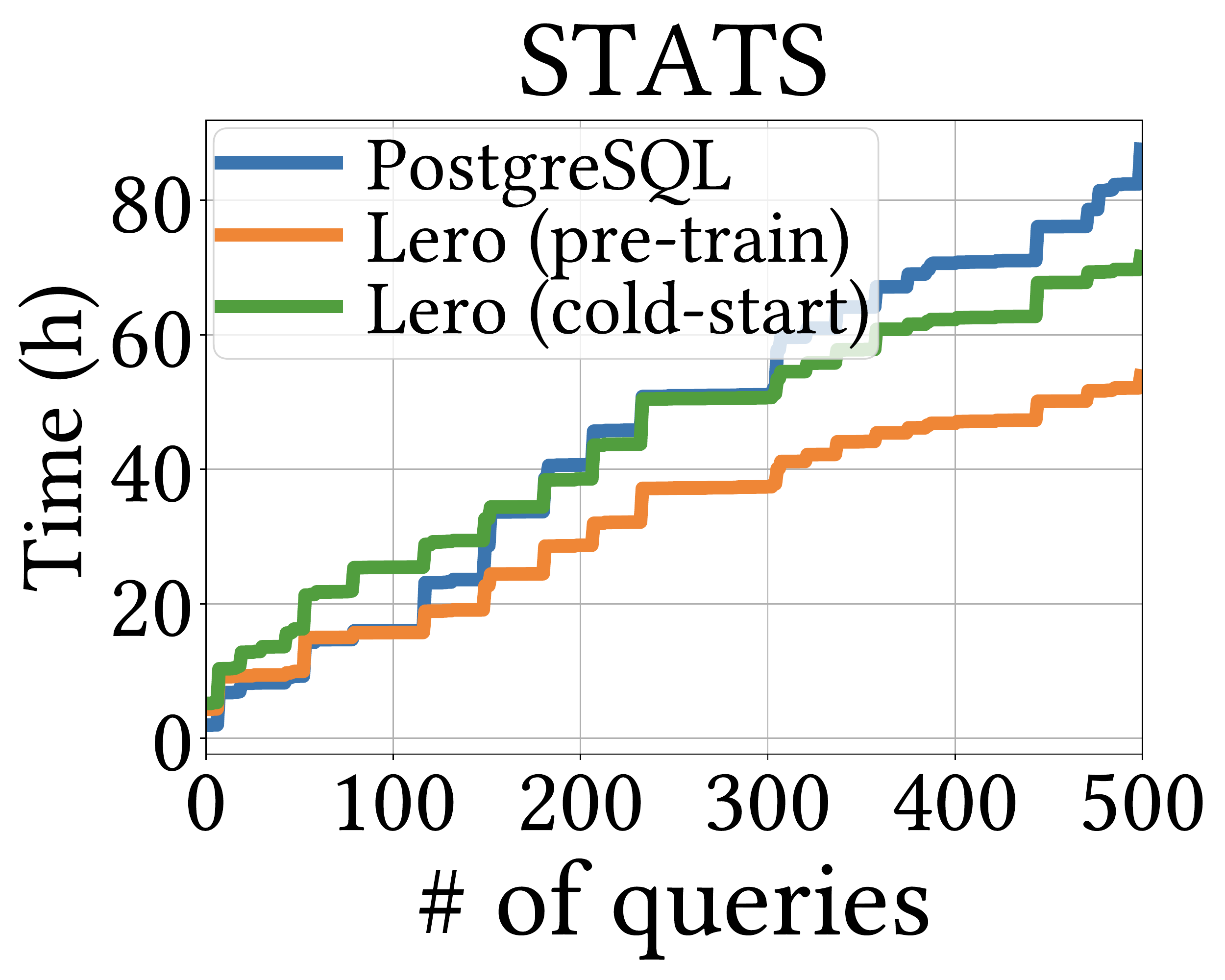}}
    \vspace{-1.9em}
	\caption{Effects of pre-training on \rao's performance curve.}
	\label{fig: exp-detail-pretrain}
	\vspace{-2.1em}
\end{figure}

In this experiment, we evaluate the impact of the pre-training procedure on \rao's performance. We start with either a pre-trained comparator model, or a cold-start model with random parameters; both are continuously trained/updated as more and more queries are executed. Figure~\ref{fig: exp-detail-pretrain} illustrates their performance curves on IMDB and STATS for the first 500 queries (the results on TPC-H are similar, so the figure is omitted due to the space constraint).

\eat{It can be seen that the pre-training procedure is important to \rao. }With pre-training, for the initial 100 queries, \rao could generate almost the same plans as PostgreSQL since its comparator is pre-trained to fit PostgreSQL's native cost model; thus, their performances are very close (this can be also observed in our former experiments where \rao is pre-trained). Note that \rao does not need to execute any query during the pre-training procedure, as introduced in Section~\ref{sec: train: train}. Without pre-training, \rao's performance can be worse than PostgreSQL for the first 200 queries before it eventually catches up. {After executing nearly 300 queries, the slope of the two methods are almost the same. This indicates that \rao with a cold-start could catch up but with more training time.}

Proper pre-training from the knowledge of the native query optimizer gives \rao a good starting point and also accelerates its model's convergence. With pre-training, \rao could consistently outperform PostgreSQL after the initial 100 queries, while with a cold-start comparator, \rao consistently outperforms PostgreSQL only after seeing nearly 200 queries.

% 1) The pre-training method could eliminate the cold-start problem. For the initial 100 queries, \rao could generate almost the same plans as PostgreSQL so their performance is very similar (this is also observed in our former experiments). However, without pre-training, \rao's performance is worse than PostgreSQL for the former 200 queries.

% 2) The pre-training method could accelerate the model's converge speed. \rao with pre-training could consistently outperform PostgreSQL after the initial 100 queries while the cold-start method  outperforms PostgreSQL after seeing nearly 200 queries. 
% 
% This is because without pre-training, the model parameters are set randomly. However, by pre-training the model to capture the cost computation rules, it is a rough proxy to approach the execution time (although with some errors), so it converges much faster.

% These results indicate that pre-training could help our training process and makes \rao more practical.

\eat{
\sstitle{Remarks on Training Time.}
After pre-training, the model training cost of \rao is also very low. Each time the model is trained or updated, the average time needed in \rao is only $11.3$, $7.6$ and $5.3$ minutes on IMDB, STATS and TPC-H, respectively. This ensures the fast update of the models to keep track of the data changes.
}

% We examine the effects of pre-training on \rao's comparator model. We compare the performance of \rao with and without per-training. Figure~\ref{fig: exp-detail-pretrain} illustrates their performance curves on the training workload for the first 500 queries. We observe that:

\begin{table}[t]
\centering
\vspace{0.1in}
\scalebox{0.75}{
    \begin{tabular}{|c|c|ccc|}
    \hline
    \rowcolor{mygrey}
     & & {\sf Average} & \sf Plans Faster & \sf Plans Slower \\ 
    \rowcolor{mygrey}
    \multirow{-2}{*}{\sf Dataset} & \multirow{-2}{*}{\sf Strategy} & {\sf $\#$ of Plans} & \sf than PostgreSQL & \sf than PostgreSQL \\ \hline
    \multirow{2}{*}{IMDB} & \rao's plan explorer & 9 & 47\% & 42\% \\
    & Hint set tuning & 16 & 36\% & 57\% \\ \hline
    \multirow{2}{*}{STATS} & \rao's plan explorer & 5 & 31\% & 49\% \\
    & Hint set tuning & 19 & 25\% & 69\% \\ \hline
    \multirow{2}{*}{TPC-H} & \rao's plan explorer & 7 & 24\% & 64\% \\
    & Hint set tuning & 15 & 6\% & 88\% \\ \hline
    \end{tabular}}
\caption{{Number and quality of {unique} candidate plans generated by different plan exploration strategies.}}
    \label{tab:exp-detail-strategy}
    \vspace{-0.4in}
\end{table}

\begin{figure}[t]
    \centering 
    \subfigure{\includegraphics[width=0.4\linewidth]{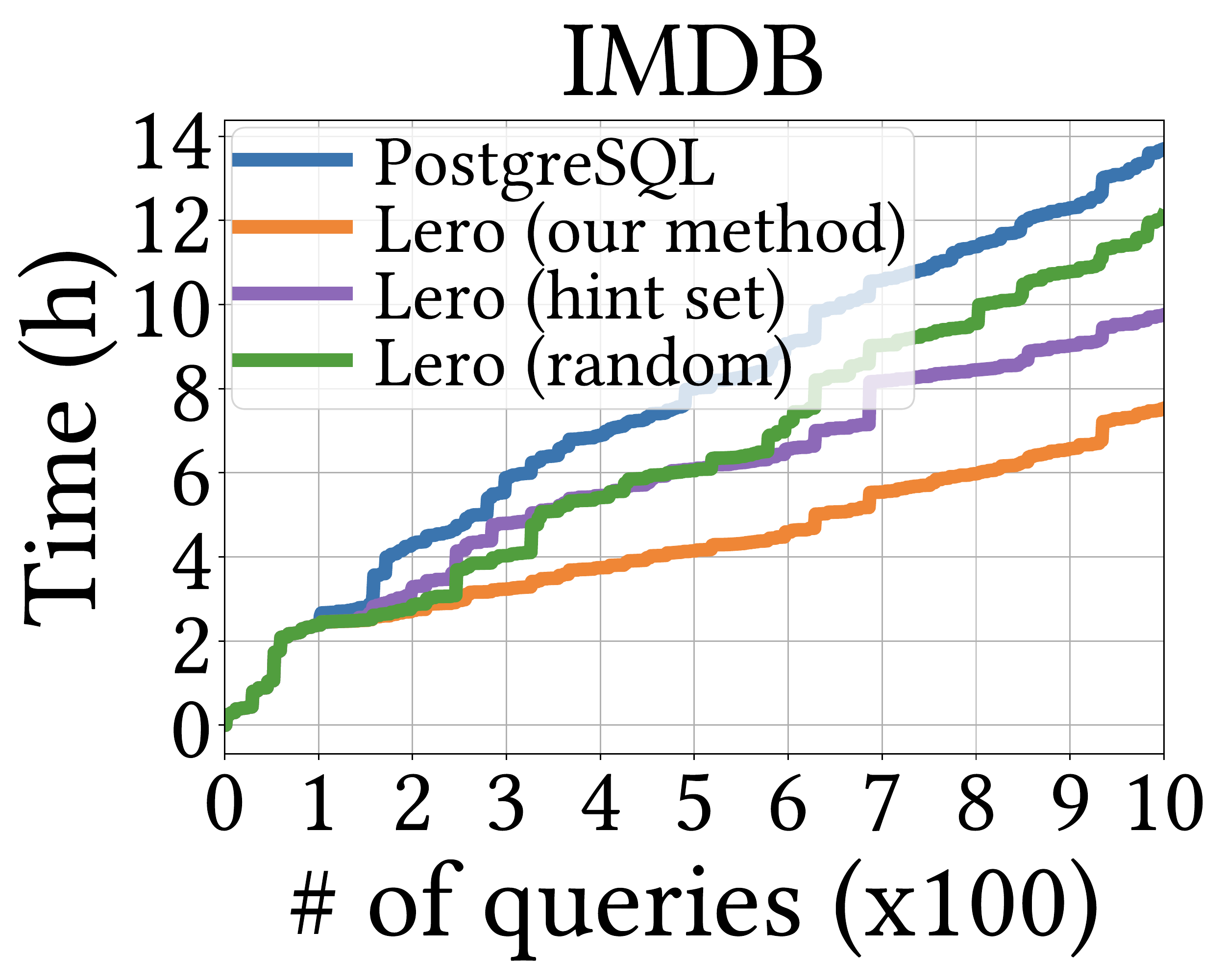}}
    \hspace{0.05\linewidth}
    \subfigure{\includegraphics[width=0.4\linewidth]{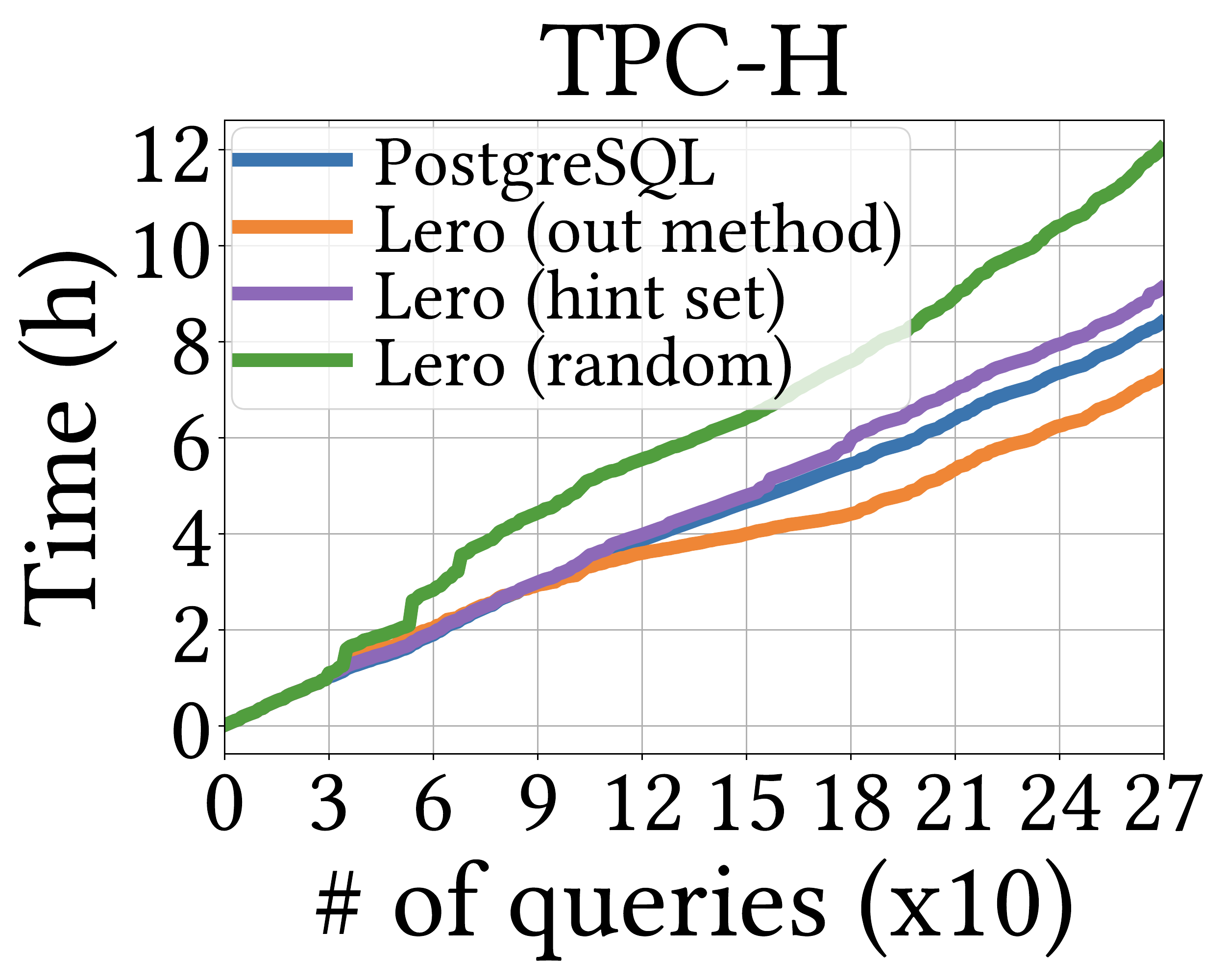}}
    % \subfigure[\normalsize IMDB Test Workload]{\includegraphics[width=0.48\linewidth]{policy-test-imdb.pdf}}
    % \hspace{0.01\linewidth}
    % \subfigure[\normalsize TPC-H Test Workload]{\includegraphics[width=0.48\linewidth]{policy-test-tpch.pdf}}
    \vspace{-1.5em}
    \caption{Performance curves of \rao since deployment with different plan exploration strategies.}
    \label{fig: exp-detail-policy}
\vspace{-1em}
\end{figure}

\subsection{Plan Exploration Strategies}
\label{sec: eval-policy}
Recall that, for each query, the plan explorer in \rao generates a list of candidate plans, which need to include some truly good plans (for query optimization) and be diversified (for model to learn new knowledge).
% 
% The plan exploration strategy plays a central role in \rao. In this subsection, 
% 
Besides the plan exploration strategy described in Figure~\ref{alg: plangen}, we also implement two other alternative approaches in \rao to compare their performance:
% . Specifically, we compare our plan exploration strategy, which uses the cardinality as the tuning knob, with two other strategies: 
1) the hint set-based strategy introduced in Bao and 2) a random strategy which randomly generates a number of plans. \eat{from the plan space the same number of candidates as our plan explorer.}

Table~\ref{tab:exp-detail-strategy} compares the quality and diversity of candidate plans generated by our plan explorer with those generated by the hint set-based strategy. {On the third column, we report the number of unique plans without duplication.}
On one hand, even with better performance, \rao's plan explorer actually generates fewer candidate plans per query on average than the hint set-based strategy (duplicate plans may be output in different iterations of lines~4-8 in Figure~\ref{alg: plangen}), which helps bring down the exploration cost. On the other hand, a higher percentage of candidate plans generated by \rao's plan explorer run faster than the plan generated by PostgreSQL's native optimizer, which proves such candidate plans worthy to explore and learn further. 
% \eat{For both strategies, there are also a number of plans that are worse than the PostgreSQL's plan to encourage the diversity in candidates.}
With cardinality tuning, \rao's plan explorer is able to generate more diversified plans while the hint set-based strategy typically generates plans with minor difference.

Figure~\ref{fig: exp-detail-policy} reports performance curves of Lero using different plan exploration strategies on IMDB and TPC-H (performance on STATS is similar to the one on IMDB, and thus is omitted due to the space limits). Both \rao's plan explorer and the hint set-based tuning strategy perform much better than the random strategy.
% 
% \eat{This is simply because the truly good plans are often rare, the random method can not easily find them to facilitate model training.}
%
We also observe that replacing \rao's plan explorer with the hint set-based tuning strategy performs worse.
\eat{We also observe that \rao with the hint set-based plan exploration strategy performs worse than PostgreSQL on TPC-H. So do Bao and Bao+ in the evaluation of Section~\ref{sec: eval-gain: stable} (see Figure~\ref{fig: exp-overall-static}) and Section~\ref{sec: eval-gain: curve} (see Figure~\ref{fig: exp-overall-train}).}
The hint set-based strategy has some intrinsic limitations, as analyzed at the beginning of Section~\ref{sec: policy}. \rao's plan explorer generates good and diversified plans for the pairwise comparator model to explore and learn more effectively.
\eat{We can see that only $6\%$ of its generated plans are faster than PostgreSQL on TPC-H, and the majority of the candidates are slower. Thus, the model may not be fully trained due to the lack of diversity in the candidate plans on TPC-H.}

% This phenomenon also happens in Section~\ref{sec: eval-gain: stable} (see Figure~\ref{fig: exp-overall-static}) and Section~\ref{sec: eval-gain: curve} (see Figure~\ref{fig: exp-overall-train}), where Bao and Bao+ using hint set tuning policy perform worse than PostgreSQL on TPC-H, respectively. This indicates that the hint set-based strategy  tuning can not perform well on TPC-H no matter we use latency-based or rank-based models, which is due to its intrinsic limitation. By Section~\ref{sec: policy}, tuning the global hint set could find better operators for some sub-queries but also possibly affect some other sub-queries using right operations. The TPC-H dataset is synthetically generated under the uniform distribution assumption so the cardinality estimation errors are smaller for most sub-queries. Disable some operations would easily affect these sub-queries and thus decrease the plan quality. We can see that only $6\%$ of its generated plans are faster than PostgreSQL, so the model is not well trained. 
% Whereas, our method is more adaptive. It only corrects a subset of sub-queries without affecting others, so it could work well on IMDB and TPC-H having large and small cardinality estimation errors, respectively.

\begin{figure}[t]
    \centering
    \subfigure{\includegraphics[width=0.4\linewidth]{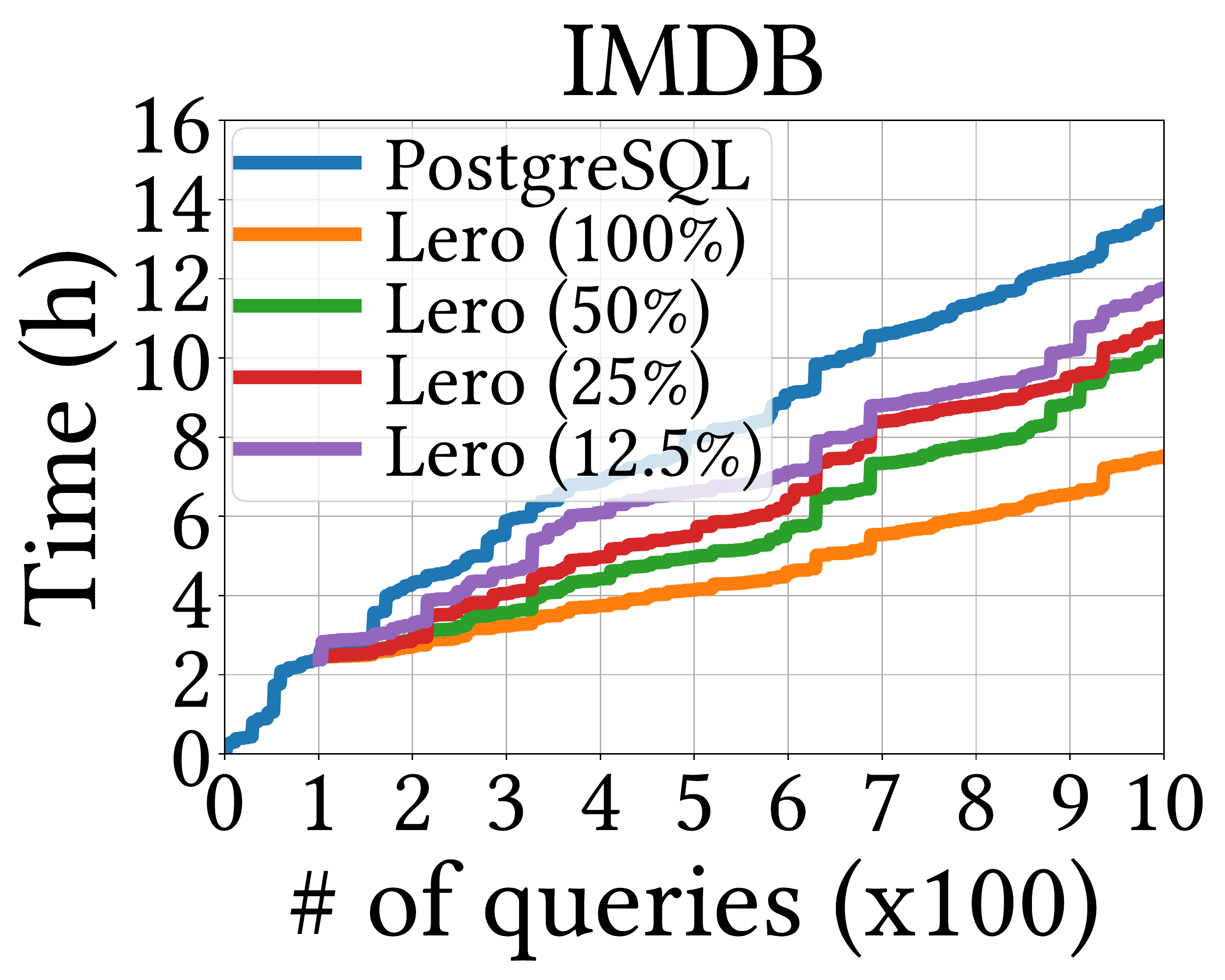}}
    \hspace{0.05\linewidth}
\subfigure{\includegraphics[width=0.4\linewidth]{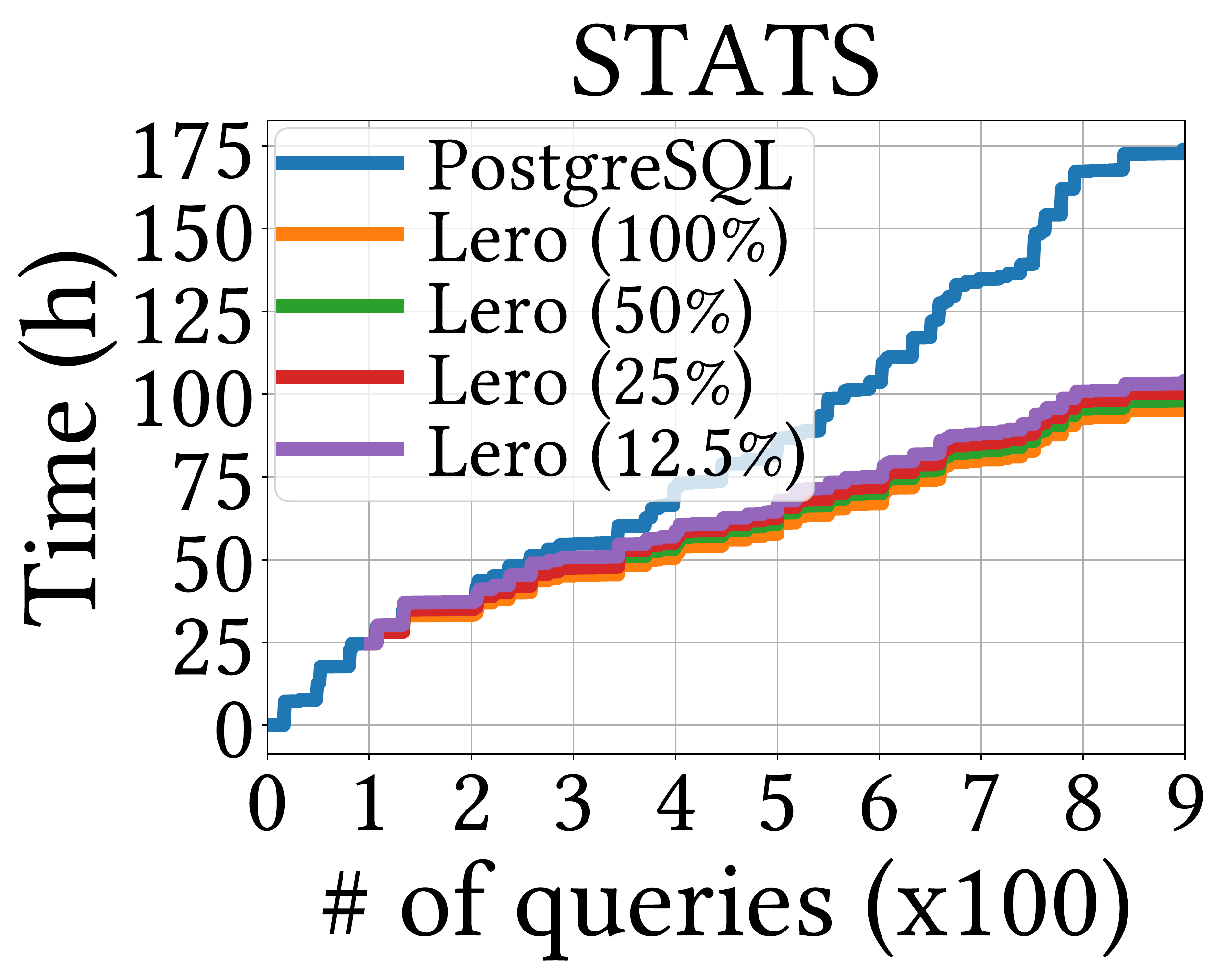}}
	\vspace{-1.5em}
	\caption{Performance curves of \rao since deployment with different amounts of idle resource.}
	\label{fig: exp-overall-idle}
	\vspace{-2.2em}
\end{figure}

\subsection{Effects of Idle Resource}
\label{sec: eval-idle}
% 
%As \rao uses idle resource to execute candidate plans to accumulate training data, we examine the performance of \rao with different amount of idle computation resource.  
% 
We examine the performance curve of \rao since deployment with different amounts of idle computation resource. By default, when an incoming query $Q_{t+1}$ is to be processed, the comparator model in \rao is trained and updated using execution statistics of different candidate plans
% \eat{ (run in parallel with idle computation resource) }
for earlier queries $Q_1, Q_2, \dots, Q_t$. Such statistics are collected by running candidate plans in idle resource.
When the idle resource is limited, we may only be able to finish executing candidate plans for a fraction of earlier queries $Q_1, Q_2, \dots, Q_t$. With only the first $50\%$, $25\%$, and $12.5\%$ earlier queries having all of their candidates executed on idle resource, we re-run the experiments in Section~\ref{sec: eval-gain: curve}. We demonstrate in Figure~\ref{fig: exp-overall-idle} that the performance of \rao can still be improved over time and converge eventually.
% \eat{ (the case for TPC-H is similar to that for STATS, and thus is omitted here).}

% all of the idle computation resource in the machine is assigned to execute candidate plans to accumulate training data. Then, we re-run the experiments in Section~\ref{sec: eval-gain: curve} but assigns only $50\%$, $25\%$ and $12.5\%$ of the idle resource for executing candidate plans in each test instance. In other words, for each incoming query $Q_{t+1}$, \rao only witness the former $50\%$, $25\%$ and $12.5\%$ plans for earlier queries $Q_1, Q_2, \dots, Q_t$ in each instance.

Even with limited idle resource, the initial performance of \rao is no worse than PostgreSQL, thanks to the pre-training procedure. However, at the beginning, the quality of the selected plans is indeed worse than the default setting (when $100\%$ of earlier queries have their candidate plans executed on idle resource), because the comparator model is not fully trained. We observe that the model still converges fast even with very limited idle resource. For instance, when only $12.5\%$ queries have their candidate plans executed and explored, the model converges after the first 500 and 200 queries on IMDB and STATS, respectively. 
% \eat{ (gaps between different performance curves become stable)}
% 

Regardless of under which configuration, \rao becomes stable eventually and performs almost the same (curves become parallel towards the right end).
% \eat{ for settings with different amounts of available idle resource}.
% 
This implies that limited idle resource would definitely slow down the model's learning and convergence, but has little impact on the ultimate performance of \rao.
% 
% \revise{As idle resource is often limited, it is an important direction in our future work to investigate how to prioritize plans for execution to accelerate model convergence. 
% }

% Figure~\ref{fig: exp-overall-idle} reports the performance curves of \rao with different amount of idle computation resource on IMDB and STATS. We find that even with less idle resource, \rao could still consistently performs better than PostgreSQL. However, with less idle resource, \rao may not witness enough plans to train the comparator model. The quality of the selected plans may become worse before the model converges. On IMDB, the model does not converge till the end of running the last training query, so the execution time of selected plans get longer with less amount of idle resource. Whereas, on STATS, the model converges after seeing around 200 queries, so the performance curves of \rao makes little difference with the amount of idle resource.

% For each testing instance, we then wait the model to trained to be stable after all candidate plans are executed and test its performance on the corresponding test workload. The performance of models in \rao trained with different amount of idle resource is very similar. This indicates that the amount of idle resource would slow down the model converge speed, but has little impact on the ultimate accuracy of the trained models. 

\begin{table}[t]
    \centering
    \vspace{0.1in}
    \scalebox{0.92}
    {
    {
    \begin{tabular}{|c|cc|cc|} 
    \hline
    \rowcolor{mygrey}
    & \multicolumn{2}{c|}{Random Split} & \multicolumn{2}{c|}{Slow Split} \\ \cline{2-5}
    \rowcolor{mygrey}
    \multirow{-2}{*}{\sf Speedup Ratio}  & \sf Train & \sf Test & \sf Train & \sf  Test \\ \hline
    Bao & $1.62$ & $1.49$  & $1.17$ & $1.05$\\ \hline
    Bao+ & $2.93$ & $1.58$  & $4.01$ & $1.74$\\ \hline
    Balsa & $2.46$ & $1.55$ & $1.23$ & \textbf{2.31}\\ \hline
    \rao & \textbf{3.54} & \textbf{1.59} & \textbf{4.38} & 1.34\\ \hline
    \end{tabular}
    }}
    \caption{{Learned optimizers on IMDB with original JOB.}}
    \label{tab:exp-speedjob}
    \vspace{-3.3em}
\end{table}

\subsection{Comparison with Balsa}
\label{sec: eval-balsa}

We compare \rao with Balsa on IMDB with the original JOB workload. We use its open-source implementation in~\cite{BalsaImp, yang2022balsa} and prepare the dataset in the same way as~\cite{yang2022balsa}. Specifically, the 113 queries in JOB are split into a training set (with 94 queries) and a testing set (with 19 queries) in two ways: 1) random split and 2) the test set consists of the 19 slowest-running queries.
% when planned by an expert optimizer. 
We run Balsa using its default settings and compare it with \rao, Bao, and Bao+. 

Table~\ref{tab:exp-speedjob} exhibits the speedup ratio of each learned query optimizer in comparison to PostgreSQL's native optimizer. The performance of Balsa matches or outperforms Bao, which is consistent with the results in~\cite{yang2022balsa}. Both Bao and Balsa try to predict plan latency to guide plan search, while Balsa has higher freedom of exploration (Bao's search space is limited by hint set tuning)~\cite{yang2022balsa}. Thus, it performs better than Bao on small and stable workload. However, even after Balsa's model is comprehensively trained on this small JOB workload, it still performs worse than Bao+ and \rao.
% This implies the inherent shortcomings of the latency prediction model, which is not as accurate and effective as our pairwise learning-to-rank model in identifying high-quality plans.

%\vspace{-1.2em}
\section{Related Work}
\label{sec: related}
%\note{Discuss other related works, e.g., learned cardinality.}

{\sstitle{Learning to Optimize Queries.}}
Recently, there is a flurry of research to apply machine learning 
% to various components 
in query optimization~\cite{zhu2022learned}. The majority of them focus on learned cardinality estimation, using either query-driven or data-driven approaches. Query-driven methods~\cite{kipf2018learned, dutt2019selectivity, liu2021fauce} apply learned models to map featurized query to its cardinality. Data-driven methods~\cite{yang2019deep, yang2020neurocard, tzoumas2011lightweight, wu2020bayescard, hilprecht2019deepdb, zhu2020flat} use different generative models to directly learn the underlying data distribution. 
% Some benchmark evaluations~ have examined 
Their superiority and limitations have been evaluated \cite{wang2020ready, han2021CEbenchmark}.
Others focus on refining traditional cost models and plan enumeration algorithms. For learned cost models,~\cite{sun2019end} and~\cite{zhou2020query, zhi2021efficient} utilize TreeLSTM and convolution models to learn cost of single and concurrent queries, respectively. 
Plan enumeration is often modelled as a reinforcement learning problem on deciding the best join order of tables.~\cite{hester2018deep, krishnan2018learning} and~\cite{yu2020reinforcement} use simple neural networks and TreeLSTM model as value networks for join order selection, respectively.~\cite{trummer2019skinnerdb} considers how to adjust the join order on the fly.
These works only optimize an individual component in query optimizer, which does not necessarily improve the overall performance. 

Besides them, recent works~\cite{marcus2019neo,yang2022balsa} provide end-to-end learned solutions for query optimization, and~\cite{marcus2021bao} learns to steer a native query optimizer using hint set tuning. However, as analyzed in Section~\ref{sec: intro-challenge}, they suffer from lots of deficiencies arising from predicting the cost or latency.
% 
% \revise{
Based on plan exploration and pairwise plan comparison, \rao 
% is the first learning-to-rank query optimizer, which 
learns the difference between plan pairs and learns to improve the end-to-end quality of query optimization.
% }
% 
% \revise{Compared with them, \rao adopts a pairwise learning approach to learn the difference between plan pairs. It is a learning-to-rank query optimizer which directly learns to improve the end-to-end quality of query optimization.} 
\eat{It is the first learning-to-rank query optimizer which directly learns to improve the end-to-end quality of query optimization.}

{\sstitle{Learning-to-Rank Paradigm.}} \rao follows a learning-to-rank paradigm, which is a class of learning techniques to train models for ranking tasks~\cite{KaratzoglouBS13ltr, vargas2011rank, liu2009ltr}. It has been widely applied for, e.g., document retrieval, collaborative filtering, and recommendation systems. Based on how ranking is generated, the learning-to-rank techniques could be classified into pointwise~\cite{fuhr1989optimum, li2007mcrank}, pairwise~\cite{freund2003efficient, liu2020personalized} and listwise~\cite{pang2020setrank, swezey2021pirank} approaches.
Among them, pairwise approach learns a classifier on a pair of items to identify which one is better.
To our best knowledge, we are the first to apply a {\em pairwise learning-to-rank} paradigm to develop a learned query optimizer.

\sstitle{Learning to Tune Indexes.} The task of indexing tuning is to find the set of indexes that fits in a given storage budget and results in the lowest execution cost for a given workload of queries. A traditional index tuner \cite{ChaudhuriN97, ValentinZZLS00, AgrawalCKMNS04} first searches for the optimal index configuration for each query (query-level search), and then enumerates different sets of those index configurations to find the optimal index set for the workload under the budget (workload-level search). Both phases need to compare the execution costs of two plans of the same query given different index configurations. Traditional index tuners rely on optimizer’s estimates for such comparisons, while \cite{ding2019ai} trains a classifier to this end with higher accuracy.

While the query-level search in \cite{ding2019ai} can be regarded as an application of learning-to-rank in index tuning, it has two fundamental differences in comparison to \rao.
% 
% We observe that comparing the execution cost of two plans of the same query corresponding to different index configurations is a key step during index tuning. Instead of using optimizer’s estimates for such comparison, our key insight is that formulating it as a classification task in machine learning results in significantly higher accuracy
% 
% \cite{ChaudhuriN98}
% 
% In the literature, learning-to-rank represents a class of machine learning techniques to construct ranking models for information retrieval systems~~\cite{KaratzoglouBS13ltr, vargas2011rank, liu2009ltr}. It is useful for document retrieval, collaborative filtering, recommendation system and many other applications. Based on the model structure, the learning-to-rank techniques could be classified into pointwise~\cite{fuhr1989optimum, li2007mcrank}, pairwise~\cite{freund2003efficient, liu2020personalized}, and listwise~\cite{pang2020setrank, swezey2021pirank} approaches.
% Among them, pairwise approach learns a classifier on a pair of items to identify which one is better.~\cite{ding2019ai} applies the pairwise learning-to-rank paradigm to optimize the index tuning problem for the first time.
% 
% \revise{
First, the index tuner in \cite{ding2019ai} is similar to traditional tuners \cite{ChaudhuriN97, ValentinZZLS00, AgrawalCKMNS04}, except that the classifier, as a pairwise comparator model, is invoked during search procedures to determine whether the plan of a query under a new index configuration is improved in comparison to the one under the initial one;
for a completely different task, query optimization, \rao is equipped with two carefully designed components, plan explorer and plan comparator model, which work together to explore the plan space.
% and collect valuable execution statistics online to improve the comparator model continuously.
% 
% These two components are coupled closely in our pairwise learning-to-rank paradigm. 
% 
% They work together to explore the plan space for reasonable candidate plans and, meanwhile, collect valuable execution statistics online to improve the comparator model continuously.
% 
% For a pair of plans $P_1$ and $P_2$ of the query $Q$ under different index configurations $C_1$ (initial configuration) and $C_2$ (new configuration), respectively, 
% 
Second, the comparator models in \cite{ding2019ai} and \rao both compare the performance of two plans; however, with different goals in the two tasks, the ways how features are encoded and the model architectures in the two comparators are different. 
%Our technical report~\cite{fullversion}
Appendix~\ref{app:addexp} compares the two models experimentally.

\eat{
We review the most relevant related work in this section, including the \ltr techniques and machine learning techniques used in query optimizer.

\sstitle{Learning-to-Rank Techniques.}
Learning-to-rank is a class of algorithms to solve ranking problems in search relevancy by machine learning models. It is a central part of many information retrieval problems~\cite{KaratzoglouBS13ltr, vargas2011rank}. By a comprehensive survey~\cite{liu2009ltr}, the techniques could be classified into pointwise~\cite{fuhr1989optimum, li2007mcrank}, pairwise~\cite{freund2003efficient, liu2020personalized} and listwise~\cite{pang2020setrank, swezey2021pirank} approaches.
Among them, pairwise approach learns a binary classifier on a pair of items to identify which one is better. It outperforms pointwise approach in terms of model accuracy while runs much faster than listwise approach. 

\highlight{
Previous works have applied ranking or pairwise comparing techniques for accuracy test~\cite{gu2012testing, upadhyaya2011latency},
performance regression detection~\cite{ding2018plan} or optimizing specific functions related to query optimization~\cite{hellerstein1998optimization, liu2008grid, ding2019ai}. However, they never target to build a complete learned query optimizer.} In this paper, we apply pairwise \ltr to design a new learned query optimizer for the first time.
}

\eat{
%\smallskip
\sstitle{Learned Query Optimizer.}
The representative works on learning the whole query optimizer have already been analyzed in Section~\ref{sec: intro-challenge}. For individual query optimizer components, namely cardinality estimation, cost model and plan enumeration, we review them as follows. 
Learned cardinality estimation methods mainly include query-driven and data-driven approaches. Query-driven methods~\cite{kipf2018learned, dutt2019selectivity, liu2021fauce} apply learned models to map featurized query to its cardinality. Data-driven methods~\cite{yang2019deep, yang2020neurocard, tzoumas2011lightweight, wu2020bayescard, hilprecht2019deepdb, zhu2020flat} use different generation models to directly learn the underlying data distribution. 
Some benchmark evaluations~\cite{wang2020ready, han2021CEbenchmark} have verified the superiority of learned methods. For learned cost model,~\cite{sun2019end} proposed a TreeLSTM based method to directly learn cost of single query. GPredictor~\cite{zhou2020query} and Prestroid~\cite{zhi2021efficient} utilize the convolution models to estimate the performance of concurrent queries. The plan enumeration problem is often modelled as a reinforcement learning instance on selecting the best join order. DQ~\cite{hester2018deep} and ReJoin~\cite{krishnan2018learning} use the simple neural networks to predict the value of each join selection. RTOS~\cite{yu2020reinforcement} adapts it by using TreeLSTM. SkinnerDB~\cite{trummer2019skinnerdb} considers how to adjust the join order on the fly. 

Besides, there also exist some attempts on deploying learned query optimizer into real-world scenarios, including the cost model component~\cite{siddiqui2020cost} and the whole query optimizer module~\cite{negi2021steering}.~\cite{zhu2022learned} presents a comprehensive review on all topics related to learned query optimizer. All of the above works are orthogonal to our \rao under \ltr paradigm. Our \rao could be deployed on top of any query optimizer with learned or traditional components.

Except query optimizer, machine learning techniques have also been widely used in other tasks in DBMS. By a comprehensive review in~\cite{zhou2020database}, the main works include but not limited to learn systems configurations, such as knob tuning~\cite{zhang2019end, li2019qtune} and view advisor~\cite{yuan2020automatic, liang2019opportunistic}, learn to design the database, such as indexing~\cite{wu2019designing, kraska2018case, galakatos2019fiting} and data layout~\cite{idreos2019learning, idreos2019design}, and learn to monitor, such as system diagnosis~\cite{ma2020diagnosing, taft2018p}.
}

%\smallskip
\eat{\sstitle{Rank and Comparison Techniques in Query Optimizer.} 
For the rank technique,~\cite{gu2012testing} ranks the plans by their exact and estimated costs to validate the accuracy of the query optimizer.~\cite{upadhyaya2011latency} analyzes the sensitivity of plan ranks w.r.t.~I/O cost and \CE errors.~\cite{hellerstein1998optimization} tries to rank the predicates in a query to optimize its generated plan.~\cite{liu2008grid} ranks each host machine to decide how to assign relation tables onto them. Obviously, these works use rank techniques for either accuracy test or optimizing a specific function in query optimizer, which are totally different from our rank based paradigm for learned query optimizer. 

Some work~\cite{ma2020active} tries to apply the comparison technique in query optimizer.~\cite{ding2018plan} detects the performance regression between a pair of plans with changes.~\cite{ding2019ai} leverages query execution difference to tune index recommendations. Their goals are different from us as we apply the comparator model to learn to preserve the relative order between multiple plans.
}

% \vspace{-1em}
\section{Conclusions}
\label{sec: con}
% \note{Conclusions and future work.}
% 
We propose \rao, a {\em \underline{le}arning-to-\underline{r}ank} query \underline{o}ptimizer. First, \eat{\rao is the first to apply learning-to-rank machine learning techniques to query optimization.} {\rao applies learning-to-rank machine learning techniques to query optimization}. We argue that it is an overkill to develop machine models to predict the exact execution latency or cardinalities in terms of query optimization. Instead, \rao adopts a {\em pairwise} approach to train a binary classifier to compare any two plans, which is proven to be more efficient and effective. Second, \rao takes advantage of decades of wisdom of database research and jointly works with the native query optimizer to improve optimization quality. Third, \rao is equipped with a plan exploration strategy, which enables \rao to explore new optimization opportunities more effectively. Finally, an extensive evaluation on our implementation on top of PostgreSQL demonstrates \rao's superiority in query performance and its ability to adapt to changing data and workload. 
 
% {With its non-intrusive design, \rao can be implemented on top of any existing DBMSs. We are in the process of implementing \rao on top of a commercial DBMS. Our preliminary results also confirm \rao's significant improvements over its native query optimizer.}

%\clearpage

\bibliographystyle{ACM-Reference-Format}
\bibliography{0-main}

% !TeX spellcheck = en_US

%\begin{color}{red}

%\clearpage 

%\nobalance

\appendix

\section*{Appendix}

\section{Comparator with $d$-$dim$ Embedding}
\label{app:multidim}

Our plan comparator model $\CP$ (described in Section~\ref{sec: train: general}) could be easily extended from 1-$dim$ plan embedding to use $d$-$dim$ ($d > 1$) plan embedding. We use the same way on encoding plans and slightly modify the plan comparator model (shown in Figure~\ref{fig: PairLearn}) as follows:

\begin{enumerate}
\item In the plan embedding layers, we change the output dimension of last embedding layer to $d$ so that the plan embedding $\PRR(P)$ becomes a $d$-$dim$ vector for any plan $P$.

\item We change the comparison layer to a learnable linear layer. It takes two $d$-$dim$ vectors $\PRR(P_1)$ and $\PRR(P_2)$ as inputs, and outputs $0$ ($P_1$ is better) or $1$ ($P_2$ is better). 
\end{enumerate}

Next, we discuss how to select the best plan under $d$-$dim$ embedding (in~\ref{app:multidim-select}) and the performance of plan comparator w.r.t.~the dimensions of plan embedding (in~\ref{app:multidim-performance}).

\eat{
\sstitle{Remarks.} 
It is worth mentioning that our model could be easily extended to learn a $d$-$dim$ embedding $\PRR(P)$ of a plan $P$ where $d > 1$. However, by our testing, higher dimensions of embeddings do not bring any benefit. On the contrary, on our evaluation benchmarks, the execution time of \rao with a larger $d \in \{2, 4, 8, 16\}$ is $1.4$ times to $2.2$ times as long as the case with $d = 1$. The reason could be that, with $d > 1$, the embedding model $\PRR$ tries to summarize more sophisticated information in plan embeddings, but requires larger training dataset and more training time to obtain more accurate embeddings. More over, when $d > 1$, it is not convenient to select the best plan as $\PRR(P)$ does not define a total order on all plans $P$. Instead, we need to use a more complex randomized approach. We defer the detailed evaluation results and method discussion on $d > 1$ into the extended version~\cite{fullversion}. 
}

\subsection{Selecting the Best Plan using Comparators}
\label{app:multidim-select}

Recall that the plan comparator $\CP(P_1, P_2)$ is an oracle comparing any two plans $P_1$ and $P_2$ of a query. Given a list of candidate plans $P_1, P_2, \dots, P_n$ of a query, for $d$-$dim$ embeddings, let
\begin{equation}\label{equ: wins}
\WP(P_i) = |\{ P_j \mid \CP(P_i, P_j) = 0, j \neq i \}|,
\end{equation}
be the number of plans whose latencies are worse than $P_i$.
$\WP(P_i)$ defines a full order of all candidate plans.
Obviously, the best candidate plan $P^{*}$ has $\WP(P^{*}) = n - 1$.

% In Section~\ref{sec: train: general}, our comparator model is to learn the above oracle. We also use $\CP$ to denote the learned comparator model. 
% 
A comparator with $d$-$dim$ plan embeddings can be trained in a pairwise way similar to the one introduced in Section~\ref{sec: train: train} (without pre-training though).
After the comparator is fully trained, we can use it to pick the best plan among candidates $P_1, \ldots, P_n$ of a possibly unseen query $Q$. In the general design where we have $d \geq 1$ for the plan embeddings, we do not require a fully trained $\CP$ to preserve transitivity, i.e., $P_1$ is better than $P_2$ and $P_2$ is better than $P_3$ imply that $P_1$ is better than $P_3$. 
In order to pick the best plan among candidates $P_1, \ldots, P_n$, we can invoke $\CP$ $n(n-1)$ times to compare all pairs of candidates. In general, the output $\CP(P_i,P_j) \in (0,1)$ gives a soft prediction (i.e., the predicted probability of whether $P_i$ or $P_j$ is better). We use the algorithm $\mathcal{A}$ in Eq.~\eqref{equ: cmp_algo} to pick the better one, and define {\em randomized wins} as
\vsshrink
\[
\RWP(P_i) = |\{ P_j \mid \mathcal{A}(P_i, P_j) = P_i, j \neq i \}|.
\vsshrink
\]
% \rzhu{The definition of wins is different and inconsistent with former one.}
Accordingly, in the $d$-$dim$ setting, we choose the best plan as
\vsshrink
\[
P^* = \argmax_{P_i \in \{P_1, \ldots, P_n\}} \RWP(P_i)
\vsshrink
\]
with the most randomized wins--ties are broken arbitrarily.
% 
% we compare all the $n(n-1)$ pairs of candidates with $\CP$, and choose the one with the max $\WP$ defined in Eq.~\eqref{equ: wins}

In the more practical design of $\CP$ with $1$-$dim$ plan embedding (Section~\ref{sec: train: general}), we do not need to derive $\RWP(P_i)$ as above.
% , i.e., the number of plans less preferable by $\CP$ than $P_i$ (defined in Eq.~\eqref{equ: cmp}) of each plan $P_i$ and then choose the plan maximizing it. 
Instead, the output of the sub-model $\PRR$ defines a {\em total order} on all candidate plans. For any pair of plans $(P_i, P_j)$, if $\PRR(P_i) > \PRR(P_j)$, we always have $\CP(P_i, P_j) > 0.5$, i.e., $P_j$ is more preferable than $P_i$ with higher probability than vice versa.
Therefore, we could select $P^{*} = \argmin_{P_i} \PRR(P_i)$ in the $1$-$dim$ setting as the best plan for execution. 

We can show that, $P^*$ chosen in this way is indeed the one with the most randomized wins {\em in expectation} (thus, the choice is equivalent to the one for the general design, in expectation). 
%Due to space limits, we put all proofs in Appendix~B of the full version~\cite{fullversion}.
% 
\begin{proposition}\label{prop:ddim:select}
$P^{*} = \argmin_{P_i} \PRR(P_i)$ chosen with the lowest $1$-$dim$ embedding also wins the most: it satisfies
\vsshrink
\[
P^* = \argmax_{P_i \in \{P_1, \ldots, P_n\}} \ep{\RWP(P_i)}.
\vsshrink
\]
\end{proposition}

\begin{proof}
By the model construction, for each $P_i$, we have
\begin{align}
    & \ep{\RWP(P_i)} = \sum_{j \in [n], j \neq i} \pr{\mathcal{A}(P_i, P_j) = P_i} \nonumber 
    \\
=   & \sum_{j \in [n], j \neq i} \!\!\!\!\!\! (1 - \CP(P_i, P_j)) = \!\!\!\!\!\! \sum_{j \in [n], j \neq i} \!\!\!\!\!\! \CP(P_j, P_i) \nonumber
    \\
=   & \left(\sum_{j \in [n]} \!\! \phi(\PRR(P_j) - \PRR(P_i))\right) - \phi(0). \label{equ: epwins}
\end{align}

From Eq.~\eqref{equ: epwins}, it is straightforward that $P^* \in \{P_1, \ldots, P_n\}$ with the lowest $\PRR(\cdot)$ would maximize $\ep{\RWP(\cdot)}$.
\end{proof}

\subsection{Performance of Comparator with Different Dimension of Plan Embedding}
\label{app:multidim-performance}

We test the performance of \rao with different dimensions of plan embedding. We set $d = \{1, 2, 4, 8, 16\}$, train \rao on the training workload and then test its performance on the test workload. Table~\ref{tab:exp-plandim} reports the execution time of \rao with different dimensions of plan embedding on IMDB and STATS benchmarks. Similar  results are also observed on TPC-H and TPC-DS benchmarks. 

\begin{table}[h]
    \centering
    \vspace{0.1in}
    \scalebox{0.95}
    {
    %\revise
    {
    \begin{tabular}{|c|cc|} 
    \hline
    \rowcolor{mygrey}
    \sf Time (in hour) & \sf IMDB & \sf STATS \\ \hline
    PostgreSQL & 1.15 & 20.19\\ \hline
    \rao ($d = 1$) & 0.35 & 11.32 \\ \hline
    \rao ($d = 2$) & 0.77 & 20.74\\ \hline
    \rao ($d = 4$) & 0.69 & 21.2\\ \hline
    \rao ($d = 8$) & 0.49 & 21.38\\ \hline
    \rao ($d = 16$) & 0.59 & 21.19\\ \hline
    \end{tabular}
    }}
    \vspace{0.05in}
    \caption{{Performance of \rao with different dimensions of plan embedding.}}
    \label{tab:exp-plandim}
    \vspace{-1.5em}
\end{table}

Interestingly, we observe that the performance of \rao does not necessarily increase with higher dimensional embeddings. On the contrary, \rao with complex embedding ($d > 1$) has much worse performance than using the simplest 1-$dim$ plan embedding. Specifically, on IMDB, the the execution time of \rao with a larger $d \in \{2, 4, 8, 16\}$ is $1.4$ times to $2.2$ times as long as the case with $d = 1$. On STATS, the  execution time of \rao with $d > 1$ is around $1.8$ times longer than $d = 1$, even longer than the execution time of the native PostgreSQL. 
The reason could be that, with $d > 1$, the embedding model $\PRR$ tries to summarize more sophisticated information in plan embeddings, but also requires larger training dataset and more training time to obtain more accurate embeddings and to converge. Therefore, the training workload may be not enough to comprehensively capture enough information for plan comparison, so the selected plan may be not the best one. 

In our experiments, \rao using 1-$dim$ plan embedding already attains significant performance improvement in comparison to the native query optimizer. Moreover, when $d > 1$, it is not as convenient to select the best plan as when $d=1$ (although Proposition~\ref{prop:ddim:select} says that there is an equivalent way for plan selection), since $\PRR(P)$ cannnot define a total order on all plans $P$ when $d>1$. Thus, we use the simplest 1-$dim$ plan embeddings in \rao. 
%In principal, $d$-$dim$ could be better than 1-$dim$ embedding, but it does require more data to train and take longer to converge. We consider it in the future work.

\section{Exploration Heuristic based on Plan Diagram}
\label{app:diagram}
Errors in estimating predicate selectivities can be propagated from bottom to top during the plan search, and incur a wrong join order and sub-optimal choices of join types.
A {\em plan diagram} \cite{DDH08identifying, DeyBDH08approximating} can be constructed by varying values of parameters in one or more predicates of the query $Q$, so that the optimizer generates a set of different plans.
% 
% (e.g., replacing $20 \leq {\sf Age} \leq 30$ with $10 \leq {\sf Age} \leq 40$)
% 
For example, for a query with a parameterized predicate ``${\sf @lower}$ $\leq$ ${\sf Age}$ $\leq$ ${\sf @upper}$'', the optimizer may generate different plans by setting $({\sf @lower}, {\sf @upper})$ as $(20,30)$ or $(10,40)$.
These plans form a diagram of different regions on the 2D space $({\sf @lower}, {\sf @upper})$: within each region, the optimizer outputs the same plan.

It is shown in \cite{DDH08identifying, DeyBDH08approximating} that replacing selectivity error-sensitive plan choices with alternatives in the plan diagram provides potentially better performance. Therefore, we can use plans in the plan diagram as the list of candidates to explore the uncertainty from estimating predicates' selectivities in the native optimizer.

The drawbacks of this strategy are obvious. 
It is not affordable to generate a plan diagram by varying parameters on more than two tables, as the number of candidate plans would be too large~\cite{DeyBDH08approximating}. If we tune parameters on two tables, the candidates may not be diversified enough to include plans with quality sufficiently higher than the plan generated by the native optimizer.
In our evaluation on the STATS benchmark~\cite{han2021CEbenchmark}, even the best candidate in a plan diagram does not have significant performance improvement. Specifically, the average performance improvement of the best candidate is less than $3\%$, and the best candidates of only less than $5\%$ of queries are faster than those generated by PostgreSQL's optimizer.
% However, if we tune cardinality on more tables, the number of candidate plans would be too large and not affordable for execution.

\eat{
\section{Additional Experiments}

We present some additional experimental results in this section.
}

\eat{
\subsection{Query Optimization Cost}
\label{sec: eval-app-opt}
While achieving significant improvement in query execution performance, \rao spends extra query optimization time in generating a list of candidate plans for an input query and applying the comparator model to pick the best candidate. The average query optimization time per-query of different optimizers is reported in Table~\ref{tab:exp-optcost}. We observe that this extra cost in query optimization is very low. In particular, the total extra cost is only at most $2.4\%$ of the total query execution time on IMDB and less than $0.1\%$ on STATS, TPC-H and {TPC-DS}. Whereas, \rao saves $13\%$ to $70\%$ execution time in comparison with PostgreSQL on these benchmarks.

\begin{table}[h]
    \centering
    \vspace{0.1in}
    \scalebox{0.92}
    {
    {
    \begin{tabular}{|c|ccc|} 
    \hline
    \rowcolor{mygrey}
    \sf Time (in millisecond) & \sf PostgreSQL & \sf Bao/Bao+ & \sf \rao \\ \hline
    IMDB & $842$ & $856$  & $1,736$\\ \hline
    STATS & $7$ & $8.1$  & $16$\\ \hline
    TPC-H & $5.3$ & $6.2$ & $12.6$\\ \hline
    {TPC-DS} & {$6.7$} & {$8.5$} & {$15.8$}\\ \hline
    \end{tabular}
    }}
    \vspace{0.05in}
    \caption{{Average query optimization time per query.}}
    \label{tab:exp-optcost}
    \vspace{-0.1in}
\end{table}
}

\eat{
\subsection{Impact of Concurrency}
\label{sec: eval-app-conc}

In the main experiments described in Section~\ref{sec: eval}, we test \rao and other (learned) query optimizers in the single thread environment. That is, after deployment, each query $Q_1, Q_2, \ldots$ in the workload is executed in sequential. In this experiment, we test the performance of \rao under concurrent query execution.

Specifically, we deploy \rao for plan exploration and selection and invoke 32 threads on our machine to execute multiple queries in parallel. Each query $Q_t$ in the workload $Q_1, Q_2, \ldots$ is fed to the machine in order. For query $Q_t$, \rao explores all of its candidate plans, selects the best plan and sets the rest of plans to the idle workers. The best plan of $Q_t$ is executed whenever there exists a free thread. As a result, the 
first 32 queries are executed by all threads in parallel. Whenever a thread finishes executing the assigned query, the next query is fed to this thread to be executed (with the plan selected by \rao). Meanwhile, \rao is trained in background on other idle workers with runtime stats of other candidate plans of these queries. We keep the same amount of idle workers as the single thread experiments and update the models in the same pace as the single thread setting in Section~\ref{sec: eval-set-scenarios}. 

Figure~\ref{fig: exp-overall-conc} illustrates the performance curves of \rao and PostgreSQL on IMDB and TPC-H. The observations on STATS and TPC-DS are similar to those on IMDB and TPC-H, respectively. We find that, \rao could still bring performance gain under concurrent query execution. Its total execution time is $12\%$ and $20\%$ less than PostgreSQL on IMDB and TPC-H, respectively. In comparison to the single thread execution (shown in Figure~\ref{fig: exp-overall-train}), the improvement ratio on IMDB (and STATS) is less. Whereas, on TPC-H (and TPC-DS), the improvement ratio is comparable. 

The reason could be that, under the single thread setting, for query $Q_t$, all queries $Q_1, Q_2, \ldots, Q_{t-1}$ are finished and their candidate plans are set to idle workers. However, under multi-thread setting, some queries in $Q_1, Q_2, \ldots, Q_{t-1}$ may not finish before the execution of query $Q_t$. As a result, \rao witnesses less training data for model training and updating. On complex real-world dataset (e.g., IMDB and STATS), \rao's model can not be comprehensively trained to identify better plans, so the performance loss is more obvious. However, on simple synthetic dataset (e.g., TPC-H and TPC-DS), the model may converge more easily and earlier using less volume of training data and thus the performance gain is not reduced too much.

\begin{figure}[h]
    \centering
    \subfigure{\includegraphics[width=0.48\linewidth]{CardEst 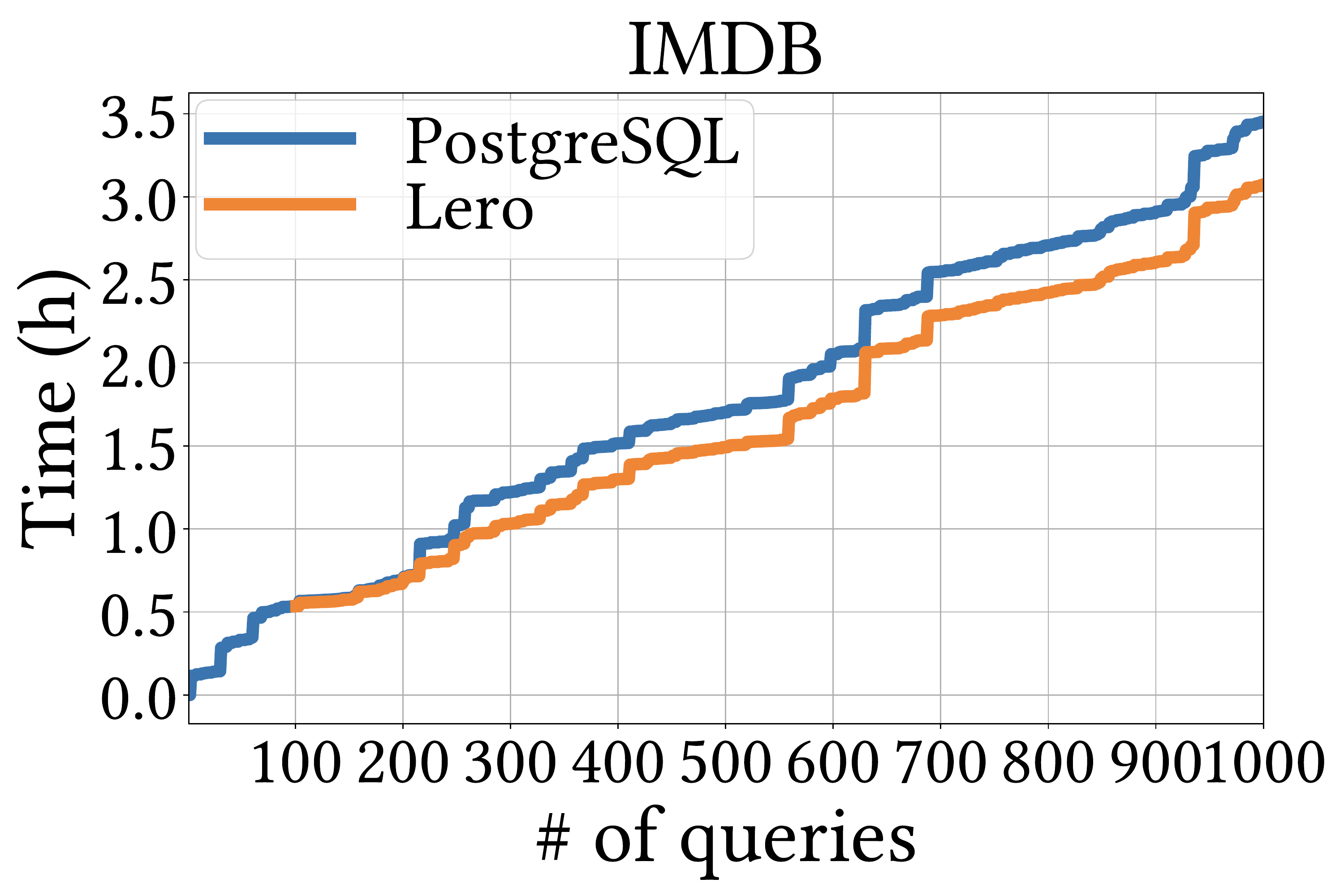}}
    \hspace{0.01\linewidth}
\subfigure{\includegraphics[width=0.48\linewidth]{CardEst 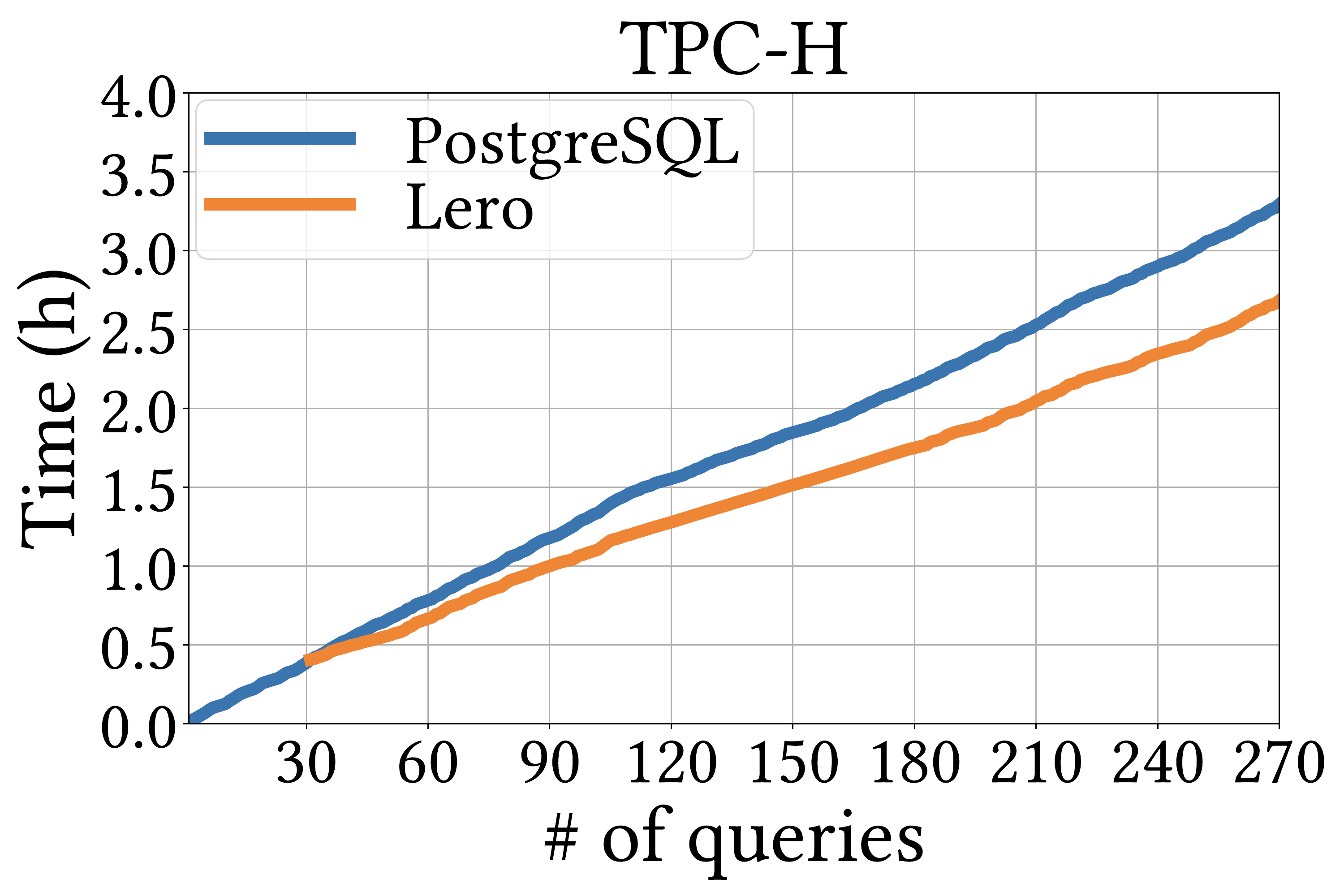}}
	\vspace{-1.5em}
	\caption{Performance curves of \rao since deployment with concurrently executed queries.}
	\label{fig: exp-overall-conc}
	\vspace{-2em}
\end{figure}
}

\section{Additional Experiments}
\label{app:addexp}
 
\subsection{Rank-Specific Analysis}
\label{sec:addexp-rank}
% Next, we analyze the plan quality in terms of the rank-specific metrics. 
For a  query optimizer, what we care about is whether the selected plan $P^{*}$ is truly the best one; if not, whether it is among the top-$k$ best plans. To this end, let $\mathcal{P}_{k}$ be the set of top-$k$ fastest plans generated by exhaustively search for each query (in terms of execution time). For different benchmarks, we record the ratio of queries for which $P^{*}$ falls into $\mathcal{P}_{k}$. Table~\ref{tab:exp-gain-rank} lists results for each query optimizer for $\mathcal{P}_{1}$ and $\mathcal{P}_{5}$. 
% 
%We find that more than $40\%$ of plans found by \rao is the exact best plan, and more than $70\%$ of plans is in the top-$5$ best plans. 
The percentage of plans selected by \rao falling into $\mathcal{P}_{1}$ and $\mathcal{P}_{5}$ is much higher than that by other query optimizers. This verifies that \rao could find plans having higher quality in terms of the rank-specific metrics.

\begin{table}[!t]
    \centering
    \scalebox{0.92}
    {
    \begin{tabular}{|c|ccc|ccc|}
    \hline
    \rowcolor{mygrey}
     \text{\textsf{Query}} & \multicolumn{3}{c|}{\text{\textsf{Plans in $\mathcal{P}_{1}$}}} & \multicolumn{3}{c|}{\text{\textsf{Plans in $\mathcal{P}_{5}$}}}\\ %\cline{3-6}
     \rowcolor{mygrey}
     \text{\textsf{Optimizer}} & \text{\textsf{STATS}} & \text{\textsf{IMDB}} & \text{\textsf{TPC-H}} & \text{\textsf{STATS}} & \text{\textsf{IMDB}} & \text{\textsf{TPC-H}} \\ \hline
     
     {PostgreSQL} & 0.137 & 0.000 & 0.167 & 0.459 & 0.336 & 0.867  \\     \hline
     {Bao} & 0.144 & 0.159 & 0.000 & 0.226 & 0.230 & 0.167  \\ \hline
     {Bao+} & 0.096 & 0.221 & \bf 0.833 & 0.171 & 0.319 & \bf 1.000 \\ \hline
     {\rao} & \bf 0.425 & \bf 0.442 & 0.733 & \bf 0.925 & \bf 0.708 & \bf 1.000  \\ \hline 
    \end{tabular}
    }
    \caption{Rank-specific metrics for different query optimizers.}
    \label{tab:exp-gain-rank}
 \end{table}
 
\begin{figure}[!h]
    \centering
    \subfigure{\includegraphics[width=0.45\linewidth]{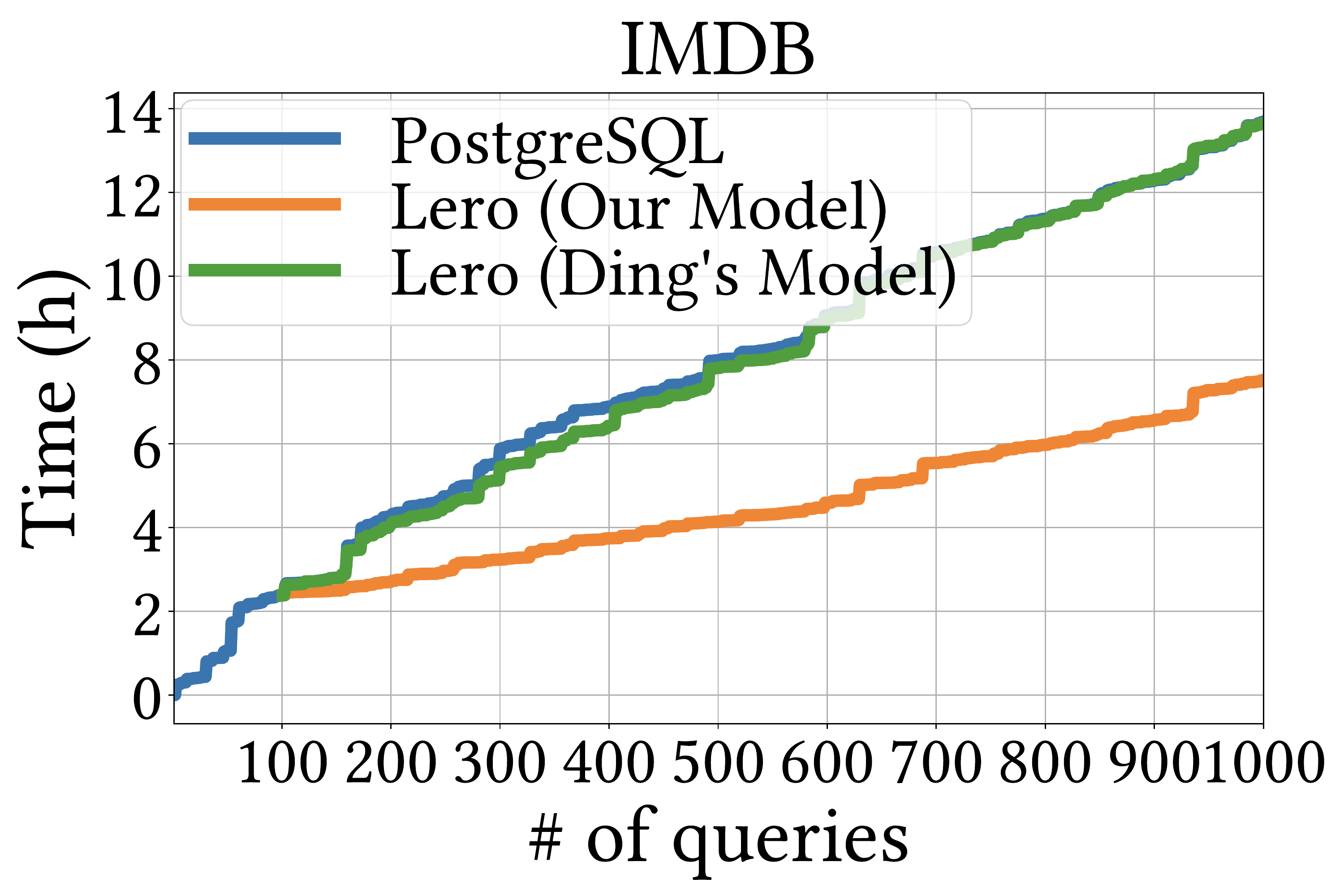}}
    \hspace{0.05\linewidth}
    \subfigure{\includegraphics[width=0.45\linewidth]{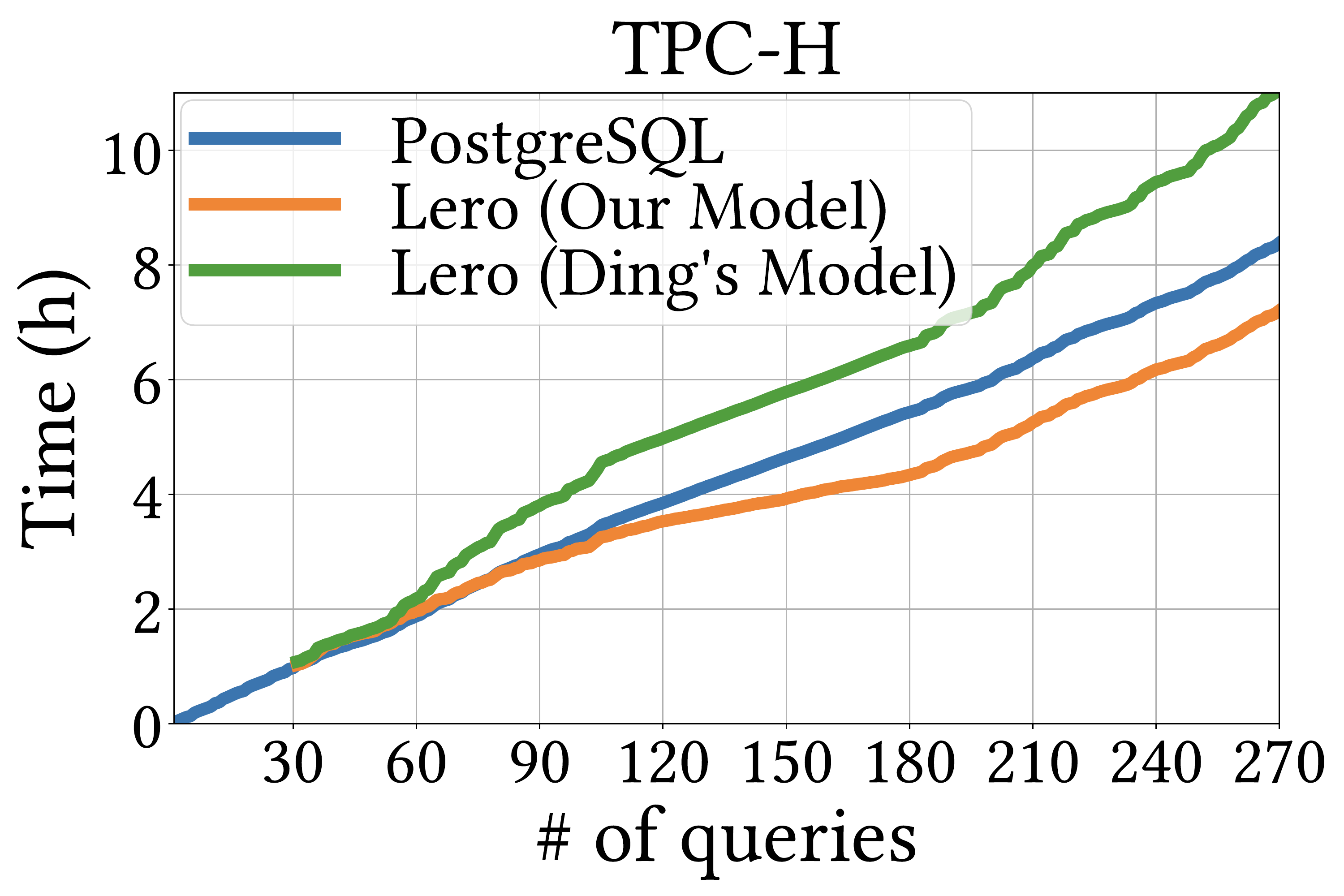}}
    \vspace{-1em}
\caption{Performance curves of \rao with different comparator models.}
\label{fig: exp-overall-embmodel}
\vspace{-1em}
\end{figure}

\subsection{Effects of Comparator Model}
\label{sec:addexp-model}

To verify the effectiveness of our comparator model in \rao, we conduct an experiment to replace \rao's plan encoding approach and comparator model with the ones in~\cite{ding2019ai}, and compare their performance. The results on IMDB and TPC-H are reported in Figure~\ref{fig: exp-overall-embmodel}. \rao with the comparator model from~\cite{ding2019ai} performs much worse than ours, which is not surprising as the model from~\cite{ding2019ai} does not consider the structural information in the plan.

Specifically, the goal of~\cite{ding2019ai} is index selection, what really matters is the amount of work done in the plan by the each type of operators, with or without indexes. Thus, a plan is flattened as a vector encoding the total estimated costs and amount of data processed by each type of operator. In \rao, the goal is plan selection, and thus we need finer-grained information on each node (e.g., operator type, estimates, tables involved) unfolded and use the tree convolution operation to capture more structural information and correlation between tables. 

\end{document}